\documentclass[11pt]{article}
\usepackage{amsmath,amsthm,amssymb,amsfonts, amsbsy, epsfig, subfig,fullpage,xr}
\usepackage[dvipsnames,usenames]{xcolor}
\usepackage{rotating}
\usepackage{multirow}
\usepackage{verbatim}
\usepackage{threeparttable}
\usepackage{booktabs}
\usepackage{bm}
\usepackage{soul}
\usepackage{graphicx}
\usepackage{wrapfig,lipsum}
\usepackage{natbib}
\usepackage{lineno,hyperref}
\usepackage[affil-it]{authblk}
\usepackage{lipsum}
\modulolinenumbers[5]

\newtheorem{theorem}{Theorem}
\newtheorem{corollary}{Corollary}

\newtheorem{proposition}{Proposition}
\newtheorem{lemma}{Lemma}
\newtheorem{example}{Example}
\theoremstyle{definition}
\newtheorem{remark}{Remark}
\newtheorem{definition}{Definition}
\newcommand{\beq}{\begin{equation}}
\newcommand{\eeq}{\end{equation}}
\newcommand{\beas}{\begin{eqnarray*}}
\newcommand{\eeas}{\end{eqnarray*}}
\newcommand{\be}{\begin{eqnarray*}}
\newcommand{\ee}{\end{eqnarray*}}
\newcommand{\bea}{\begin{eqnarray}}
\newcommand{\eea}{\end{eqnarray}}
\newcommand{\bei}{\begin{itemize}}
\newcommand{\eei}{\end{itemize}}
\newcommand{\ben}{\begin{enumerate}}
\newcommand{\een}{\end{enumerate}}
\newcommand{\bet}{\begin{theorem}}
\newcommand{\eet}{\end{theorem}}
\newcommand{\bel}{\begin{lemma}}
\newcommand{\eel}{\end{lemma}}
\newcommand{\bep}{\begin{proposition}}
\newcommand{\eep}{\end{proposition}}
\newcommand{\bed}{\begin{definition}}
\newcommand{\eed}{\end{definition}}
\newcommand{\bec}{\begin{corollary}}
\newcommand{\eec}{\end{corollary}}
\newcommand{\bex}{\begin{example}}
\newcommand{\eex}{\end{example}}

\newcommand{\Bg}{\B^{(g)}}

\newcommand{\cuxk}{\textbf{X}_{k}}

\newcommand{\cuxgk}{\textbf{X}^{(g)}_{k}}
\newcommand{\hek}{\sum\limits_{k=1}^{n}}
\newcommand{\hekg}{\sum\limits_{k=1}^{n_{g}}}
\newcommand{\hel}{\sum\limits_{l=1}^{q}}

\renewcommand{\vec}{{\rm vec}}

\def\0{\boldsymbol{0}}
\def\u{\boldsymbol{u}}
\def\v{\boldsymbol{v}}
\def\U{\boldsymbol{U}}

\def\A{\boldsymbol{A}}
\def\B{\boldsymbol{B}}

\def\H{\boldsymbol{H}}

\def\x{\boldsymbol{x}}

\def\be{\boldsymbol{\beta}}

\def\X{\boldsymbol{X}}
\def\R{\boldsymbol{R}}

\def\Z{\boldsymbol{Z}}
\def\S{\boldsymbol{\Sigma}}

\def\a{\boldsymbol{a}}
\def\W{\boldsymbol{W}}

\def\D{\boldsymbol{D}}
\def\I{\boldsymbol{I}}

\def\hA{\hat{\boldsymbol{A}}}

\newcommand{\RR}{\mathbb{R}}

\newbox\TempBox \newbox\TempBoxA

\def\pr{\textsf{P}} 
\def\Cov{\textsf{Cov}} 
\def\Var{\textsf{Var}} 

\def\text#1{\mbox{\rm #1}}

\def\underwiggle 1{
\ifmmode\setbox\TempBox=\hbox{$ 1$}\else\setbox\TempBox=\hbox{
1}\fi \setbox\TempBoxA=\hbox to \wd\TempBox{\hss\char'176\hss}
\rlap{\copy\TempBox}\smash{\lower9pt\hbox{\copy\TempBoxA}} }

\renewcommand{\baselinestretch}{1.5}

\newcommand{\myfootnote}[1]{\footnote{\renewcommand{\baselinestretch}{.75}\scriptsize #1}}

\newcommand\yx[1]{{#1}}
\newcommand\dyP[1]{{#1}}

\newcommand\yin[1]{{\color{black} #1}}

\begin{document}
	
	

\title{{\Large Testing and Support Recovery of Correlation Structures for Matrix-Valued Observations with an Application to Stock Market Data}}

%

\author[a]{\small Xin Chen }
\author[, b]{\small Dan Yang  \footnote{Corresponding authors \\ Email address: xchen528@uw.edu (Xin Chen), dyanghku@hku.hk (Dan Yang), yanxuj@hku.hk (Yan Xu), xiayin@fudan.edu.cn (Yin Xia), dongwangunc@gmail.com (Dong Wang), haipeng@hku.hk (Haipeng Shen)}}
\author[b]{\small Yan Xu}
\author[c]{\small Yin Xia $^{*, }$}
\author[d]{\small Dong Wang}
\author[b]{\small Haipeng Shen}

\affil[a]{\footnotesize Department of Statistics, University of Washington, Seattle, WA 98195, USA}
\affil[b]{\footnotesize Faculty of Business and Economics, University of Hong Kong, Pokfulam Road, Hong Kong}
\affil[c]{\footnotesize Department of Statistics and Data Science, School of Management, Fudan University, Shanghai, China}
\affil[d]{\footnotesize Department of Statistics, Rutgers University, Piscataway, NJ 08854, USA}

\date{}
\maketitle
\providecommand{\keywords}[1]{\textbf{\textit{Keywords---}} #1}

\vspace{-3em}
\begin{abstract}
Estimation of the covariance matrix of asset returns is crucial to portfolio construction. As suggested by economic theories, the correlation structure among assets differs between emerging markets and developed countries. It is therefore imperative to make rigorous statistical inference on correlation matrix equality between the two groups of countries. However, if the traditional vector-valued approach is undertaken, such inference is either infeasible due to limited number of countries comparing to the relatively abundant assets, or invalid due to the violations of temporal independence assumption. This highlights the necessity of treating the observations as matrix-valued rather than vector-valued. With matrix-valued observations, our problem of interest can be formulated as statistical inference on covariance structures under sub-Gaussian distributions, i.e., testing \yin{non-correlation} and correlation equality, as well as the corresponding support estimations. We develop procedures that are asymptotically optimal under some regularity conditions. Simulation results demonstrate the computational and statistical advantages of our procedures over certain existing state-of-the-art methods for both normal and non-normal distributions.  Application of our procedures to stock market data reveals interesting patterns and validates several economic propositions via rigorous statistical testing.



\end{abstract}




\newpage

\section{Introduction}
\label{intro.sec}

{
Understanding the covariance matrix of asset returns is of paramount
importance as asset pricing theories dictate that the distribution of
returns are related to the business cycle and consumption states, which affect the demands for holding financial
assets and generate time-varying risk premia \citep{Moskowitz-03RFS}.
However, characterizing the covariance structure of returns can be
challenging, particularly when the number of assets is large. 

We now take the perspective of a global investor and
consider an even more difficult challenge that the stock returns are from
various industries in multiple countries across the world. To enable estimation, we
employ a country-industry Kronecker structure model given the short sample
period relative to the tremendous cross section of asset
returns.
However, cautions need to be taken, as these countries can be
categorized into emerging markets and developed countries, assuming
the same covariance matrix of industry returns for these two groups
could lead to undesirable consequences in making optimal investment decision. 
As pointed out by
\citet{BekaertHarvey-95JF}, developed markets are
more financially integrated while emerging markets
are more financially segmented, thus industries may
have different amount of systematic risk depending on the level of
segmentation. 

As the market return is an aggregation of all industries
returns, therefore an industry's systematic risk is simply the value-weighted
average of its covariance with other industries, over its own
variance. Therefore, we focus on the correlation matrices and 
\dyP{conjecture} that they should be
significantly different between the two groups of countries.
Specifically, emerging markets are characterized by frequent regime
switches, and sudden changes of fiscal, monetary and trade policies \citep{AguiarGopinath-07JPE}. When these
economic policies change frequently in an unanticipated way in
emerging markets, they tend to make cyclical sectors more pro-cyclical
than those in developed markets. Furthermore, \citet{kohn2018trade}
show that emerging economies produce more commodities than they consume while developed markets do
not. Therefore we also expect to detect larger comovements of commodity industries returns with
others in emerging countries. Finally, due to different demographic patterns
\citep{DellaVignaPollet-07AER}, we also expect in emerging countries,
recreative business industries co-vary more with the market returns.

Combined, these propositions all highlight the necessity of rigourous
statistical testing of the equality of the correlation matrices from the two groups of countries.
In the literature, researchers often treat the returns of multiple
industries as vector-valued observations over time in each country
\citep{FamaFrench-97JFE,HongTorousValkanov-07JFE}. With the
vector-valued approach, as typically the number of assets far exceeds
the length of the time series and the number of countries, estimating
the correlation matrix can be so challenging that certain potentially problematic assumption of the temporal independence must be made.} In this case, adopting the vector-based approach of two sample test, such as {\citet{li2012two,cai2013two,cai2016inference,chang2017comparing,zheng2019test}}, will be either infeasible due to the small number of observations, or invalid due to the violation of temporal independence.

The goal of this article is to perform hypothesis test of the equality of the two correlation matrices by considering observations that are matrix-valued. Compared to the conventional vector-valued observations, the matrix-valued observations add one more dimension, corresponding to the time domain. The temporal dimension is allowed to have {a wide range of} dependence, which is more {flexible} than in the vector-based approach. The addition of the temporal dimension also alleviates the problem of small sample size and insufficient length of the time series as seen later.

{To be specific, let $g=1,2$ correspond to the two groups of countries, emerging and developed, respectively.} There are $n_1$ (resp. $n_2$) countries in the emerging (resp. developed) group. Denote $\X_k^{(g)}$, for $k=1,...,n_g$, the matrix of returns for country $k$ in group $g$. Each matrix is of size $p \times q$ when there are $p$ industries and $q$ time points. 
We are interested in the inference on the {correlation} matrix {$\text{Cor}(\vec(\X_k^{(g)}))$}, 
where $\vec(\cdot)$ is the vectorization operation that stacks the columns of a matrix into a long vector.

\dyP{As the {correlation} matrix is of enormous size $pq\times pq$ while the sample size is only $n_g$, it is necessary to make further assumption.
We consider the Kronecker product model for the covariance matrix $\Sigma^{(g)} = \Cov(\vec(\X_k^{(g)})) = \B^{(g)}\otimes\A^{(g)},~g=1,2$. More information on the Kronecker product model, such as references, motivation, and applications, is given in Section \ref{sec-kronecker}. Here, $\A^{(g)}$ is the covariance matrix of size $p\times p$ for the covariances between $p$ industries in group $g$ and $\B^{(g)}$ is the $q\times q$ covariance matrix along the temporal dimension in group $g$.
As such, the covariance of the observed matrix $\X_k^{(g)}$ is decomposed into the product of the industrial covariance and temporal covariance.
This Kronecker product structure reduces the number of unknown parameters from $O(p^2q^2)$ to $O(p^2+q^2)$. Furthermore, much smaller sample size is needed: without such {structural assumption}, a sample size of $n_g\ge pq$ is necessary to make the sample covariance matrix full rank, while with the structure, the sample size is sufficient as long as $n_gp\ge q$ and $n_gq\ge p$.
}

Under the Kronecker product model, our goal is to test the equality of the {correlation} matrices of the industries by considering $\B^{(g)}$ as nuisance parameters. 
{Consider} the correlation matrices:
$\R_{A}^{(g)} = (\D_A^{(g)})^{-1/2}\A^{(g)}(\D_A^{(g)})^{-1/2}$, for $g=1,2$, where
$\D_A^{(g)}$ is the diagonal matrix consisting of the diagonal entries
of $\A^{(g)}$.  As such, we test
\begin{equation}\label{eq1:two-sample}
  H_0:
\R_{A}^{(1)}=\R_{A}^{(2)} \text{ versus } H_1:
\R_{A}^{(1)}\ne \R_{A}^{(2)}.
\end{equation}
This is referred to as the {\bf two-sample
hypothesis test} of the equality of the correlation
matrices of the two groups with matrix-valued observations.

Furthermore, it is also of interest to test whether the columns of the matrix-valued observations are \yin{uncorrelated}. This is important because, if indeed there is no temporal correlation, then the vector-based  approach can be implemented. This goal can be achieved by testing
\begin{equation}\label{eq1:one-sample-intro}
H_{0,B,g}: \B^{(g)} \text{ is diagonal versus } H_{1,B,g}: \B^{(g)} \text{ is not diagonal},
\end{equation}
within group $g$.
Similarly, \yin{non-correlation} of the industries within either
group can also be tested via $H_{0,A,g}: \A^{(g)}$ is diagonal.
These are referred to as the {\bf one-sample hypothesis test} of the
\yin{non-correlation} of the columns, or rows, respectively, of the matrix-valued observations.

Moreover, when the null hypothesis of the one-sample
hypothesis test is rejected, it is of further interest to identify
which months or which industries have non-zero correlations; similarly, when
the null hypothesis of the two-sample hypothesis test is rejected, it
is important to further identify which industries have
significantly different correlations among emerging
countries versus developed countries. These are referred to as
{\bf support recovery} problems.

As a prelude, in our real data application, 
the one-sample null hypothesis, $H_{0,B,g}: \B^{(g)}$ is diagonal, is rejected by our method introduced in Section~\ref{one.sec}, suggesting the existence of significant temporal correlation. This implies that we cannot use the aforementioned vector-based two-sample tests, and manifests the need of developing a method for the two-sample
hypothesis test directly using matrix-valued observations (Section~\ref{two.sec}). According to our analysis (Section~\ref{real.sec}), the two-sample null hypothesis, $H_0:
\R_{A}^{(1)}=\R_{A}^{(2)}$, is also rejected, so we indeed identify
significant differences in correlations across the two groups of countries.
Furthermore, our support recovery analysis finds consistent evidence with existing economic propositions \dyP{while vector-based method ignoring the temporal correlation contradicts with economic theories}.

\subsection{Literature Review}

\dyP{
For the estimation and inference on the covariance/correlation/precision matrix, there have been numerous efforts and the list below is far from being comprehensive.
A succinct summary for vector-valued observations is in place.} From the aspect of estimation, a number of methods were proposed to estimate the covariance/correlation matrix 
\citep[e.g.]{Bickel2008High,rothman2009generalized,cai2011adaptive,Cai2012Optimal,han2013optimal,cai2016inference}. Meanwhile, various methods of estimating the precision matrix have also been proposed \citep[e.g.]{meinshausen2006high,yuan2007model,friedman2008sparse,ravikumar2011high,cai2011constrained}, and some other works extend the single precision matrix to multiple precision matrices \citep[e.g.]{danaher2014joint,zhu2014structural,cai2016joint}. From the other aspect of inference, hypothesis testing procedures for vector data have been developed recently. In particular, \citet{cai2011limiting,li2012two,cai2013two,cai2013optimal,cai2016inference,chang2017comparing,zheng2019test}, for example, considered the one-sample or two-sample covariance/correlation matrix testing problem in high-dimensions. To investigate the graphical models, \citet{liu2013gaussian} and \citet{xia2015testing}, for example, proposed procedures to test the property of the precision matrix under one-sample or two-sample settings. 

For matrix-valued observations,
\dyP{
to the best of our knowledge, most of the existing works require matrix normal distribution, which combines the Kronecker product model and normality. Under this distribution, \citet{leng2012sparse,yin2012model,zhou2014gemini,qiu2016joint,han2016sparse,zhu2018multiple} proposed methods to estimate the precision matrix and inspect the graphical structure; \citet{xia2017hypothesis,xia2018matrix} studied the one-sample and two-sample hypothesis testing of the structure of the precision matrices respectively; \citet{dawid1981some, dutilleul1999mle,werner2008estimation,hoff2011separable} considered estimation of covariance matrices.
}

\dyP{Table \ref{literature} summarizes the status of literature related to the hypothesis testing on covariance or precision matrix for both vector-valued and matrix-valued data under one-sample and two-sample regimes. The table in Online Supplement \ref{table.sec} is more comprehensive with key references listed for each scenario. This article fills in the blank of hypothesis testing of correlation structures for the {matrix-valued observations}. 
There is a pressing need for such hypothesis testing in various finance and economics studies, such as the portfolio management challenge faced by a global investor.

In addition, for matrix-valued observations, the previous literature all adopted matrix normal distribution, while we extend the framework to matrix sub-Gaussian distribution (see Section \ref{theo.sec} for the details) for theoretical development.
Technically, such extension makes the article considerably different from those methods that aim at testing of precision/partial correlation matrices and are thus highly dependent on the Gaussian assumption. As a result, the technical tools are much more involved and are not directly available in the literature due to the non-equivalence of ``uncorrelated'' and ``independent''. A variation of Hanson-Wright inequality is derived, which is of separate interest to the literature of matrix-valued and tensor-valued observations in general.

Numerically, our proposed methods are more advantageous than the twist of precision matrix based methods:
under the one-sample normal setting, our methods are similar to those existing ones statistically;
under the two-sample normal setting, our methods are more powerful than them;
under the one-sample and two-sample heavy-tailed settings, existing methods cannot control the size while our methods can; lastly, we offer much better computational performance in all settings.
}


\begin{table}[h]
  \begin{center}
\begin{tabular}{|l|l|c|c|}
\hline
          &          & Vector-valued data & Matrix-valued data  \\\hline
Covariance or&One-sample&$\surd$& This article\\
Correlation matrix&Two-sample&$\surd$& This article\\\hline
\multirow{2}{*}{Precision matrix} &One-sample&$\surd$&$\surd$\\
          &Two-sample&$\surd$&$\surd$\\\hline
\end{tabular}
    \caption{Summary of the literature on the hypothesis testing for both vector-valued and matrix-valued data under one-sample and two-sample regimes.}
  \label{literature}
  \end{center}
\end{table}

%

\vspace{-2em}
\subsection{The Kronecker Product Model}
\label{sec-kronecker}

Matrix-valued or tensor-valued data are ubiquitous nowadays. When dealing with such data, and sometimes even vector-valued data, Kronecker product structure has been a powerful tool because of its ability to approximate an arbitrary matrix \citep{cai2019kopa} and reduce dimensionality. \citet{HafnerLintonTang-19JE} used Kronecker product to approximate the covariance matrix for vector-valued data and aimed to estimate the approximated covariance matrix. \citet{chen2018autoregressive} investigated matrix autoregressive models where the coefficient matrix has Kronecker product structure. For tensor-valued time series, \citet{wang2019,chen2019constrained,chen2019matrix,chen2019factor} assumed that the tensor factor model has a signal that exhibits Kronecker structure. \citet{aston2017tests,constantinou2017testing} performed a test of the separability of terms in the Kronecker product. \citet{molstad2019penalized} proposed an algorithm to fit the linear discriminant analysis model with Kronecker product. \dyP{There are countless works on tensor regression and tensor decomposition with Kronecker product.
These articles demonstrate a wide range of applications in finance, economics, engineering, neuroimaging, geophysics, and many more.

The Kronecker product structure naturally arises when data have multiplicative structure. Suppose for a generic country, its return follows the interactive effect model of \citet{bai2009panel}, $x_{ij}=u_i v_j$, where $x_{ij}$ is the return of industry $i$ at time $j$. This is equivalent to matrix $\X=\u\v'$ where $\u= (u_1,\ldots,u_p)'$ and $\v = (v_1,\ldots,v_q)'$, which implies that $\vec(\X) = \v\otimes\u$. Suppose both $\u$ and $\v$ are random, mean zero, and independent of each other, then $\Cov(\vec(\X)) = \Cov(\v)\otimes \Cov(\u)$. Here $\u$ is the industry factor and $\v$ is the time series factor. Hence the covariance of $\vec(\X)$ is separable and consists of the product of cross-sectional industry dependence and temporal dependence. The argument above applies to other applications whenever the multiplicative random factor model holds. More generally, there might be multiple independent components/factors, in which case the covariance matrix can be decomposed into the summation of Kronecker products. Such extension remains an interesting and open question.

The following two concrete examples offer more insights on the applicability of the Kronecker product model in asset pricing studies.
}
\yx{\citet{BrandtSanta-Clara-06JF} consider augmented asset space in optimal portfolio problem.
Then solutions to the optimal portfolio weights naturally involve estimates of Kronecker product of covariance of returns and covariance of
firm characteristics. \citet{BrandtSanta-ClaraValkanov-09RFS} specify portfolio weights as linear functions of firm characteristics.  To accommodate possible time variation in the coefficients of the portfolio
policy, the impact of the characteristics on the portfolio weight may be allowed to vary with the realization of the state variables. Again, with the independence between firms
characteristics and business cycle variables, estimating the portfolio weights also involves Kronecker product of covariances of these two sets of variables.}

The random multiplicative factor model is only a sufficient, not necessary, condition. For example, \citet{HafnerLintonTang-19JE} estimates a Kronecker product of covariance matrix of S\&P500 stock returns 
{where an apparent Kronecker structure is clearly missing. Still, it} is interesting to see that they find 
{an under-specified model with Kronecker structure} performs as well as conventional shrinkage estimator's \citep{LedoitWolf-04JMA} 
when building minimum variance portfolios. In such case, the Kronecker product of the covariance matrix can be viewed as an approximation of the true yet high dimensional covariance matrix.

\dyP{The Kronecker product assumption in the stock market application mentioned above is indeed verified via the hypothesis testing method on separability} of \citet{aston2017tests}. See Section \ref{real2.sec} for the details.


\subsection{Roadmap}
The rest of the article is organized as follows. Section \ref{one.sec} is devoted to the one-sample global hypothesis testing on the \yin{non-correlation} of the columns or the rows of matrix-valued observations and the recovery of the dependent entries when the global hypothesis test is rejected. Section \ref{two.sec} is dedicated to the two-sample global hypothesis testing of the equality of two correlation matrices along one dimension of the matrix-valued observations (in two groups), and the support recovery of the difference of the two correlation matrices. Section \ref{theo.sec} establishes the theoretical properties of these procedures \dyP{on hypothesis tests} for both one-sample and two-sample settings.  The numerical comparison of our procedures with existing ones via simulation is provided in Section \ref{sim.sec} and the real data analysis of the aforementioned stock returns data is given in Section \ref{real.sec}. \dyP{Section \ref{sec-conclusion} concludes.
More theorems on support recovery methods, the proofs, and additional simulations are delegated to \yin{the Online Supplement}.}

\section{One-Sample Testing of \yin{Non-correlation}}
\label{one.sec}
To formulate the stock return example in terms of the matrix-valued two-sample hypothesis testing problem as introduced in \eqref{eq1:two-sample}, we shall first check whether the \yin{ non-correlation} assumption holds for the temporal dimension. Hence,  we start with the one-sample testing of \eqref{eq1:one-sample-intro}, which is easier to comprehend due to its simple structure and notation. 
We omit the superscript that denotes the group membership.

Suppose there are $n$ i.i.d. centered random matrix-valued observations $\{\X_1,...,\X_{n}\}$, each with dimension $p\times q$, {from the matrix sub-Gaussian distribution as defined in Section \ref{theo.sec} with the $p\times p$ matrix $\A$ and $q\times q$ matrix $\B$ being the covariance matrices associated with the rows and columns respectively.}
The vectorization $\vec(\X_k)$ is a vector of length $pq$ following a multivariate sub-Gaussian distribution with mean zero and a covariance matrix of the form $\S={\sigma^2}\B\otimes\A$. Denote $\A=(a_{i,j})_{p\times p}$ and $\B=(b_{i,j})_{q\times q}$. Without loss of generality (WLOG), we derive the testing procedure below for testing \yin{non-correlation} relating to the matrix $\A$ and assume that $\sigma = 1$. Note that we can simply transpose the observation $\X_k$ so that the roles of $\A$ and $\B$ are switched and the procedure to test $\A$ can be used to test $\B$ after the transpose.

Our goals are to test the null hypothesis globally
\begin{equation}\label{eq:one-sample-global}
H_0: ~\A  \text{  is diagonal versus  } H_1:~\A \text{ is not diagonal},
\end{equation}
and to identify nonzero entries $a_{i,j}\ne 0$, both of which are invariant up to a constant. As such, even though $\A$ and $\B$ are not identifiable as $c^{-1}\A$ and $c\B$ will lead to the same {distribution} for any positive scalar $c$, this has no effect on the global hypothesis testing procedure of Section \ref{test.sec} and the support recovery approach of Section \ref{support.sec}. Throughout the paper, we use $c,c',c_0,c_1$ to denote constants whose values may change from line to line.

\subsection{Global Testing Procedure}
\label{test.sec}

To test the property of $\A$, it is necessary to construct the test statistic based on an estimate of $\A$. A naive estimate of $\A$ is $\tilde{\A}=\frac{1}{nq}\sum_{k=1}^n\X_k\X_k'$
$=\frac{1}{nq}\sum_{k=1}^n\sum_{l=1}^q\X_{k,\cdot l}\X_{k,\cdot l}'$, where $\X_{k,\cdot l}$ denotes the $l$-th column of matrix $\X_k$. 
This naive estimate is the same as the sample covariance matrix for vector-valued observations if we treat $\X_{k,\cdot l}$, for $k=1,\ldots,n$ and $l=1,\ldots,q$, as $nq$ i.i.d. observations. But note that these $nq$ observations are only uncorrelated when $\B$ is a multiple of an identity matrix, which requires no temporal correlation.
According to the {structural assumption}, the covariance matrix of any column $\X_{k,\cdot l}$ is proportional to the matrix $\A$. 
It follows that, for the naive estimate $\tilde{\A}$, there exists a constant $c>0$ such that 
$\tilde{\A}/c$ is an unbiased estimate of $\A$. Similarly, $\tilde{\B}/c'$ is an unbiased estimate of $\B$ with a proper $c'$, where
\beq
\label{eq:Btilde}
\tilde{\B}= \frac{1}{np}\sum_{k=1}^n\X_k'\X_k.
\eeq
However, the above naive estimation is not efficient and can be improved further as follows.

Consider $\Z_k = \X_k\B^{-1/2}$, for $k=1,...,n$, where right-multiplying matrix $\B^{-1/2}$ can be thought of as {a step of pre-whitening}.
Because of {the assumed structure}, we obtain that all of the columns of $\Z_k$ are 
\yin{uncorrelated} with covariance $\A$.
Therefore, when $\B$ is known, $\frac{1}{nq}\sum_{k=1}^n\Z_k\Z_k'$ is the most efficient and oracle estimate of $\A$. In practice, $\B$ is often unknown, in which case, plugging in a legitimate estimate of $\B$, such as $\tilde{\B}/c'$, is a natural approach, which leads to the following estimate of $\A$,
\begin{equation}
\label{eq:ahat-one}
(\hat{a}_{i,j})=:\hat{\A}=\frac{1}{nq}\sum_{k=1}^n\X_k(\tilde{\B}/c')^{-1}\X_k',
\end{equation}
where $\tilde\B$ is defined in \eqref{eq:Btilde}. {Note that when $np>q$, $\tilde\B$ defined above is invertible with probability one.}

To test whether $\A$ is diagonal in \eqref{eq:one-sample-global}, it is tempting to consider the magnitudes of all the off-diagonal entries of $\hA$ in \eqref{eq:ahat-one}. However, they cannot be used directly because of different levels of variability. To make them comparable, it is necessary to standardize.
Since \eqref{eq:ahat-one} can be re-expressed as
$
\hat{\A} = \frac{1}{nq}\sum_{k,l}\Big(\X_k(\tilde{\B}/c')^{-1/2}\Big)_{\cdot l}\Big(\X_k(\tilde{\B}/c')^{-1/2}\Big)_{\cdot l}',
$
it has the oracle counterpart when $\B$ is known
$
\hat{\A}^{o} = \frac{1}{nq}\sum_{k,l}\Big(\X_k\B^{-1/2}\Big)_{\cdot l}\Big(\X_k\B^{-1/2}\Big)_{\cdot l}',
$
whose entries are
$
\hat a_{i,j}^{o} = \frac{1}{nq}\sum_{k,l}\Big(\X_k\B^{-1/2}\Big)_{i,l}\Big(\X_k\B^{-1/2}\Big)_{j,l}'.
$
Then, it is natural to define the relevant population variances as
\beq
\label{eq:theta-one}
\theta_{i,j}  = \Var\Big((\X_k\B^{-1/2})_{i,l}(\X_k\B^{-1/2})_{j,l}\Big) = \Var\Big((\Z_k)_{i,l}(\Z_k)_{j,l}\Big),
\eeq
for all $i,j$. Note that the definition of $\theta_{i,j}$ above does not depend on $l=1,...,q$ nor $k=1,...,n$.
The sample estimates of these variances can be obtained by
\begin{equation}
\label{eq:thetahat-one}
\hat{\theta}_{i,j}=\frac{1}{nq}\sum_{k=1}^n\sum_{l=1}^q\Big{[}\big(\X_k(\tilde{\B}/c')^{-1/2}\big)_{i,l}\big(\X_k(\tilde{\B}/c')^{-1/2}\big)_{j,l}-\hat{a}_{i,j}\Big{]}^2.
\end{equation}
So the variance of $\hat a_{i,j}$ can be estimated by $\hat{\theta}_{i,j}/(nq)$.
Similar spirit of the estimation of the variances has been used in \citet{cai2011adaptive} and \citet{cai2013two}, where the observations are vector-valued and do not need pre-whitening or the plugged-in estimate $\tilde{\B}/c'$, while ours are matrix-valued and the estimation is more involved.

The standardized statistics are readily defined as
\beq\label{eq:Mij-one}
M_{i,j}=\frac{\hat{a}_{i,j}^2}{\hat{\theta}_{i,j}/(nq)}, \text{ $1\leq i<j \leq p$},
\eeq
where $\hat{a}_{i,j}$ and $\hat{\theta}_{i,j}$ are defined in \eqref{eq:ahat-one} and \eqref{eq:thetahat-one} respectively.
The $M_{i,j}$'s are on the same scale and can be compared together. It is also seen that $M_{i,j}$ doesn't depend on $c'$ as the constant $c'$ in the numerator and denominator of \eqref{eq:Mij-one} is cancelled. WLOG, we set $c' = 1$ for the rest of the article.

Note that the null hypothesis $H_0:~ \A$ is diagonal is equivalent to $H_0:$ all of the off-diagonal entries of $\A$ are zero, and hence further equivalent to $H_0:$ the maximum of all the off-diagonal entries is zero, i.e., $H_0:~ \max_{1\leq i< j\leq p}|a_{ij}|=0$. Therefore, it is natural to construct the following test statistic,
\begin{equation}
\label{eq:Mn-one}
M_n=\max_{1\leq i<j\leq p}M_{i,j},
\end{equation}
where $M_{i,j}$ is the standardized statistic for the $i,j$-th entry in \eqref{eq:Mij-one}. Under the alternative hypothesis, there exists at least one non-zero off-diagonal entry $a_{i,j}\ne 0$, whose associated statistic $M_{i,j}$ is large, and the maximum test statistic $M_n$ will be large. Therefore, the null hypothesis should be rejected for large value of the test statistic $M_n$.

To perform hypothesis test based on the test statistic $M_n$, we further need to establish its null distribution. The exact theoretical property of its limiting behavior will be discussed in details in Section \ref{theo.sec}. For now, we can still obtain some intuition of the critical value. Roughly speaking, under the null hypothesis, each $M_{i,j}$ is approximately the square of a standard normal random variable due to standardization, and under certain conditions, the $M_{i,j}$'s are only weakly correlated with each other. So loosely speaking, the test statistic $M_n$ is the maximum of $\binom{p}{2}$ squared normals that are weakly dependent. Since the extreme value of {the square of} $n$ i.i.d. normal random variables is close to $2\log n$, $M_n$ is close to $2\log \binom{p}{2}\approx 4\log p$ under $H_0$. To be precise, theorems in Section \ref{theo.sec} will show rigourously that under the null distribution $H_0$ and certain regularity assumptions, $M_n-4\log p+\log \log p$ converges to a Gumbel distribution. Due to this limiting distribution, for any significance level $0<\alpha < 1$, we can define the global test $\Phi_\alpha$ by
\beq
\label{eq:phi-alpha-one}
\Phi_{\alpha}= I( M_{n}\geq q_{\alpha}+4\log p-\log\log p),
\eeq
where $I(\cdot)$ is the indicator function. Here, the quantity
\beq
\label{eq:q.alpha}
q_\alpha = -\log (8\pi) - 2 \log\log(1-\alpha)^{-1},
\eeq
is the $1-\alpha$ quantile of the Gumbel distribution with the cumulative distribution function (cdf)
$\exp(-(8\pi)^{-1/2}\exp(-{x}/{2}))$.
The null hypothesis $H_0:~ \A$ is diagonal is rejected whenever $\Phi_\alpha =1$.

\begin{remark}
\label{rmk-other-estimation-B}
There are many appropriate choices for the estimation of $\B$ besides the simple sample estimator as long as it satisfies the equation \eqref{eq:key-proof} in our proof. This may lead us to use, for example, the banded estimator in \cite{rothman2010new}, the adaptive thresholding estimator in \cite{cai2011adaptive}, etc., if we have the prior information on the structure of $\B$.
\end{remark}

\dyP{
\begin{remark}
Note that we define the test statistic $M_n$ based on the standardized statistics $M_{i,j}$. The standardization achieves two purposes simultaneously: first, it eliminates the necessity of estimation of $c'$, even though the Kronecker product is only identifiable up to a constant; second, the statistics $M_{i,j}$'s are scale-free and maximum of them can be taken. The current definition of $M_{i,j}$ is the ratio of the squared covariance estimate $\hat a_{i,j}^2$ and its associated variance estimate. An alternative definition is the ratio of the squared correlation estimate $\hat r_{i,j}^2 = \hat a_{i,j}^2/(\hat a_{i,i}\hat a_{j,j})$ and its corresponding variance estimate $\hat{\vartheta}_{i,j}/(nq) = \hat\theta_{i,j}/(\hat a_{i,i}\hat a_{j,j})/(nq)$, which is exactly the same as the covariance version of the definition due to cancellation of $\hat a_{i,i}\hat a_{j,j}$ in both the numerator and denominator. Therefore, for the one-sample case, the covariance version and the correlation version are equivalent. Nevertheless, as a prelude, in Section 3 on the two-sample test of the correlation matrix equality, the standardized statistics are defined with the correlation version, not the covariance version. This is because we are interested in the equality of the two correlation matrices, and in the two-sample case the two versions of definitions are not equivalent any more since difference is taken before standardization.
\end{remark}
}


\begin{remark}
\label{rmk-sparse-alternative}
Since $M_n$ is the maximum of $M_{i,j}$, the test $\Phi_\alpha$ is best suited for the case when the alternative hypothesis is sparse, that is, when only a small number of the off-diagonal entries of the covariance matrix are large. As long as one of the off-diagonal entries is large enough, the test will reject the null hypothesis. This test does not assume any other structure of the alternative hypothesis. In Section \ref{theo.sec}, we will show that this test is optimal against sparse alternatives. Note that, when the alternative is dense and many small off-diagonal entries exist, the proposed test $\Phi_\alpha$ is less capable of rejecting the null.  Nevertheless, the large body of literature on portfolio construction typically assumes i.i.d excess returns and all serial correlations are zero (for a survey, see \citet{Brandt-09}). In practice, the temporal correlations are more apparent in daily or even weekly returns due to non-synchronous trading or the bid-ask bounce effect, but much less so at monthly frequency so most of them may not be different from zero \citep{CampbellLoMacKinlaly97}.
\end{remark}

\dyP{
\begin{remark}
For simplicity of notation, we have assumed that the observations $\{\X_1,\cdots,\X_{n}\}$ are from distributions with mean zero. The proposed methods in this article can be easily extended to the cases with nonzero means  \citep[e.g.]{chen2019graph}.
\end{remark}
}


\subsection{Support Recovery Procedure}
\label{support.sec}
We have focused on the test of the \yin{non-correlation} of the rows of $\X_k$ by testing globally whether all of the off-diagonal entries of the row covariance matrix $\A$ are zeros in Section \ref{test.sec}. If the null hypothesis is rejected, it is of great value to locate the places where the covariances are not zero. Taking the stock return data for example, if the \yin{ non-correlation} of the months is rejected (the matrix-valued observations need to be transposed before feeding into the testing procedure), one may want to identify which months are highly correlated, and if the \yin{non-correlation} of industries is rejected, it might be interesting to know which industries are correlated. Another example is brain imaging analysis, where the matrix-valued observations for patients are spatial-temporal data \citep[e.g.]{xia2017hypothesis,xia2018matrix}, and it is worthwhile investigating further how voxels of the brains are correlated after the rejection of \yin{non-correlation} of voxels. 

This problem of support recovery can be thought of as simultaneous testing of whether the off-diagonal entries of the covariance matrix $\A$ are zero. Let the support of $\A$, neglecting the diagonal entries, be
\beq
\label{eq:support.one.true}
\Psi=\Psi(\A)=\{(i,j): a_{i,j}\neq0, 1\leq i<j\leq p\}.
\eeq
{The same intuition that leads to \eqref{eq:phi-alpha-one} can be used to construct the support estimate with the following threshold, }
\beq
\label{eq:support.one}
\hat{\Psi}(\tau)= \{(i,j):M_{i,j}\geq \tau\log p,~~1\leq i<j\leq p\},
\eeq
where the $M_{i,j}$'s are previously defined in \eqref{eq:Mij-one}, and $\tau$ is a threshold constant. Online supplement Section \ref{sec-support} will show that: when $\tau=4$, the probability of exact recovery goes to 1 asymptotically if the nonzero entries are large enough; 
a smaller choice $\tau<4$ will fail to recover the support under certain conditions;  therefore $\tau=4$ is optimal.

\begin{remark}
{We aim for the asymptotic exact recovery of the support in this and the following sections for the matrix-valued scenarios. The direct application of vector-valued support recovery methods in existing literatures will lead to poor results as shown in the simulation section. }
\end{remark}



\section{Two-Sample Testing of Correlation Matrix Equality}
\label{two.sec}
Having derived the procedure for the (one-sample) testing of \yin{non-correlation}, we can extend the approach to the two-sample scenario of testing the equality of two correlation matrices.
Following the same notation as in the introduction, {we have i.i.d. matrix-valued observations from matrix
	sub-Gaussian distribution for two groups $g = 1,2$.} Considering the definition of the correlation matrices for the two groups in the introduction, we wish to test
\begin{equation}\label{eq:two-sample}
H_0^*:
\R_{A}^{(1)}=\R_{A}^{(2)} \text{ versus } H_1^*:
\R_{A}^{(1)}\ne \R_{A}^{(2)}.
\end{equation}
Hereafter, we use the superscript $^*$ to distinguish the quantities that are of relevance to the two-sample case from the one-sample case. 

Given the {centered} observations $\{\X_1^{(1)},\cdots,\X^{(1)}_{n_{1}}\}$ and $\{\X_1^{(2)},\cdots,\X^{(2)}_{n_{2}}\}$, as discussed for the one-sample case in Section \ref{one.sec}, we can 
construct the estimates of the covariance and correlation matrices for the two groups separately,
\begin{align}
\label{eq:ahat-two}
&(\hat{a}_{i,j}^{(g)}) =:  \hat{\A}^{(g)} = \frac{1}{n_{g}q} \sum\limits_{k=1}^{n_g} \cuxk^{(g)}(\tilde{\B}^{(g)})^{-1} (\cuxk^{(g)})^{'}, \\
\label{eq:rhat-two}
&(\hat{r}_{i,j}^{(g)}) =: \hat{\R}_{A}^{(g)} = \left(\frac{\hat{a}_{i,j}^{(g)}}{(\hat{a}_{i,i}^{(g)}\hat{a}_{j,j}^{(g)})^{1/2}}\right),
\end{align}
where $\tilde{\B}^{(g)}=\frac{1}{n_gp}\sum_{k=1}^{n_g}(\X_k^{(g)})'\X_k^{(g)}$ is the naive estimate of $\B^{(g)}$.
Again, we cannot directly make inference based on $\hat{r}_{i,j}^{(1)} - \hat{r}_{i,j}^{(2)}$, because they are heteroscedastic. To make them homoscedastic, define the entry-wise population variance and the sample counterpart similarly as in \eqref{eq:theta-one} and \eqref{eq:thetahat-one},
\begin{align*}
&\theta_{i,j}^{(g)} = \Var\Big((\cuxgk (\Bg)^{-1/2})_{i,l}(\cuxgk (\Bg)^{-1/2})_{j,l}\Big), \\ &\hat{\theta}_{i,j}^{(g)}=\frac{1}{n_{g}q}\sum\limits_{k=1}^{n_g}\sum\limits_{l=1}^q
\Big{[}\big(\X_k^{(g)}(\tilde{\B}^{(g)})^{-1/2}\big)_{i,l}\big(\X_k^{(g)}(\tilde{\B}^{(g)})^{-1/2}\big)_{j,l}-\hat{a}_{i,j}^{(g)}\Big{]}^2.
\end{align*}
As such, the variance of $\hat{r}_{i,j}^{(g)}$ can be estimated by $\hat{\vartheta}_{i,j}^{(g)}/(n_{g}q)$, where
$\hat{\vartheta}_{i,j}^{(g)} = \frac{\hat{\theta}_{i,j}^{(g)}}{\hat{a}_{i,i}^{(g)}\hat{a}_{j,j}^{(g)}}.$
Consequently, the variance of $\hat{r}_{i,j}^{(1)} - \hat{r}_{i,j}^{(2)}$ can be estimated by $\hat{\vartheta}_{i,j}^{(1)}/(n_{1}q) + \hat{\vartheta}_{i,j}^{(2)}/(n_{2}q)$.
{Note that, for vector-valued observations, to test the equality of the correlations from two populations, 
	\citet{cai2016inference} estimated the variance by a careful investigation of the Taylor expansion in the calculation of correlation from covariance, and \citet{cai2016large} introduced a variance stabilization method based on Fisher's $z$-transformation. Our approach is different from both methods.}

When we focus on a single entry of the hypothesis in \eqref{eq:two-sample} such as $\hat{r}_{i,j}^{(1)} = \hat{r}_{i,j}^{(2)}$, in accordance with the two-sample $t$-test with unequal variances for i.i.d. random variables, it is natural to define the standardized statistic as
\beq
\label{eq:Mij-two}
M_{i,j}^* = \frac{\left(\hat{r}_{i,j}^{(1)} - \hat{r}_{i,j}^{(2)}\right)^{2}}{\hat{\vartheta}_{i,j}^{(1)}/(n_{1}q) + \hat{\vartheta}_{i,j}^{(2)}/(n_{2}q)},
\eeq
and the maximum test statistic as
\beq
\label{eq:Mn-two}
M_{n}^* = \max\limits_{1\leq i < j\leq p}M_{i,j}^*.
\eeq
Because the diagonal entries of the correlation matrix are all 1, the maximum is only taken over off-diagonal entries.
The rest of the two-sample case proceeds exactly the same as the one-sample case.
The $M_{n}^*$ in the two-sample scenario has similar properties as the $M_{n}$ \eqref{eq:Mn-one} in the one-sample scenario. 
Section \ref{theo.sec} proves that $M_n^*-4\log p+\log \log p$ also converges to a Gumbel distribution under $H_0^*$ and certain regularity assumptions. Therefore, for a given significance level $0<\alpha < 1$, the test $\Phi_\alpha^*$ can be defined in parallel as \eqref{eq:phi-alpha-one},
\beq
\label{eq:phi-alpha-two}
\Phi_{\alpha}^*= I( M_{n}^*\geq q_{\alpha}+4\log p-\log\log p).
\eeq
The hypothesis $H_0^*: \R_{A}^{(1)} = \R_{A}^{(2)}$ is rejected whenever $\Phi_\alpha^* =1$.

To find which industries have correlations that are significantly different between emerging countries and developed countries, we need to recover the support of the difference of the correlation matrices between the two groups of countries.
Denote the support of
$\R_{A}^{(1)} - \R_{A}^{(2)}$ by
\beq
\label{eq:support.two.true}
\Psi^*=\Psi^*(\R_{A}^{(1)}, \R_{A}^{(2)})=\{(i,j): r_{i,j}^{(1)}\neq r_{i,j}^{(2)}, 1\leq i<j\leq p\}.
\eeq
We threshold the entry-wise statistic $M_{i,j}^*$ in \eqref{eq:Mij-two} at an appropriate level to obtain the estimated support as
\beq
\label{eq:support.two}
\hat{\Psi}^*(\tau)= \{(i,j):M_{i,j}^* \geq \tau\log p,~~1\leq i<j\leq p\},
\eeq
where the optimal choice of the threshold constant is $\tau=4$.


\section{Theoretical Properties}
\label{theo.sec}
We present the theoretical properties of the \dyP{testing} procedures for the one-sample case in Section \ref{theo1.sec} and the two-sample case in Section \ref{theo2.sec}. \dyP{The theorems on support recovery are delegated to Online Supplement.}

The following notational conventions are adopted. 
For a length $p$ vector $\a={(a_{1},\dotsc,a_{p})^{'}}\in \RR^{p}$, denote its Euclidean norm by $\|\a\|_{2}=\sqrt{\sum_{j=1}^{p}a^{2}_{j}}$. For a size $p\times q$ matrix $\A=(a_{i,j})\in\RR^{p\times q}$, denote its Frobenius norm by $\|\A\|_{F}=\sqrt{\sum_{i,j}a^{2}_{i,j}}$ and its spectral norm by $\|\A\|_{2}=\sup_{\|\x\|_2\leq  1}\|\A\x\|_2$. For a matrix $\A\in\RR^{p\times p}$, let $\lambda_{\min}(\A)$ and $\lambda_{\max}(\A)$ be its largest and smallest eigenvalues respectively. Denote its matrix 1-norm as $\|\A\|_{L_1} = \max_{1\le j\le p}\sum_{i=1}^{p}|a_{i,j}|$. 
For two sequences of real numbers $\{a_{n}\}$ and $\{b_{n}\}$, write $a_{n} = O(b_{n})$ (respectively $a_n \asymp b_n$) if there exists a constant $c$ such that $|a_{n}| \leq c|b_{n}|$ (respectively $1/c \le |a_{n}|/|b_n| \leq c$) holds for all sufficiently large $n$ and write $a_{n} = o(b_{n})$ if $\lim_{n\rightarrow\infty}a_{n}/b_{n} = 0$.

\dyP{To deal with matrix-valued observations, an intuitive assumption of matrix normal distribution has been predominantly adopted; see for example \citet{leng2012sparse,yin2012model,zhou2014gemini,qiu2016joint,han2016sparse,zhu2018multiple}. Under such assumption, the two key ingredients are normality and the Kronecker product structure of the covariance matrix $\Sigma^{(g)} = \Cov(\vec(\X_k^{(g)}))= \sigma^2 \B^{(g)}\otimes\A^{(g)},~g=1,2$. The motivation and applicability of Kronecker product are stated in Section \ref{sec-kronecker}.
The matrix normal distribution is equivalent to assuming that the pre-whitened matrix $(A^{(g)})^{-1/2}\X_{k}^{(g)}(B^{(g)})^{-1/2}$ has i.i.d. Gaussian entries.
Unlike the aforementioned papers, this article relaxes the Gaussian assumption and studies more general settings such that the pre-whitened matrix has i.i.d. sub-Gaussian entries. 
Under such matrix sub-Gaussian distributions, the covariance matrix preserves the Kronecker product structure. Furthermore, the matrix sub-Gaussian distribution will reduce to the matrix normal distribution if the pre-whitened entries are normally distributed.
}

\subsection{Theoretical Properties for Testing of \yin{Non-correlation}}
\label{theo1.sec}

The theoretical properties of the one-sample global testing procedure \eqref{eq:phi-alpha-one} will be established from two perspectives: the size and the power. Specifically, to study the asymptotic size of the test, we prove the asymptotic distribution of the test statistic under the null hypothesis; to analyze the power, we consider the sparse alternatives where only a small subset of the entries are nonzero.

The regularity conditions are as follows.
\noindent
\begin{itemize}
\item[{(C1)}] Assume that $\log p=o((nq)^{1/5})$, $np>q$, and
there are some constant $c_0, c_1>0$ such that, $c_0^{-1}\leq \lambda_{\min}(\A)\leq \lambda_{\max}(\A)\leq c_0$, and $c_1^{-1}\leq \lambda_{\min}(\B)\leq \lambda_{\max}(\B)\leq c_1$.

{\item [{(C2)}] Let $\X_k \stackrel{d}{=} \X =: (x_{i,j})$. Assume that the entries $x_{i,j}$ satisfy the sub-Gaussian-type assumption, i.e., there exist some constants $\eta > 0$ and $K>0$ such that  }
\begin{align*}
{\mathbb{E}\exp(\eta x_{i,j}^2) \leq K, ~~~\text{for all } i,j.}
\end{align*}
\item[{(C3)}] \yin{Assume that $q^{3} \log q \log^{3}\max(p,q,n) = o(np)$.}
\end{itemize}

\dyP{
\begin{remark}
\label{rmk-C3}
Condition (C1) on the eigenvalues of the covariance matrices is commonly assumed in the high-dimensional setting. It implies that the majority of the variables are not highly correlated with the others in either the row direction or the column direction. Intuitively, to make a valid inference for $p$ variables of interest, the dimension $p$ should not grow too fast compared to $nq$, and that is why we need the condition $\log p = o((nq)^{1/5})$ to obtain rates of similar form as $\{(\log p)/(nq)\}^{1/2}$.
Condition (C2) is the moment condition on X which is much weaker than the Gaussian assumption.

\yin{A few comments on Condition (C3) are in order. Firstly, Condition (C3) requires that $q$ does not grow too fast compared to $np$. Secondly, it is assumed to {ensure that $\tilde{\B}^{-1}$ defined in \eqref{eq:Btilde}, as the estimator of the inverse of the nuisance covariance $\B^{-1}$, is reasonably accurate}. As such, the oracle estimate $\hat\A^{o}$ will be close to the estimate $\hat\A$ in \eqref{eq:ahat-one} as shown in the proof. Thirdly, this is applicable to any other estimator of $\B$ as long as it satisfies $\|\tilde\B-\B\|_{\infty} = O_p[\{\log q/(np)\}^{1/2}]$. Hence we can use, for example, the methods in Remark \ref{rmk-other-estimation-B}, given certain prior knowledge on the structure of $\B$. One can also estimate $\B^{-1}$ directly and all results still hold as long as $\|\tilde{\B}^{-1}-\B^{-1}\|_2$ = $O_{p}[q\{\log q/(np)\}^{1/2}]$ or $\|\tilde{\B}^{-1}-\B^{-1}\|_{\infty}$ = $O_{p}[\{\log q/(np)\}^{1/2}]$. This can be satisfied by many precision matrix estimates such as the CLIME estimator in \cite{cai2011constrained}. Finally, if we have prior information on the structure of $\B$, say the AR(1) model where the off-diagonal elements decay exponentially as they get further away from the diagonal as in the simulation section (i.e. $\B^{-1}$ is banded), or the moving average model with banded $\B$, then $\|\tilde{\B}^{-1}-\B^{-1}\|_2$ has a faster rate of convergence and Condition (C3) can be further relaxed to $q \log q \log^{3}\max(p,q,n) = o(np)$. } 
\end{remark}
}

Under Conditions (C1)-(C3), as mentioned in Section \ref{one.sec}, Theorem \ref{ther1} shows that $M_{n}-4\log p+\log\log p$ indeed converges weakly to a Gumbel distribution under the null hypothesis.

\begin{theorem}
\label{ther1}
Suppose that the regularity conditions (C1)-(C3) hold.
Then under $H_{0}$,  for any $t \in\RR$,
\begin{eqnarray}\label{th1}
\pr\Big{(}M_{n}-4\log p+\log\log p\leq t \Big{)}\rightarrow \exp\Big{(}-\frac{1}{\sqrt{8\pi}}\exp\Big{(}-\frac{t}{2}\Big{)}\Big{)},
\end{eqnarray}
as $nq,~p\rightarrow\infty$. Furthermore,  under $H_0$, the convergence in (\ref{th1}) is uniform for all  $\{\X_k,k=1,\ldots, n\}$ satisfying (C1)-(C3).
\end{theorem}

We next turn to the power analysis of the test $\Phi_{\alpha}$. In order to perform the power analysis, we focus on sparse alternative hypothesis, as explained in Section \ref{test.sec}. Define the following class of covariance matrices associated with the row direction of the matrix-valued observations:
\beq
\label{eq:powerclass1}
\mathcal{U}(c)=\Big{\{}\A=(a_{i,j})_{p\times p}:~ \max_{1\leq i < j\leq p}\frac{|a_{i,j}|}{\sqrt{\theta_{i,j}/(nq)}}\geq c\sqrt{\log p}\Big{\}},
\eeq
where $\theta_{i,j}$ was defined previously in \eqref{eq:theta-one}. Note that this class of covariance matrices only requires one element to be large enough, $|a_{i,j}|/{\sqrt{\theta_{i,j}/(nq)}}\geq c\sqrt{\log p}$.
As $\theta_{i,j}=O(1)$, it essentially requires only one off-diagonal entry of $\A$ to be larger than $c\sqrt{\log p/(nq)}$. For such matrices with $c=4$ as the alternative hypothesis, Theorem \ref{power} shows that $\Phi_\alpha$ can distinguish the alternative hypothesis from the null hypothesis, where the off-diagonal entries of $\A$ are all zero, asymptotically. In other words, $H_0$ is rejected by $\Phi_{\alpha}$ with probability tending to 1 if $\A\in \mathcal{U}(4)$.

\begin{theorem}\label{power}
Suppose that Conditions (C1)-(C3) hold. As $nq,~p \rightarrow \infty$, we have
\[
\inf_{\A\in \mathcal{U}(4)}\pr(\Phi_{\alpha}=1)\rightarrow 1.
\]
\end{theorem}

Theorem \ref{optimal} further demonstrates that the lower bound of $4\sqrt{\log p}$ in the definition of the class of covariance matrices is rate optimal. Let $\mathcal{T}_{\alpha}$ be the set of level $\alpha$ tests, i.e., we have $\pr(T_{\alpha}=1)\leq \alpha$ under the null hypothesis for any test $T_{\alpha}\in \mathcal{T}_{\alpha}$.

\begin{theorem}\label{optimal}
Suppose that $\log p=o(nq)$. Let $\alpha,\beta>0$ and $\alpha+\beta<1$. There exists some constant $c_0>0$ such that for all sufficiently large $nq$ and $p$,
\[
\inf_{\A\in \mathcal{U}(c_0)}\sup_{T_{\alpha}\in \mathcal{T}_{\alpha}}\pr(T_{\alpha}=1)\leq 1-\beta.
\]
\end{theorem}
The above theorem implies that, when $c_0$ is small enough, with probability going to one, any level $\alpha$ test cannot reject the null hypothesis uniformly over $\mathcal{U}(c_0)$. As a consequence, the rate $\sqrt{\log p}$ as the lower bound of ${|a_{i,j}|}/{\sqrt{\theta_{i,j}/(nq)}}$ cannot be improved.

To sum up, Theorems \ref{ther1}-\ref{optimal} suggest that the test $\Phi_\alpha$ defined in Section \ref{one.sec} has asymptotic level $\alpha$, it has power one asymptotically under certain sparse alternative hypothesis, and the rate requirement on the sparse alternative is the weakest possible one.

\subsection{Theoretical Properties for Testing of Correlation Matrix Equality}
\label{theo2.sec}

For the two-sample testing of correlations, we assume the sample sizes from the two groups are comparable, $n_{1} \asymp n_{2}$, and write $n=\max(n_1,n_2)$ in this section.

The Conditions (C1)-(C3) in the one-sample case need to be replaced by the following conditions for the two-sample case.
\begin{itemize}
	\item[{(C1$^{*}$)}] Assume that $\log p=o((nq)^{1/5})$, $n_gp>q$, and
	there are some constant $c_0, c_1>0$ such that, $c_0^{-1}\leq \lambda_{\min}(\A^{(g)})\leq \lambda_{\max}(\A^{(g)})\leq c_0$, and $c_1^{-1}\leq \lambda_{\min}(\B^{(g)})\leq \lambda_{\max}(\B^{(g)})\leq c_1$, for $g = 1, ~2$.
	
	\item[{(C2$^{*}$)}] {Let $\X_k^{(g)} \stackrel{d}{=} \X^{(g)} =: (x_{i,j}^{(g)})$. Assume that the entries $x_{i,j}^{(g)}$ satisfy the sub-Gaussian-type assumptions, i.e., there exist some constants $\eta > 0$ and $K>0$ such that  }
	\begin{align*}
	&\mathbb{E}\exp(\eta x_{i,j}^{(1)})^2 \leq K\\
	&\mathbb{E}\exp(\eta x_{i,j}^{(2)})^2 \leq K, ~~~\text{for all }i,j.
	\end{align*}
	
		\item[{(C3$^{*}$)}] \yin{Assume that $q^{3} \log q \log^{3}\max(p,q,n) = o(np)$.} 

	{\item[{(C4$^{*}$)}] Let $\X_{k}^{(g)}(\B^{(g)})^{-1/2} =: \Z_{k}^{(g)} \stackrel{d}{=} \Z^{(g)} =: (z_{i,j}^{(g)})$. Assume that there exist $\kappa_1,\kappa_2 \geq\frac{1}{3}$ such that for any $j,k,l,m \in \{1,2,\cdots,p\}$ and $i \in \{1,2,\cdots,q\}$,}
	\begin{align*}
	&\mathbb{E}z_{j,i}^{(1)}z_{k,i}^{(1)}z_{l,i}^{(1)}z_{m,i}^{(1)} = \kappa_1(a_{j,k}^{(1)}a_{l,m}^{(1)}+a_{j,l}^{(1)}a_{k,m}^{(1)}+a_{j,m}^{(1)}a_{k,l}^{(1)}),\\
	&\mathbb{E}z_{j,i}^{(2)}z_{k,i}^{(2)}z_{l,i}^{(2)}z_{m,i}^{(2)} = {\kappa_2}(a_{j,k}^{(2)}a_{l,m}^{(2)}+a_{j,l}^{(2)}a_{k,m}^{(2)}+a_{j,m}^{(2)}a_{k,l}^{(2)}).
	\end{align*}
	\item[{(C5$^{*}$)}] {There exists some $\gamma>0$ such that $|A_{\gamma}| = o(p^{1-\nu})$ for any sufficiently small constant $\nu>0$, where the set is defined as
	\begin{equation*}
	A_{\gamma} = \{(i,j): |r_{i,j}^{(g)}| \ge (\log p)^{-1-\gamma}, 1 \leq i < j \leq p, \text{ for } g=1 \text{ or }2\}.
	\end{equation*}}
\end{itemize}

Note that, Conditions (C1$^{*}$) and (C2$^{*}$) are the two-sample analogue of the one-sample conditions (C1) and (C2), and Condition (C3$^{*}$) is the same as Condition (C3).  {Condition (C4$^{*}$) holds for the elliptically contoured distributions with $\kappa_g = \text{Kurtosis}(z_{i,j}^{(g)})/3$, $g=1,2$.} (C5$^{*}$) ensures that most of the variables are not highly correlated with each other.

Under appropriate regularity conditions, Theorems \ref{ther5}-\ref{optimal*} are the two-sample counterparts of the one-sample Theorems \ref{ther1}-\ref{optimal}. In particular, Theorem \ref{ther5} shows the limiting distribution of $M_n^*$ \eqref{eq:Mn-two} under the null hypothesis and proves that $\Phi_\alpha^*$ \eqref{eq:phi-alpha-two} has level $\alpha$ asymptotically, Theorem \ref{power*} provides the power analysis of $\Phi_\alpha^*$, and Theorem \ref{optimal*} demonstrates the optimality of the test.

\begin{theorem}\label{ther5} Suppose that Conditions (C1$^{*}$)-(C5$^{*}$) hold.
	Then under $H_{0}^*$ in \eqref{eq:two-sample},  for any $t \in\RR$,
	\begin{eqnarray}\label{th1*}
	\pr\Big{(}M_n^*-4\log p+\log\log p\leq t \Big{)}\rightarrow \exp\Big{(}-\frac{1}{\sqrt{8\pi}}\exp\Big{(}-\frac{t}{2}\Big{)}\Big{)},
	\end{eqnarray}
	as $nq,~p\rightarrow\infty$. Furthermore,  under $H_0^{*}$, the convergence in (\ref{th1*}) is uniform for all  $\{\X^{(1)}_k,~k=1,\ldots, n_1\}$ and $\{\X^{(2)}_k,k=1,\ldots,n_2\}$ satisfying (C1$^{*}$)-(C5$^{*}$).
\end{theorem}

To analyze the power of $\Phi_\alpha^*$, in parallel with \eqref{eq:powerclass1}, define the following class of matrices:
\beq
\label{powerclass2}
\mathcal{U}^*(c)=\Big{\{}(\R_{A}^{(1)}, \R_{A}^{(2)}):~ \max_{1\leq i< j\leq p}\frac{|r_{i,j}^{(1)} - r_{i,j}^{(2)}|}{\sqrt{{\vartheta_{i,j}^{(1)}}/(n_{1}q)+{\vartheta_{i,j}^{(2)}}/(n_{2}q)}}\geq c\sqrt{\log p}\Big{\}},
\eeq
{where $\vartheta_{i,j}^{(g)} = \theta_{i,j}^{(g)}/(a_{i,i}^{(g)}a_{j,j}^{(g)})$.}
We have the following result.
\begin{theorem}\label{power*}
	Suppose that Conditions (C1$^{*}$) - (C3$^{*}$) hold. As $nq,~p \rightarrow \infty$, we have
	\[
	\inf_{(\R_{A}^{(1)}, \R_{A}^{(2)})\in \mathcal{U}^*(4)}\pr(\Phi_{\alpha}^*=1)\rightarrow 1.
	\]
\end{theorem}
Note that $\Phi_\alpha^*$ is able to distinguish the alternative from the null so long as one entry satisfies the requirement ${|r_{i,j}^{(1)} - r_{i,j}^{(2)}|} / \big({\vartheta_{i,j}^{(1)}}/(n_{1}q)+{\vartheta_{i,j}^{(2)}}/(n_{2}q)\big)^{1/2} \geq 4\sqrt{\log p}$.

The above rate is optimal because of the next theorem. Let $\mathcal{T}_{\alpha}^*$ be the set of all $\alpha$-level tests, i.e., $\pr(T_{\alpha}=1)\leq \alpha$ under $H_0^*$ for any $T_{\alpha}\in \mathcal{T}_{\alpha}^*$.
\begin{theorem}\label{optimal*}
	Suppose that $\log p=o(nq)$. Let $\alpha,\beta>0$ and $\alpha+\beta<1$. There exists some constant $c_0>0$ such that for all large $nq$ and $p$,
	\[
	\inf_{(\R_{A}^{(1)}, \R_{A}^{(2)})\in \mathcal{U}^*(c_0)}\sup_{T_{\alpha}\in \mathcal{T}_{\alpha}^*}\pr(T_{\alpha}=1)\leq 1-\beta.
	\]
\end{theorem}

\section{Simulation Studies}
\label{sim.sec}

\dyP{In this section, we investigate numerical performance of the proposed approaches and compare them with relevant existing procedures via simulation studies. We examine the global hypothesis test $\Phi_{\alpha}$ \eqref{eq:phi-alpha-one} for the one-sample case under matrix normal distribution in Section \ref{sim1.sec} and the global test $\Phi_{\alpha}^*$ for the two-sample case under matrix normal distribution in Section \ref{sim2.sec}. Online Supplement \ref{add.sim.normal} provides additional hypothesis test experiment under normal distribution;  Online Supplement \ref{add.sim.t} lists the simulation results under $t$ distribution; Online Supplement \ref{add.sim.supp} is devoted to the simulation results for support recovery including $\hat{\Psi}(\tau)$ \eqref{eq:support.one} and $\hat{\Psi}^*(\tau)$ \eqref{eq:support.two}; Pseudo simulation based on real data application is given in Online Supplement \ref{add.sim.pseudo}.}

\subsection{One-Sample Testing of \yin{Non-correlation}}
\label{sim1.sec}

\dyP{\bf Our estimators and competing methods.} To perform the one-sample global hypothesis testing $H_0: \A$ is diagonal in \eqref{eq:one-sample-global}, we compare seven methods, three of which are ours and the remaining four are based on certain twists of existing methods. To the best of our knowledge, there is no existing literature that targets directly at the same goal as ours.

The first three are our methods described in Section \ref{one.sec} and their variants. To implement the test $\Phi_\alpha$ in \eqref{eq:phi-alpha-one}, the statistics $M_{i,j}$ and $M_n$ in \eqref{eq:Mij-one} and \eqref{eq:Mn-one} need to be constructed and they depend on $\hat a_{i,j}$ and $\hat\theta_{i,j}$ in \eqref{eq:ahat-one} and \eqref{eq:thetahat-one}. Based on the definitions of $\hat a_{i,j}$ and $\hat\theta_{i,j}$, it is required to plug in an estimate of the covariance matrix $\B$. We experiment with three choices: (i) the oracle procedure when the true $\B$ is known and plugged in (denoted as ``One sample cov: oracle''), (ii) the procedure when the sample estimate $\tilde\B$ \eqref{eq:Btilde} is plugged in (denoted as ``One sample cov: sample-est''), and (iii) the procedure when a banded estimate of $\B$ \citep{rothman2010new} is plugged in (denoted as ``One sample cov: banded-est''). The ``One sample cov: oracle'' method can serve as a benchmark to see the effectiveness of the other two methods. When $\B$ is the covariance matrix associated with temporal or spatial dimension, those measurements that are far apart in the ordering are most often weakly correlated and $\B$ tends to exhibit some banded structure. For this type of data, a banded estimate of $\B$ is more accurate than the naive sample estimate $\tilde\B$ \citep{rothman2010new,bickel2008regularized}, so ``One sample cov: banded-est'' is expected to perform better or no worse than ``One sample cov: sample-est''.

The remaining four methods are related to existing approaches. Although they are not designed to test the covariance matrix of {matrix-valued observations} in \eqref{eq:one-sample-global}, they can be tweaked to achieve the same goal and are hence potential competitors. 

Three of them are originally designed to test the precision matrix, the inverse of the covariance matrix, of \dyP{matrix-valued observations with matrix normal distribution} \citep{xia2017hypothesis}. They are relevant because of the following observation: when the covariance matrix $\A$ is diagonal, its inverse $\A^{-1}$, the precision matrix, is also diagonal. So we can make inference of $\A$ by considering an equivalent hypothesis testing problem $H_{0,precision}: \A^{-1}$ is diagonal versus $H_{1,precision}: \A^{-1}$ is not diagonal. The latter problem was solved by \citet{xia2017hypothesis}. In their approach, to construct the test statistic that is related to $\A^{-1}$, it is also necessary to plug in an estimate of $\B$, which then leads to three versions as well: (iv) ``One sample pre: oracle'', (v) ``One sample pre: sample-est'' and (vi) ``One sample pre: banded-est''.

The last method, (vii) ``One sample vector'', is based on the covariance matrix testing approach \citep{cai2013two}, for \dyP{the vector-valued observations, where we simply ignore the temporal correlation captured by the column covariance matrix $\B$} and treat the $n$ matrix-valued observations $\X_k$ as $nq$ vector-valued observations $\X_{k,\cdot l}$ that are i.i.d. random vectors.

\noindent\dyP{\bf Simulation setup.} We simulate the data according to the following model, which is also adopted in \citet{xia2017hypothesis}. The data follow {matrix normal distribution} $\mathcal{MN}_{pq}(\0, \A,\B )$. The nuisance covariance matrix $\B$ has the structure of a time series autoregressive model where the off-diagonal elements decay exponentially as they get further away from the diagonal, that is, $b_{i,j} = 0.4^{|i-j|}, 1 \leq i,j \leq q$. The targeted covariance matrix $\A$ is set up differently under the null and alternative hypotheses. Under the null hypothesis, to evaluate the size of the test, we set $\A=\I$. Under the alternative hypothesis, to evaluate the power of the test, we set $\A = (\I + \U + \delta\I)/(1 + \delta)$, where $\delta$ is a scalar with value $\delta = |\lambda_{\min}(\I + \U)| + 0.05$, and $\U$ is a sparse and symmetric matrix with eight nonzero entries. Four of these nonzero entries are located randomly in the lower triangle of the matrix,  have random magnitudes that follow uniform distribution on $[2\{\log p/(nq)\}^{1/2}, 4\{\log p/(nq)\}^{1/2}]$, and possess random positive or negative signs. This definition of $\A$ can be thought of as a perturbation of $\I$, which is diagonal under the null, with $\U$ as the perturbation. The terms related to $\delta$ are to ensure the positive-definiteness of the covariance matrix. We examine a range of matrix dimensions and sample sizes. Specifically, combinations of $p = \{50, 200\},~ q = \{50, 200\}$, and $n = \{10, 50\}$ are considered. 1000 replications are conducted in each configuration.

\noindent\dyP{\bf Simulation results and interpretation.} Table \ref{Tab02} summarizes the empirical size and power, respectively, of the aforementioned seven methods with significance level $\alpha = 0.05$. Overall, the six methods (i)-(vi), that are based on matrix-valued observations, can control the sizes no larger than 0.05, but the vector-based approach (vii) has serious size distortion across all configurations. Regarding to the power performance, no matter whether covariance matrix based or precision matrix based methods are considered, the ``banded-est'' methods are as powerful as the ``oracle'' ones, which is unsurprising as the covariance matrix $\B$ indeed exhibits banded structure. The ``sample-est'' ones are typically slightly worse than the ``banded-est'' and ``oracle'' ones, which is expected as the estimation of $\B$ is worse. 

\dyP{It is seen that ``One sample cov: sample-est'' is sometimes undersized. This is because the size depends on the accuracy of the estimation of $B$, which is driven by how large $np$ is compared with $q$ (recall Remark \ref{rmk-C3} on Condition (C3)). It is hence expected that when $q$ is smaller, or $n$ and $p$ are bigger, the size is closer to the nominal level. For instance, when $q=50, ~ p=200, ~n=50$, the size is close to 0.05 while when $q=200,~ p=50, ~ n=10$, the size is smaller.
Additional simulation result for even larger sample size $n=500$ is given in Table \ref{Tab02-2} in Online Supplement, which shows that the ``one sample cov:sample-est'' is no longer undersized.
Note that the theoretical condition (C3) may be relaxed for ``one sample cov: banded-est'', whose size is close to the oracle and nominal level. Moreover, this phenomenon occurs for ``one sample pre:sample-est'' as well because of the same reason.}

It is also observed that our methods based on covariance matrix (i)-(iii) are comparable as those methods \citep{xia2017hypothesis} based on precision matrix (iv)-(vi). This phenomenon of comparable performance can be understood as follows. The power of the covariance matrix based methods depends on the largest magnitude of the correlations, whereas the power of the precision matrix based methods depends on the largest magnitude of the partial correlations. In the model setting, the perturbation matrix $\U$ is extremely sparse with only eight nonzero entries out of $50\times 50$ or $200\times 200$ entries with magnitude $[2\{\log p/(nq)\}^{1/2}, 4\{\log p/(nq)\}^{1/2}]$. With very high probability, the random locations of the nonzero entries will guarantee the largest magnitudes of the correlation and the partial correlation being similar to each other. Put these altogether, the levels of difficulty in detecting the nonzero off-diagonal entries of the correlation matrix and the partial correlation matrix are nearly identical, which makes (i)-(iii) and (iv)-(vi) similar in statistical performance.

\begin{table}[]
	\scriptsize
	\centering
	\begin{tabular}{clcccc}
		\toprule
		\multicolumn{2}{c}{}&	
		\multicolumn{2}{c}{n = 10} & \multicolumn{2}{c}{n = 50} \\
		
		\cmidrule(r){3-4} \cmidrule(r){5-6}
		
		$p$      &  methods
		
		&  $q=50$   &  $q=200$
		
		&  $q=50$    &  $q=200$  \\
		
		\midrule
		& &  \multicolumn{4}{c}{Empirical size}\\
		\midrule
		& One sample cov: oracle  &	3.9(0.6)&	4.5(0.7)&		4.2(0.6)&	5.3(0.7)
		   \\
		
		 &One sample cov: sample-est&	1.9(0.4)&	0.1(0.1)&	3.9(0.6)&	1.8(0.4) \\

		&One sample cov: banded-est&	3.3(0.6)&	4.7(0.7)&		5.0(0.7)&	4.6(0.7)
		\\
		
		50 &One sample pre: oracle	&	4.4(0.6)&	5.0(0.7)&		4.8(0.7)&	5.6(0.7)
		  \\
		
		&One sample pre: sample-est&	1.5(0.4)&	0(0)&		3.8(0.6)&	1.8(0.4) \\
		
		&One sample pre: banded-est&	3.6(0.6)&	4.6(0.7)&		4.8(0.7)&	4.9(0.7)
		\\

		&One sample vector&	35.8(1.5)&	41.1(1.6)&		38.3(1.5)&	44.6(1.6)\\
		
		\midrule
		
		&One sample cov: oracle&		3.6(0.6)&	5.1(0.7)&		5.1(0.7)&4.8(0.7)
		\\
		
		&One sample cov: sample-est&	3.4(0.6)&	1.0(0.3)&		5.3(0.7) & 3.3(0.6)
		\\
		
		&One sample cov: banded-est&		3.6(0.6)&	4.9(0.7)&		5.5(0.7) & 4.9(0.7)
		\\
		
		200&One sample pre: oracle&		4.3(0.6)&	4.5(0.7)&		5.0(0.7)& 4.7(0.7)
		\\
		
		&One sample pre: sample-est&		2.7(0.5)&	0.9(0.3)&		4.8(0.7)& 3.3(0.6)
		\\
		
		&One sample pre: banded-est&		3.6(0.6)&	4.2(0.6)&		5.2(0.7)& 4.8(0.7)
		\\
		
		&One sample vector&		58.5(1.6)&	66.9(1.5)&		68.4(1.5)& 69.5(1.5)
		\\
		\midrule
		& &  \multicolumn{4}{c}{Empirical power}\\
		\midrule
		
		&One sample cov: oracle &	84.9(1.1)&	70.0(1.4)  &	67.0(1.5) &	61.1(1.5) \\
		
		&One sample cov: sample-est &	65.2(1.5)&	1.8(0.4)&		63.5(1.5)&	40.8(1.6)\\
		
		&One sample cov: banded-est &	83.0(1.2)&	67.0(1.5)&		66.1(1.5)&	59.7(1.6)\\

		50&One sample pre: oracle &	87.4(1.0) &	70.9(1.4) &		67.9(1.5) &	60.9(1.5) \\
		
		&One sample pre: sample-est&		67.8(1.5) & 	2.0(0.4) &	64.0(1.5) &	41.9(1.6)\\
		
		&One sample pre: banded-est&		84.7(1.1)&	67.3(1.5)&		67.2(1.5)&	60.0(1.5)\\

		&One sample vector &	90.3(0.9)&	82.7(1.2)& 	81.6(1.2)&	80.6(1.3)\\
		
				\midrule

		&One sample cov: oracle&	92.5(0.8) &	72.8(1.4) &		70.4(1.4) &	58.6(1.6)
		\\
		
		&One sample cov: sample-est&		89.1(1.0)&	46.1(1.6)&		68.5(1.5)&	53.8(1.6)
		\\
		
		&One sample cov: banded-est&	92.1(0.9)&	72.0(1.4) &		70.4(1.4)&	58.8(1.6)
		\\
		
		200&One sample pre: oracle&		94.9(0.7) &	74.1(1.4) &		70.8(1.4) &	58.7(1.6)
		\\
		
		&One sample pre: sample-est &	92.6(0.8)&	48.4(1.6)&		69.6(1.5) &	54.1(1.6)
		\\
		
		&One sample pre: banded-est&		94.3(0.7) &	72.7(1.4)&		71.2(1.4)&	58.5(1.6)
		\\
		
		&One sample vector  &	97.1(0.5)&	91.7(0.9)&		91.6(0.9)&	87.5(1.0)
		\\

		\bottomrule
	\end{tabular}
	\caption{\small The empirical size and power of the \dyP{testing procedures for the one-sample case under normal distribution based on 1000 replications. The percentages are shown with the standard errors provided in parentheses.} The significant level is $\alpha=5\%$. The number of observations and the dimensions of the matrices vary: $p = \{50, 200\},~ q = \{50, 200\}$, and $n = \{10, 50\}$. }
	
	\label{Tab02}
	
\end{table}

However, our methods (i)-(iii) and the methods by \citet{xia2017hypothesis} (iv)-(vi) have different computational performance as shown in Table \ref{Tab03}. When comparing (i) vs (iv), similarly (ii) vs (v) and (iii) vs (vi), we can exclude the computation time of estimating $\B$ because the same estimation strategies are adopted by \citet{xia2017hypothesis} and our proposed methods. We only need to compare the computation time of the testing procedures with given $\B$, either known or estimated.  As such, Table \ref{Tab03} only presents the average computation times of (i) and (iv)  in seconds per replication.
Our method is much faster than the method of \citet{xia2017hypothesis}. The computational advantage becomes even more pronounced as the dimensions increase. This is expected as \citet{xia2017hypothesis} performs column-wise LASSO which is time-consuming. Given the similar statistical performance and different computation performances, our covariance matrix based methods are preferred.

\begin{table}[]
	
	\scriptsize
	
	\centering
	
	\begin{tabular}{clcccc}
		\toprule

		\multicolumn{2}{c}{}&	
		\multicolumn{2}{c}{n = 10} & \multicolumn{2}{c}{n = 50} \\
		
		\cmidrule(r){3-4} \cmidrule(r){5-6}
		
		p      &  methods
		
		&  $q=50$   &  $q=200$
		
		&  $q=50$    &  $q=200$ \\
		
		\midrule

		50& One sample cov: oracle  &	0.04&0.09&	0.12	&	0.41
		\\
		
		\vspace{2ex}
		
		 &One sample pre: oracle	& 0.24&	0.46&		0.58&1.90
		\\

		200&One sample cov: oracle &	0.70&	1.85&		2.22 &7.38\\

		&One sample pre: oracle&	6.61&	10.70&		12.85&35.46\\

		\bottomrule
	\end{tabular}
	\caption{\small The \dyP{average computation time of method (i) ``One sample cov: oracle'' and (iv) ``One sample pre: oracle'' in seconds per replication for the one-sample case under normal distribution.} The number of observations and the dimensions of the matrices vary: $p = \{50, 200\},~ q = \{50, 200\}$, and $n = \{10, 50\}$.}
	
	\label{Tab03}
	
\end{table}

\subsection{Two-Sample Testing of Correlation Matrix Equality}
\label{sim2.sec}

\dyP{\bf Our estimators and competing methods.}  To perform the two-sample global hypothesis testing $ H_0^*: \R_{A}^{(1)}=\R_{A}^{(2)}$ in \eqref{eq:two-sample}, we compare seven methods. These include our method $\Phi_\alpha^*$ proposed in Section \ref{two.sec}, where the covariance matrices $\tilde\B^{(g)}$ from both groups can be the ground truth, the sample estimate, or the banded estimate, which again lead to three approaches to be compared: (i)``Two sample cov: oracle'', (ii) ``Two sample cov: sample-est'', and (iii) ``Two sample cov: banded-est''.

Similar to the one-sample case in Section \ref{sim1.sec},
we can also twist the method of \citet{xia2018matrix} which was initially designed for the two-sample partial correlation matrix test under matrix normal distribution. 
To draw the connection, recall the model setup $\X_k^{(g)}\sim \mathcal{MN}_{pq}(\0, \A^{(g)},\B^{(g)} )$. \citet{xia2018matrix} is interested in the precision matrices $\boldsymbol\Omega_A^{(g)}=\big(\A^{(g)}\big)^{-1}$. The precision matrices are not identifiable either under the \dyP{Kronecker product model}. So the attention was given to the partial correlation matrix $\boldsymbol{P}_A^{(g)} = \big(\D_{A,precision}^{(g)}\big)^{-1/2}\boldsymbol\Omega_A^{(g)}\big(\D_{A,precision}^{(g)}\big)^{-1/2}$, where $\D_{A,precision}^{(g)}$ is a diagonal matrix consisting of the diagonal entries of the precision matrix $\boldsymbol\Omega_A^{(g)}$. The partial correlation matrix under the \dyP{Kronecker product model} is identifiable. \citet{xia2018matrix} eventually tested $H_{0,precision}^*: \boldsymbol{P}_{A}^{(1)}=\boldsymbol{P}_{A}^{(2)}$ versus $H_{1,precision}^*: \boldsymbol{P}_{A}^{(1)}\ne \boldsymbol{P}_{A}^{(2)}$.
Because the correlation matrix and partial correlation matrix fully determine each other, $ H_0^*$ in \eqref{eq:two-sample} is equivalent to $H_{0,precision}^*$. By analogy, (iv)``Two sample pre: oracle'', (v) ``Two sample pre: sample-est'', and (vi) ``Two sample pre: banded-est'' correspond to the three choices of the nuisance matrices $\B^{(g)}$ in \citet{xia2018matrix}. Lastly, (vii) ``Two sample vector'' based on the modified correlation version of \citet{cai2013two} is implemented in a similar fashion as in Section \ref{sim1.sec}.

\noindent\dyP{\bf Simulation setup.} We next describe the model to generate the data. Again, the matrix-valued observations from two groups follow {matrix normal distribution}, $\X_k^{(g)}\sim \mathcal{MN}_{pq}(\0, \A^{(g)},\B^{(g)} )$. The
covariance matrices $\B^{(g)}$ are the autocorrelation matrices of AR(1) process with coefficient $0.8$ and $0.9$ respectively. Under the null hypothesis, the two covariance matrices are identical $\A^{(1)}=\A^{(2)}$, and they are set to be the $\boldsymbol\Sigma^{(1)}$ in Model 1 of \citet{cai2013two}, {i.e., $\S^{*(1)}=(\sigma^{*(1)}_{i,j})$, where $\sigma^{*(1)}_{i,i}=1$, $\sigma^{*(1)}_{i,j}=0.5$ for $5(k-1)+1\leq i \neq j \leq 5k$ with $k=1,\ldots,p/5$ and $\sigma^{*(1)}_{ij}=0$ otherwise, and $\S^{(1)}=\D^{1/2}\S^{*(1)}\D^{1/2}$, where $\D=(d_{i,j})$ is a diagonal matrix with diagonal elements $d_{i,i}=\text{Unif}(0.5,2.5)$ for $i=1,...,p$.} Under the alternative hypothesis, we set $(\A^{(1)})^{{-1}} = (\boldsymbol{\Sigma}^{(1)} + \delta\I)/(1 + \delta)$ and $(\A^{(2)})^{{-1}} = (\boldsymbol{\Sigma}^{(1)} + \U + \delta\I)/(1 + \delta)$, where $\delta = |\min\{\lambda_{\min}(\boldsymbol{\Sigma}^{(1)}), \lambda_{\min}(\boldsymbol{\Sigma}^{(1)} + \U)\}| + 0.05$. Here, the perturbation matrix $\U$ is symmetric and has ten random nonzero entries, five of which are located randomly in the lower triangle with random sign and random magnitude on the interval $[3\{\log p/(nq)\}^{1/2}, 5\{\log p/(nq)\}^{1/2}]$.

\noindent\dyP{\bf Simulation results and interpretation.} Under the two-sample model setup, Table \ref{Tabtwopower} presents the empirical size and power, respectively, of Methods (i-vii) based on 1000 replications with significance level $\alpha = 0.05$. Considering the sizes, Table \ref{Tabtwopower} for the two-sample case shares the same message as Table \ref{Tab02} for the one-sample case, where the vector-based method (vii) cannot control the Type I error, while our methods (i-iii) and the precision matrix based methods (iv-vi) in \citet{xia2018matrix} behave well.

\dyP{Note that ``two sample cov: sample-est'' and ``two sample pre: sample-est'' are undersized sometimes. The reason for this phenomenon is similar to that of the one-sample case. 
In essence, the size will be closer to the nominal level when the temporal covariance estimation is more accurate.
        Moreover, it is observed in \citep[Proposition 1]{cai2013two} that the empirical sizes can be smaller than the nominal level due to the correlation among the variables. Hence, the sizes may vary depending on the model setting (the correlations among the variables).
Table \ref{normal-two-test-M2} in Online Supplement \ref{add.sim.normal} provides the simulation results when $\boldsymbol\Sigma^{(2)}$, Model 2, of \citet{cai2013two} is used instead of $\boldsymbol\Sigma^{(1)}$, Model 1; see A.4.1 for the details of the simulation setup. Both Tables 4 and 6 are the two-sample test results for normal distribution. The difference between them is that Table 4 is based on data generating with $\Sigma^{(1)}$ while Table 6 with $\Sigma^{(2)}$. The sizes in Table 6 are generally larger that those in Table 4 and closer to 0.05. This demonstrates that the sizes vary depending on the model setting, although both our methods and \citet{xia2018matrix} can control the size by 0.05. 
}

In contrast, considering the powers, the messages from Tables \ref{Tab02} and \ref{Tabtwopower} are not always the same between the one-sample and two-sample cases. First of all, the relative performances of the procedures with different plugged-in estimates of $\B$ are consistent. That is, both tables show that the ``oracle'' methods and the ``banded-est'' methods perform similarly, which dominate the ``sample-est'' methods most of the time. 

However, the relative performances of the covariance matrix based methods and the precision matrix based methods are different between the two cases. In particular, in the one-sample case (Table \ref{Tab02}), our methods are similar as the precision matrix based methods; but in the two-sample case (Table \ref{Tabtwopower}), our tests are more powerful than those based on the precision matrix. Here is the fundamental reason for this phenomenon: for the two {precision} matrices, $(\A^{(1)})^{-1} = (\boldsymbol{\Sigma}^{(1)} + \delta\I)/(1 + \delta)$ and $(\A^{(2)})^{-1} = (\boldsymbol{\Sigma}^{(1)} + \U + \delta\I)/(1 + \delta)$, which differ by a sparse matrix, the magnitude of the maximum difference between the corresponding {correlation} matrices can be dramatically different from the magnitude of the maximum difference between the corresponding {partial correlation} matrices, primarily because of the non-diagonal structure of $\boldsymbol{\Sigma}^{(1)}$. This is different from the one-sample case, where the two covariance matrices, $\I$ and $(\I + \U + \delta\I)/(1 + \delta)$, also differ by a sparse matrix, but the above mentioned two magnitudes are about the same primarily because of the diagonal structure under the null hypothesis.

The comparison of the computation time of our methods and \citet{xia2018matrix} is not provided for the two-sample case because the takeaway is the same as in the one-sample case.

\begin{table}[]
	
	\scriptsize
	
	\centering

	\begin{tabular}{clcccc}
		\toprule
		\multicolumn{2}{c}{}&	
		\multicolumn{2}{c}{n = 10} & \multicolumn{2}{c}{n = 50} \\
		
		\cmidrule(r){3-4} \cmidrule(r){5-6}
		
		$p$      &  methods
		
		&  $q=50$   &  $q=200$
		
		&  $q=50$    &  $q=200$  \\
		
		\midrule
		& &  \multicolumn{4}{c}{Empirical size}\\
		\midrule
		
		& Two sample cov: oracle  &	2.1(0.5)&	2.8(0.5)&		2.4(0.5)&	2.0(0.4)
		\\
		
		&Two sample cov: sample-est&		0.3(0.2)&	0(0)&		2.0(0.4)&	0.8(0.3) \\

		&Two sample cov: banded-est&		0.9(0.3)&	3.1(0.5)&		2.0(0.4)&	1.8(0.4)
		\\
		
		50 &Two sample pre: oracle	& 	2.1(0.5)&	2.0(0.4)&		2.6(0.5)&	2.1(0.5)
		\\
		
		&Two sample pre: sample-est&	2.3(0.5)&	0(0)&		2.9(0.5)&	2.9(0.5) \\
		
		&Two sample pre: banded-est&	3.1(0.5)&	4(0.6)&		2.8(0.5)&	2.8(0.5)
		\\

		&Two sample vector &		100(0)&	100(0)&	100(0)&		100(0)\\
		
		\midrule

		&Two sample cov: oracle&		3.0(0.5)&	2.3(0.5)&		2.3(0.5)&	2.4(0.5)
		
		\\
		
		&Two sample cov: sample-est&		1.7(0.4)&	0.2(0.1)&	2.5(0.5)&	1.2(0.3)
		
		\\
		
		&Two sample cov: banded-est&		1.5(0.4)&	1.0(0.3)&	2.4(0.5)&	2.2(0.5)
		
		\\
		
		200&Two sample pre: oracle&		1.8(0.4)&	3.3(0.6)&	2.4(0.5)&	2.9(0.5)
		
		\\
		
		&Two sample pre: sample-est&		1.9(0.4)&	2.7(0.5)&	2.4(0.5)&	3.2(0.6)
		
		\\
		
		&Two sample pre: banded-est&  2.0(0.4)&	3.3(0.6)&	2.5(0.5)&	3.4(0.6)
		
		\\
		
		&Two sample vector &	100(0)&	100(0)&	100(0)&100(0)
		
		\\

		\midrule
		& &  \multicolumn{4}{c}{Empirical power}\\
		\midrule
		
		& Two sample cov: oracle & 	91.0(0.9)&	91.2(0.9)&	90.6(0.9)&	91.7(0.9)
		
		\\
		
		&Two sample cov: sample-est&		70.4(1.4)&	1.6(0.4)&		88.0(1.0)&	76.2(1.3)
		\\

		&Two sample cov: banded-est&		75.4(1.4)&	57.9(1.6)&		88.0(1.0)&	87.4(1.0)
		
		\\
		
		50 &Two sample pre: oracle	& 	85.4(1.1)&	68.1(1.5)&		66.5(1.5)&	57.8(1.6)
		
		\\
		
		&Two sample pre: sample-est&		63.8(1.5)&	2.1(0.5)&		62.2(1.5)&	43.0(1.6)
		\\
		
		&Two sample pre: banded-est&		68.8(1.5)&	39.7(1.5)&		62.7(1.5)&	52.5(1.6)
		
		\\

		&Two sample vector &	100(0)&	100(0)&	100(0)&		100(0)
		\\
		
		\midrule

		&Two sample cov: oracle&		98.2(0.4)&	99.4(0.2)&		99.0(0.3)&	98.8(0.3)
		
		\\
		
		&Two sample cov: sample-est&		96.8(0.6)&	91.0(0.9)&		99.0(0.3)&	98.1(0.4)
		
		\\
		
		&Two sample cov: banded-est&		97.0(0.5)&	97.9(0.5)&		99.0(0.3)&	98.8(0.3)
		
		\\
		
		200&Two sample pre: oracle&		99.3(0.3)&	93.7(0.8)&		90.8(0.9)&	77.3(1.3)
		
		\\
		
		&Two sample pre: sample-est&		98.9(0.3)&	80.3(1.3)&		90.7(0.9)&	73.8(1.4)
		
		\\
		
		&Two sample pre: banded-est& 98.8(0.3)&	89.2(1.0)&	 91.1(0.9)&	76.7(1.3)
		
		\\
		
		&Two sample vector &	100(0)&	100(0)&	100(0)&		100(0)
		
		\\

		\bottomrule
	\end{tabular}
	\caption{\small The empirical size and power of the \dyP{testing procedures for the two-sample case under normal distribution based on 1000 replications. The percentages are shown with the standard errors provided in parentheses. The $\S^{(1)}$ matrix adopts the form of Model 1 in \citet{cai2013two}.} The significant level is $\alpha=5\%$. The number of observations and the dimensions of the matrices vary: $p = \{50, 200\},~ q = \{50, 200\}$, and $\dyP{n_1=n_2=n} = \{10, 50\}$. }
	\label{Tabtwopower}
	
\end{table}

\dyP{In addition, we have tried experiments with Laplace and gamma distributions, the simulation results are
qualitatively quite similar to the normal cases, and hence omitted. Online Supplement \ref{add.sim.t} provides the simulation results under $t$ distribution, in which case, the precision matrix based approaches (which depend on normality heavily) in both one-sample and two-sample scenarios are no longer valid, while our covariance matrix based approaches (which do not rely on normality) are. Moreover, borrowing the configuration of the stock data, Online Supplement \ref{add.sim.pseudo} further demonstrates the validity and power of our methods and the invalidity of the precision matrix related methods.
}

\section{Real Data Analysis}
\label{real.sec}

In our real data application, we employ a comprehensive sample of 30 industry sector returns from 43 countries around the world from 2001:07$\sim$2017:12.

We obtain the data on stock market returns and accounting items
for international public firms from the Thomson-Reuters
Datastream and
Worldscope databases. Our sample covers 22 emerging markets and 21 developed countries from
2001:07 up to 2017:12\myfootnote{22 emerging countries include:
Argentina, Brazil, Chile, China, Czech, Greece,	Hungary, India,
Indonesia, Israel, Malaysia, Mexico, Pakistan, Peru, Philippines,
Poland, Portugal, South Africa,	South Korea, Taiwan, Thailand, and Turkey. 21 Developed countries include:
Australia, Austria, Belgium, Canada, Denmark, Finland, France, Germany, Hong
Kong, Ireland, Italy, Japan, Netherlands, New Zealand, Norway, Singapore,
Spain, Sweden, Switzerland, the U.K., and the U.S.}. All returns are
denominated in USD and excess return is calculated after subtracting
the U.S. risk free rate. We categorize firms into 30 industry groups
by their SIC (Standard classification code)\myfootnote{These industries
  are:  1 Food, 2 Beer, 3 Smoke, 4 Games, 5 Books, 6 Hshld, 7 Clths, 8
  Hlth, 9 Chems, 10 Txtls , 11 Cnstr, 12 Steel, 13 FabPr, 14 ElcEq, 15
  Autos, 16 Carry, 17 Mines, 18 Coal, 19 Oil, 20 Util, 21 Telcm, 22
  Servs, 23 BusEq, 24 Paper, 25 Trans, 26 Whlsl, 27 Rtail, 28 Meals,
  29 Fin, 30 Other. See French
  \href{http://mba.tuck.dartmouth.edu/pages/faculty/ken.french/Data_Library/det_30_ind_port.html}{http://mba.tuck.dartmouth.edu/pages/faculty/ken.french/Data\_Library/det\_30\_ind\_port.html}.}. We
use the last December market capitalization denominated in USD as
portfolio weight, then aggregate each stocks into industry portfolios
for each country\myfootnote{There are a small number of missing
  industries in certain emerging countries. We augment our data by approximating the country-industry return as a portfolio of that country's specific value-weighted return and that industry's specific value-weighted return, and use different portfolio weights to ensure the robustness of our results. In our main results, we assume an equally weighted portfolio of country and industry returns. Alternatively, we regress all non-missing country-industry returns on both country and industry returns and obtain weights of 0.79 and 0.21 after scaling. We also conduct such a regression with constraint of weights summing up to one, and obtain weights of 0.553 and 0.447. All the results are qualitatively similar.}.

\subsection{Economic Propositions}
\label{real1.sec}


The well-documented evidence on the financial market
segmentation suggests that in general, asset returns co-move with the
market return differently across countries. In an integrated financial
market,
industry sectors have the same amount of systematic risk no matter from which
country. However, this is not true for a segmented
market. Obviously, with more rigid exchange rate policies, higher
investment barrier, and sovereign risks, emerging countries are more
segmented while developed countries are more integrated to global
financial markets \citep{BekaertHarvey-95JF}. The measure of
systematic risk, the market beta, is just the ratio of covariance
between the stock's excess return and market excess return, over the variance of the latter.
As we standardize the variances and market is just an aggregation of
all industries, we therefore conjecture that there should be
significant differences in correlations
comparing developed to emerging countries. In particular, we make the
following three propositions.

{\noindent \bf Proposition 1.} In general, developed countries are more industrialized, and
its financial markets are also more developed, usually accompanied with a
better information environment. For example, \citet{CampbellLettauMalkielXu-01JF}
show that in the U.S., there is a growing trend of idiosyncratic
volatility in stock returns. In other words, stock returns generally co-move less than in the past.
In contrast, emerging countries are
more vulnerable to disasters, sudden change in supply or demand of
natural resources, and population displacement. Stock
returns then co-move more closely, and achieve higher returns in periods of economic prosperity and
expansion and lower in periods of economic downturn and
contraction.

Therefore, we conjecture that some
industries are more pro-cyclical in emerging countries than in developed
countries.
For example, industries are
more pro-cyclical in the production of durable goods, such as raw materials and heavy equipment. Also, airline industry may prosper more at
the economic boom when people travel more. Developed economies are
more open and free market-oriented, characterized by a
relatively stable trend, and industry-specific shock tends not to
affect other industries in the economic system.
In contrast, emerging markets are characterized by frequent regime switches, a premise motivated by the dramatic reversals in fiscal, monetary and trade policies \citep{AguiarGopinath-07JPE}. When these policies change frequently in an unanticipated way, they tend to have bigger and longer impacts on the entire economy and make procyclical sectors more cyclical than those in developed markets.

{\noindent \bf Proposition 2.} \citet{kohn2018trade} show that emerging economies
produce more commodities than they consume; in contrast, developed economies
produce and consume commodities in similar amounts.
They document these sectoral imbalances in the trade of commodities and manufactures in emerging economies are positively correlated with business cycle volatility. Therefore we also expect to detect larger co-movements of commodity industries returns with others.

{\noindent \bf Proposition 3.} One more major difference between emerging markets and
developed economies is the demographic
patterns. \citet{DellaVignaPollet-07AER} argue that
different goods have distinctive age profiles of consumption, and changes in the age distribution can forecast shifts in demand for various goods. These shifts in demand induce predictable changes in profitability for industries that are not perfectly competitive. Given that emerging countries account for 90\% of the global population aged under 30, we also expect in emerging countries, recreative business industries co-vary higher with the market returns.

\subsection{Economic Interpretation of the Testing Results}
\label{real2.sec}

\dyP{The matrix sub-Gaussian distribution is reasonable for our real data application.
First of all, it is crucial for the stock return data to satisfy the Kronecker product assumption for the covariance matrix. To this end, we use the test in \citet{aston2017tests}, where the null hypothesis is that the covariance matrix indeed adopts a Kronecker product structure. The bootstrap method in \citet{aston2017tests} fails to reject the null, which supports our fundamental assumption.
Second, our observations are monthly value-weighted industry portfolio returns. As shown by \citet{CampbellLoMacKinlaly97}, monthly value-weighted index returns are less leptokurtic than their daily counterparts, and even less than daily individual stock returns. They conclude that fat-tailed distributions can be suited for shorter-horizon return, but the Central Limit Theorem applies and drives longer-horizon returns towards normality.

Moreover, the matrix sub-Gaussian distribution} \yx{is also consistent with the broad mean-variance optimal portfolio literature. For example, \citet{OkhrinSchmid-06JE} study distributional properties for minimum variance portfolio weights, where all results for finite samples are made assuming normally distributed returns. Also, in studying high-frequency returns, \citet{CaiHuLiZheng-20JE} propose two consistent estimators of the minimum risk of the global minimum variance portfolio, where the second estimator also assumes returns are i.i.d. normally distributed. Similarly, \citet{FrahmMemmel-10JE} also make i.i.d. normal assumptions in deriving two shrinking estimators of minimum variance portfolio, and concur that normality seems a reasonable assumption for monthly stock returns.}

{Recall the notations from the introduction - we have matrix-valued observations from two groups of countries: $n_1=22$ and $n_2=21$ for emerging and developed countries respectively. $\X_k^{(g)}\in\mathbb{R}^{30\times 198}$ is the matrix of returns for 30 industries over 198 months for country $k$ in group $g$. }

\noindent\dyP{\bf One-sample test of temporal \yin{non-correlation}.} We first perform one-sample hypothesis test on the temporal covariance matrix within each group of countries, $H_{0,B,g}: \B^{(g)}$ is diagonal, by feeding the transposed matrices $\Big(\X_k^{(g)}\Big)^{'},~k=1,...,n_g,$ into the algorithms. We apply three methods including ``One sample cov: sample-est'', ``One sample pre: sample-est'' \citep{xia2017hypothesis}, and ``One sample vector'' \citep{cai2013two} described in Section \ref{sim1.sec}. The oracle procedures cannot be applied, because the covariance matrices of the industries $\A^{(g)}$ are unknown. Neither is ``One sample cov: banded-est'' nor ``One sample pre: banded-est'' applied, because it is not reasonable to assume the covariance matrices of the industries $\A^{(g)}$ have banded structure. All three methods reject the null hypothesis within each group respectively. These results unanimously show strong temporal dependence in both emerging and developed countries and serve as a warning that the vector-based approaches might be problematic.

\noindent\dyP{\bf One-sample test of industrial \yin{non-correlation}.} We also implement all one-sample methods in Section \ref{sim1.sec}, except for the oracle ones, to test the covariance matrices of the industries, $H_{0,A,g}: \A^{(g)}$ is diagonal. Again, all five methods reject the null hypotheses. Furthermore, most methods reveal that almost all pairwise covariances are significantly larger than zero, conveying the fact that almost all industries are positively correlated with each other in both emerging and developed countries.

\begin{figure}
  \centering
  \includegraphics[width=\textwidth,page=1]{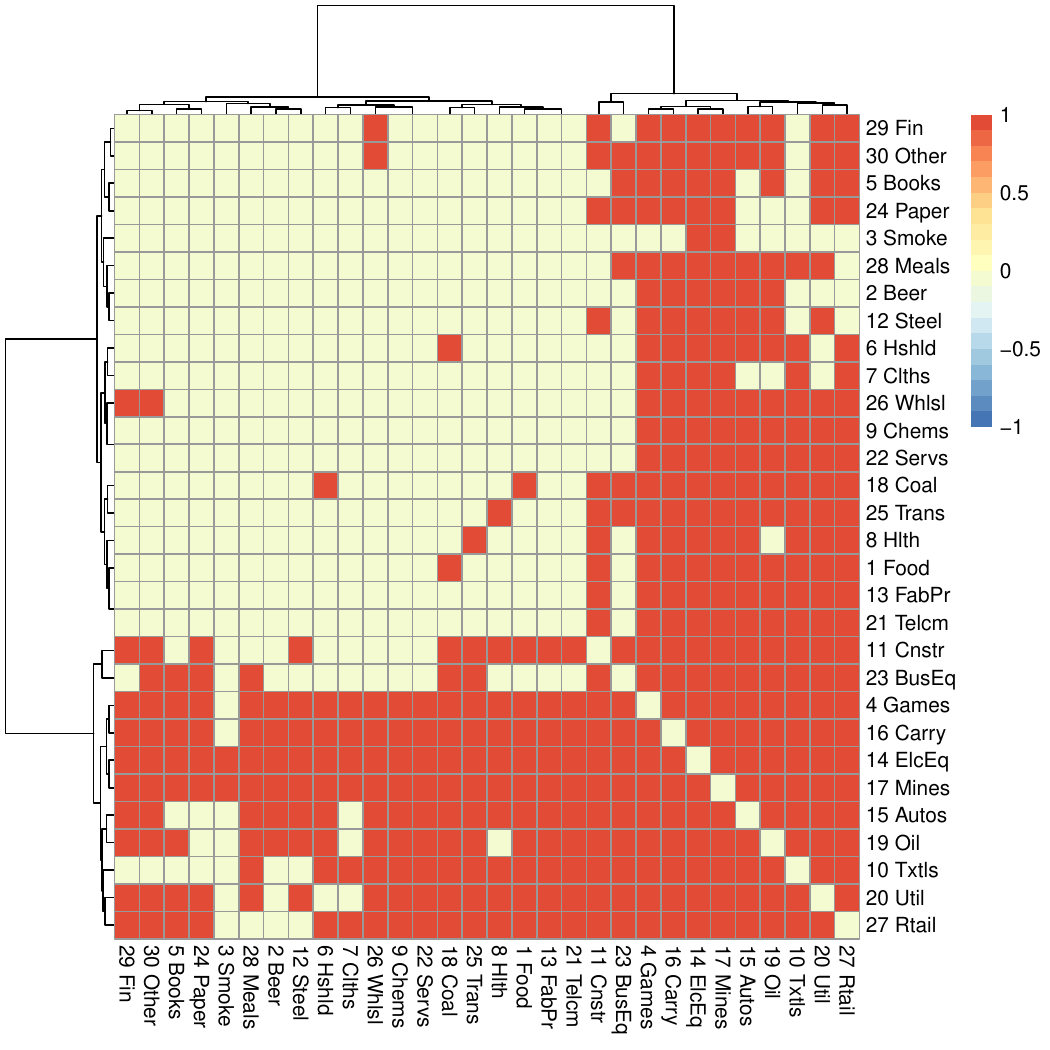}
  \caption{Support recovery result by our method ``Two sample cov: banded-est''. Heat map of the sign matrix of $\R_{A}^{(1)} - \R_{A}^{(2)}$. The columns and rows of the 30 industries are ordered according to a hierarchical clustering algorithm with the clustering result shown on the left and the top.}
  \label{fig:our}
\end{figure}

\begin{figure}
  \centering
  \includegraphics[width=\textwidth,page=2]{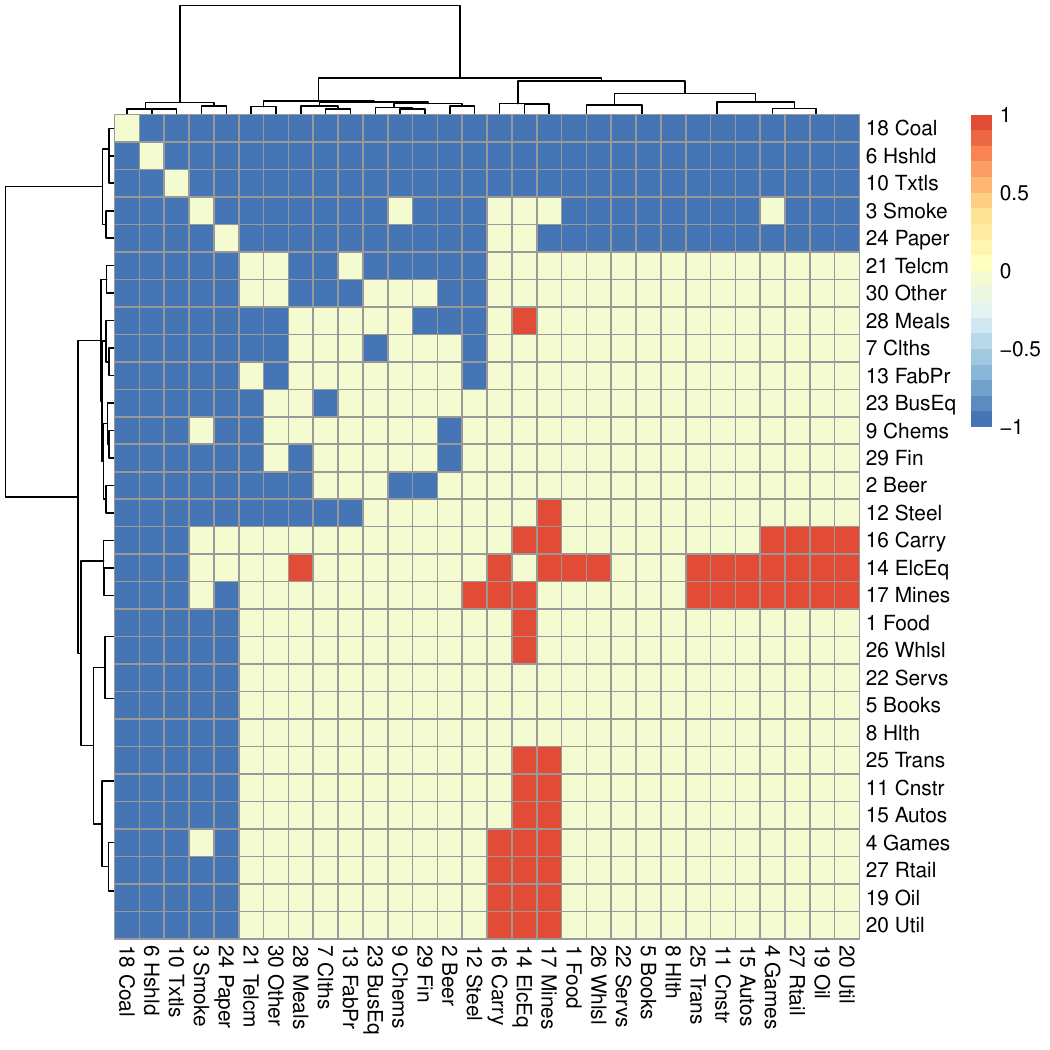}
  \caption{Support recovery result by ignoring the temporal correlation and implementing the two-sample hypothesis testing method for vector-valued data: ``Two sample vector''. Heat map of the sign matrix of $\R_{A}^{(1)} - \R_{A}^{(2)}$. The columns and rows of the 30 industries are ordered according to a hierarchical clustering algorithm with the clustering result shown on the left and the top. }
  \label{fig:cai}
\end{figure}

\noindent\dyP{\bf Two-sample test and support recovery.} To verify the economic propositions in Section \ref{real1.sec}, we apply the two-sample global hypothesis testing described in Section \ref{sim2.sec} and two-sample support recovery methods. All five methods, including ``Two sample cov: sample-est'', ``Two sample cov: banded-est'', ``Two sample pre: sample-est'', ``Two sample pre: banded-est'' \citep{xia2018matrix}, and ``Two sample vector'' \citep{cai2013two}, reject the null, $H_0^*: \R_A^{(1)}=\R_A^{(2)}$. This result suggests that the overall correlation structures of the industries between emerging and developed countries are significantly different. To dive into details of the discrepancy, two-sample support recovery techniques are carried out. The approaches based upon precision matrix are not included as they uncover the pattern of partial correlation, not correlation. Since the correlation matrix associated with the temporal dimension normally has a banded structure,  we present the support recovery results by our method ``Two sample cov: banded-est'' and ``Two sample vector'' in Figures \ref{fig:our} and \ref{fig:cai}, respectively.

In these two figures, red, yellow, and blue cells indicate whether the difference $\big(\R_{A}^{(1)}\big)_{(i,j)} - \big(\R_{A}^{(2)}\big)_{(i,j)}$ is significantly larger than, equal to, or significantly smaller than zero. Again, $g=1,2$ correspond to emerging and developed countries, respectively.
Note that the orderings of industries, columns and rows, in these two figures are different, as each is generated according to a hierarchical clustering algorithm based on the value of the sign matrix of $\R_A^{(1)}-\R_A^{(2)}$ for better visualization.

We find that by using our method, there are no blue entries, suggesting that overall correlations in the emerging countries are no smaller than those in the developed countries. In contrast,
the modified correlation version of the method in \citet{cai2013two} ignores the temporal correlation and potentially destroys the matrix structure due to vectorization.
Therefore, this method produces some counter-intuitive results in Figure \ref{fig:cai}, where quite a few blue cells appear. Among them, Household (consumer goods), a pro-cyclical industry, can be seen to significantly co-move less with many other pro-cyclical industries in emerging countries. This is also the case with Coal, a commodity industry, and Textile, another consumer goods industry. These findings are generally inconsistent with the literature and the propositions of Section 6.1, casting some doubts on the method of \citet{cai2013two}. As the hypothesis test of temporal \yin{non-correlation} is already rejected, these findings further highlight the weakness of implementing the vector-based approach for matrix-valued data.

Now we focus on Figure \ref{fig:our} and examine our economic propositions.
Based on the procedures detailed in Section~\ref{two.sec}, we find that for many industries,
correlations (off-diagonal terms) do not differ. Still, there are exceptions.
In Figure \ref{fig:our}, the clustering algorithm separates the 30 industries into three groups and 9 industries stand out in the rightmost group: \#27 Retail, \#20 Utilities, \#10 Textile, \#19 Oil, \#15 Automobiles and Trucks, \#17 Mines, \#14 Electrical Equipment, \#16 Aircraft, and \#4 Games. For these industries, all the correlations between them and the other industries are larger for the emerging countries than the developed markets.

In particular, consistent with Proposition 1 that in emerging
countries pro-cyclical industries co-move more, we find that among these 9
industries, there are indeed more pro-cyclical industries, such as \#14
Electrical Equipment, \#15 Automobiles and Trucks, \#16 Aircraft,
ships, and railroad equipment, and \#20 Utilities. The average of the estimated
correlations between each of these industries and all the other industries in the developed countries are
respectively, 0.218, 0.279, 0.235, and 0.292.
In comparison, in the emerging countries, the corresponding values
are separately, 0.402, 0.418, 0.399, and 0.423. Thus the
magnitudes of the differences are also large.

Also consistent with Proposition 2 that commodity industries may
co-move with other industries differently, we find that the
correlations are larger for the emerging countries in commodity industries
such as \#17 Mines and \#19 Oil. The average of the
correlations in the developed countries for these industries are
separately, 0.229, and 0.265.
In comparison, the average of the correlations in the emerging countries
are separately, 0.409, and 0.402.

Finally, also consistent with the argument per
\citet{DellaVignaPollet-07AER} in Proposition 3, we find industries that are more
related to discretionary consumer goods co-move differently,
reflecting demographic pattern differences, such as \#4 Games, \#10
Textile, and \#27 Retail. The average of the
correlations in the developed countries for these industries are
separately, 0.255, 0.223, and 0.267.
In comparison, the average of the correlations in the emerging countries
are separately, 0.403, 0.349, and 0.392.

Given all the statistically significantly larger correlations,
we conclude that these 9
industries command more systematic risk in emerging countries than in
developed countries.

\dyP{
\begin{remark}
Even though our real data application uses monthly industry returns, which tend to have light tail since they are low frequency and aggregated according to \citet{CampbellLoMacKinlaly97}, there are other financial and economic applications, which are heavy-tailed sometimes. To offer additional assurance, besides simulations with normal, gamma, Laplace, and $t$ distributions for different combinations of $n,p,q$, we further performed two experiments in Online Supplement \ref{add.sim.pseudo}: 1. we use the same setting of $\A^{(1)},~\A^{(2)},~\B^{(1)},\B^{(2)}$ as in Section \ref{sim.sec} with $n,p,q$ similar as the real data under $t_3$ distribution, and our methods are still valid under the null and powerful under the alternative; 2. we did a pseudo simulation with $t_3$ distribution and parameters $\A^{(1)},~\A^{(2)},~\B^{(1)},\B^{(2)}$ estimated from the real data, our methods have power one.
\yin{It is worth mentioning that our stock market data does not meet Condition (C3$^{*}$), i.e., $q^{3} \log q \log^{3}\max(p,q,n) = o(np)$, with $n\approx 20, p=30, q\approx 200$, but it is close to the weaker condition $q \log q \log^{3}\max(p,q,n) = o(np)$ for ``Two sample cov: banded-est'' under the AR model as discussed in Remark \ref{rmk-C3}.}
\end{remark}
}

\subsection{Impact on Portfolio Construction}

We now consider an investor aiming to build the global minimum variance (GMV) portfolio with 30 industry sector stock returns on the international equity market.
We show that the differences in correlation matrix between emerging and developed countries have direct implications for the GMV portfolio weights, which
are proportional to $\Sigma ^{-1}\iota$,  where
$\iota$ is a vector of ones and the covariance matrix of the return is
the only input that needs to be estimated.
As noted by
\citet{Jagannathan+Ma-RiskReduLargPort:03}, stocks that have high
covariances with other
stocks tend to receive small or even negative portfolio weights. The rationale is that such stocks may be more likely to be redundant in the efficient portfolio construction as the remaining stocks can achieve the risk reduction already at the absence of the former stocks. Put it differently, these stocks can be better hedged by other stocks, and their returns may be spanned by other returns \citep{Stevens-98JF, GotoXu-15JFQA}.

To ensure that differences on the GMV portfolio weights are not driven by differences in variances (diagonal terms of the covariance matrix), we scale the
covariance matrices and make their diagonal terms identical, so that any differences in
covariances (off-diagonal terms) must be the only source of different
portfolio weights. Specifically, we
take the estimates of the covariance matrix of the emerging countries industry returns $\A^{(1)}$ and the correlation matrix of the developed market industry returns $\R^{(2)}$ from \eqref{eq:ahat-two} and \eqref{eq:rhat-two}, and then obtain two covariance matrices that are directly comparable: $\Sigma_1=\A^{(1)}$, and
$\Sigma_2={(\D_A^{(1)})^{1/2}\R^{(2)}(\D_A^{(1)})^{1/2}}$.
Then we can build two GMV portfolios both using 30 industry returns, and calculate the portfolio weights as
$\Sigma_1^{-1}\iota/\iota^{\top}\Sigma_1^{-1}\iota$ and $\Sigma_2^{-1}\iota/\iota^{\top}\Sigma_2^{-1}\iota$, separately for
the emerging and developed markets.

In particular, we focus on the
weights of the following industries whose return correlations have
systematically exhibited differences between the two groups of countries:

\begin{center}

\begin{tabular}{lrrrrrrrrr}
Industry  & Games & Txtls & ElcEq & Autos & Carry & Mines & Oil & Util & Rtail\\
Emerging  &  0.078 &  0.025 &  0.047 &  0.005 &  0.075 &  0.009 & -0.013 &  0.023 & -0.001\\
Developed &  0.121 &  0.059 &  0.136 &  0.063 &  0.132 &  0.085 &  0.032 &  0.052 &  0.051\\
\end{tabular}

\end{center}

In the first row we present the GMV portfolio weights for the emerging markets, and in the second row the counterparts for the developed countries.
Obviously, for these 9 specific industries, their portfolio weights
are all larger for the developed markets, indicating they all contribute to
the GMV portfolio in a non-trivial way. In contrast, all the weights
are smaller for the emerging countries, including two less than 1\% (Autos and Mines)
and two negative (Oil and Rtail) weights. Therefore, these
industries can be more likely to be spanned by the remaining
industries.  When we calculate the sum of absolute
positions (bets on both long and short positions), we find emerging and developed markets GMV
portfolios obtain very similar values, 144.78\% and 144.94\%, which
suggests very similar level of leverage. This fact further highlights
that the contribution of these 9 industry returns in the emerging
markets is overshadowed by that of the others.

In practice of conducting out-of-sample tests, when an investor faces
huge estimation uncertainty, a long-only (no-short-sale) constraint
is typically assumed. In this case, $-$0.013 (Oil) and $-$0.001
(Retail) can simply be excluded in the efficient portfolio
construction. Furthermore, given the high transaction cost, trivial
weights such as 0.005 (Automobiles and trucks) and 0.009 (Mines) can
also fail the threshold due to their tiny magnitudes.

Alternatively, we could also fix the diagonals of the covariance matrix of the developed market
industry returns, then blend with the correlation matrix of the
emerging market industry returns:
$\Sigma_1={(\D_A^{(2)})^{1/2}\R^{(1)}(\D_A^{(2)})^{1/2}}$, and
$\Sigma_2=\A^{(2)}$. Similarly,
we calculate the GMV portfolio weights and single out these 9 industries:

\begin{center}

\begin{tabular}{lrrrrrrrrr}
Industry  & Games & Txtls & ElcEq & Autos & Carry & Mines & Oil & Util & Rtail\\
Emerging  &-0.064 & -0.009 & -0.079 & -0.086 & -0.033 & -0.087 & -0.024 &  0.045 & -0.035\\
Developed & 0.004 &  0.026 &  0.022 & -0.011 &  0.025 & -0.012 &  0.004 &  0.084 &  0.021\\
\end{tabular}
\end{center}

Again, the portfolio weights for these 9 specific industries are all
larger for the developed markets. This time, for the emerging market industry
portfolio, all except one
industry (Util) receive negative weights (short positions), ranging from
$-$0.9\% to $-$8.7\%. Again, these industries make contributions to the construction of the GMV portfolio only through short sellings.
As short sale is generally difficult and costly, taking
large short positions can be undesirable. Furthermore, the sum of absolute
positions is 190.24\% and 119.25\% separately for the emerging and developed markets, indicating additional leverage, thus extra risk for
the GMV portfolio using the emerging market industry returns.
Therefore, these results are consistent with the literature that in emerging countries when stocks tend to move together, diversification gains can be achieved using a smaller set of investible assets.

\section{Conclusions}
\label{sec-conclusion}

\dyP{
Under matrix sub-Gaussian distribution, we have proposed one-sample and two-sample tests of the correlation structures for matrix-valued observations. Theoretically, the proposed methods are asymptotically optimal, and empirically, they outperform the methods based on precision matrices both computationally and statistically for normal and other heavy-tailed distributions. Applying the proposed testing methods to the stock return data endorses the economic theory rigorously and the support recovery method unveils interesting structural difference of the correlations among different industries between developed and developing countries.

Further theoretical extension to include structural estimation of $\B$ is worth investigating in the future.
It will also be fruitful to propose methodology with theoretical guarantees to deal with heavy-tailed distribution by implementing robust strategy such as thresholding, Huber loss, and rank-based statistic, so that more financial and economical data are applicable. Higher-order tensorial observations, which are frequently encountered in many fields, exhibit even more challenges to analyze the theoretical properties. The current method relies upon one-step estimation of the nuisance parameter $\B$. Empirically, iterative estimations of $\B$ are more accurate and can generate more powerful tests for finite sample. However, the theoretical analysis of the iterative approach is much more involved and currently under study.

}

\section*{Acknowledgments}
\dyP{We sincerely thank the editor, associate editor, anonymous referees for their extremely insightful comments, which helped to improve the paper substantially.
We thank Mr. Lutao Dai and Jianlong Shao for their help with the computationally intensive simulation study.
We gratefully acknowledge the support from the United States NSF Grant IIS-1741390 (Yang), the GRF sponsored by the RGC in Hong Kong No. 17511716 (Xu) and No. 17301620 (Yang), Hong Kong CRF C7162-20G (Yang and Shen), }
NSFC Grants 12022103, 11771094 and 11690013 (Xia), the Ministry of Science and Technology Major Project of China 2017YFC1310903, University of Hong Kong (HKU) Stanley Ho Alumni Challenge Fund, and HKU BRC Grant (Shen).


\setlength{\bibsep}{0pt}
\bibliographystyle{apalike}
\bibliography{ref}

\newpage
\setcounter{page}{1}


\noindent
{\bf \Large Supplementary Material to ``Testing and Support Recovery of Correlation Structures for Matrix-Valued Observations with an Application to Stock Market Data''}
\setcounter{subsection}{0}
\def\thesubsection{A.\arabic{subsection}}
\vspace{0.2in}

\dyP{The Online Supplement is organized as follows: Online Supplement \ref{sec-support} states the theorems related to the support recovery for both the one-sample and the two-sample cases.
Proofs of Theorems \ref{ther1}-\ref{optimal*} in the main text and Theorems \ref{th4}-\ref{th8} in Online Supplement \ref{sec-support} are given in Section \ref{proof.sec}. 
Lemmas and their proofs are shown in Online Supplement \ref{lemma.sec}. Online Supplement \ref{add.sim} provides more simulation results. A more comprehensive version of Table \ref{literature} with key references listed is provided in Online Supplement \ref{table.sec}.}

\subsection{Theorems on Support Recovery for the One/Two-Sample Cases}
\label{sec-support}

To study the theoretical property of the support recovery procedure $\hat\Psi$ in \eqref{eq:support.one}, recall the definition of the true support of $\A$ in \eqref{eq:support.one.true} and define the following class of covariance matrices in parallel with \eqref{eq:powerclass1}:
\[
\mathcal{W}(c)=\Big{\{}\A=(a_{i,j})_{p\times p}: \min_{(i,j)\in\Psi}\frac{|a_{i,j}|}{\sqrt{\theta_{i,j}/(nq)}}
\geq c\sqrt{\log p}\Big{\}}.
\]
Note that $\mathcal{U}(c)$ requires the maximum of ${|a_{i,j}|}/{\sqrt{\theta_{i,j}/(nq)}}$ to be lower bounded by $c\sqrt{\log p}$ while $\mathcal{W}(c)$ requires the minimum of ${|a_{i,j}|}/{\sqrt{\theta_{i,j}/(nq)}}$ over the support is lower bounded by the same quantity. This requirement essentially means that all of the entries over the support are sufficiently large and thus can be distinguished from the noise. Then Theorem \ref{th4} below shows that the estimator $\hat \Psi(4)$ with threshold constant $\tau=4$ recovers the support $\Psi$ perfectly with probability going to $1$ when the magnitudes of all the nonzero off-diagonal entries are above certain thresholds as in $\mathcal{W}(4)$.

\begin{theorem}\label{th4}  Suppose that Conditions (C1)-(C3) hold. As $nq,~p\rightarrow\infty$, we have
\begin{eqnarray*}
\inf_{\A\in \mathcal{W}(4)}\pr\Big{(}\hat{\Psi}(4)=\Psi\Big{)}\rightarrow 1.
\end{eqnarray*}
\end{theorem}

\begin{remark}
With the same reasoning as in \citet{cai2013two}, it can be easily verified that the choice of the threshold constant $\tau=4$ is optimal. As a matter of fact, for any $\tau<4$, the probability of exact recovery of the support goes to zero. The failure of exact recovery is because the small threshold of $\tau\log p$ will estimate some of the zero entries by nonzero values, i.e., the estimated support will be larger than the true support. In addition, the rate of $\sqrt{\log p/(nq)}$ as the requirement of the nonzero entries of $\A$ cannot be relaxed.
\end{remark}

\dyP{Turning to the support recovery procedure $\hat{\Psi}^*$ in \eqref{eq:support.two} as an estimate of $\Psi^*$ in \eqref{eq:support.two.true} for the two-sample case,
construct} the set of matrices whose support has the rate defined in \eqref{powerclass2}, namely,
\[
\mathcal{W}^*(c)=\Big{\{}(\R_{A}^{(1)}, \R_{A}^{(2)}): \min_{(i,j)\in\Psi^*}\frac{|r_{i,j}^{(1)} - r_{i,j}^{(2)}|}{\sqrt{{\vartheta_{i,j}^{(1)}}/(n_{1}q)+{\vartheta_{i,j}^{(2)}}/(n_{2}q)}}\geq c\sqrt{\log p}\Big{\}}.
\]
Theorem \ref{th8} claims that $\hat{\Psi}^*(4)$ can recover such matrices exactly with probability tending to one.
\begin{theorem}\label{th8} Suppose that  Conditions (C1$^{*}$), (C2$^{*}$) and (C3$^{*}$) hold. As $nq,~p\rightarrow\infty$, we have
	\begin{eqnarray*}
		\inf_{(\R_{A}^{(1)}, \R_{A}^{(2)})\in \mathcal{W}^*(4)}\pr\Big{(}\hat{\Psi}^*(4)=\Psi^*\Big{)}\rightarrow 1.
	\end{eqnarray*}
\end{theorem}

\subsection{Proofs of Theorems}
\label{proof.sec}

\noindent\textbf{Proof of Theorem \ref{ther1}}

WLOG, throughout this section, we assume $a_{i,i}$ = 1 for $i = 1, \cdots, p$. Define the following oracle quantities when $\B$ is known,
\begin{align*}
	&(\hat{a}_{i,j}^o)  = \hat{\A}^o = \frac{1}{nq}\sum\limits_{k=1}^{n} \X_{k} \textbf{B}^{-1} \X_{k}^{'},	\\
	&\hat{\theta}_{i,j}^o = \frac{1}{nq}\sum\limits_{k=1}^{n} \sum\limits_{l=1}^{q}\left((\X_{k} \B^{-1/2})_{i,l}(\X_{k} \B^{-1/2})_{j,l} - \hat{a}_{i,j}^o\right)^{2},
\end{align*}
and define two maximum statistics that are based on these oracle quantities,
\begin{align*}
	\hat{M}_n^{o}& = \max\limits_{1\leq i<j \leq p} \frac{\hat{a}_{i,j}^{2}}{\hat{\theta}_{i,j}^o/(nq)},\\
	M_{n}^o &= \max\limits_{1\leq i<j \leq p} \frac{(\hat{a}_{i,j}^o)^{2}}{\hat{\theta}_{i,j}^o/(nq)}.
\end{align*}

Theorem \ref{ther1} is readily proved when combining the following facts. Under the Conditions (C1)-(C3),
\begin{align}
\label{sta1}
&P(M_{n}^o - 4\log p +\log \log p \leq t) \rightarrow \exp\left(-\frac{1}{\sqrt{8\pi}}\exp(-\frac{t}{2})\right),\\
\label{sta2}
&|M_{n} - \hat{M}_n^{o}| = o_{p}(1),\\
\label{sta3}
&|M_{n}^o - \hat{M}_n^{o}| = o_{p}(1),
\end{align}
where (\ref{sta1}) is proved by Lemma \ref{lemma4} and (\ref{sta2}) and (\ref{sta3}) are proved by Lemma \ref{lemma5} in Section \ref{lemma.sec}.~~~~$\blacksquare$

\noindent\textbf{Proof of Theorem \ref{power}}

We first prove for any constant $c>0$,
\begin{align*}
P(M_{n}^o \geq c + q_{\alpha} + 4\log p - \log \log p) \rightarrow 1.
\end{align*}
Let
\begin{align*}
M_{n}^{o,d} = \max\limits_{1\leq i < j \leq p}\frac{(\hat{a}_{i,j}^o - a_{i,j})^{2}}{\hat{\theta}_{i,j}^o/nq}.
\end{align*}
By Lemmas \ref{lemma2} and \ref{lemma3},
\begin{equation}\label{22}
P(M_{n}^{o,d} \leq 4\log p -\frac{1}{2}\log \log p)\rightarrow 1,
\end{equation}
as $nq$, $p \rightarrow \infty$. By Lemma \ref{lemma2}, the inequalities
\begin{align*}
\max\limits_{1\leq i < j \leq p}\frac{ a_{i,j}^{2}}{\hat{\theta}_{i,j}^o/nq} \leq 2M_{n}^{o,d} + 2M_{n}^o
\end{align*}
and
\begin{align*}
\max\limits_{1\leq i < j \leq p}\frac{ a_{i,j}^{2}}{\theta_{i,j}/nq} \geq 16 \log p,
\end{align*}
we conclude that $P(M_{n}^o \geq c+q_{\alpha} + 4\log p - \log \log p) \rightarrow 1$ as $nq$, $p \rightarrow \infty$.

Next, by the proof of Lemma \ref{lemma5}, it can be easily shown that, under Conditions (C1)-(C3), $|M_n - M_n^o| = o_p\left(\sqrt{\frac{M_n^o}{\log p}}\right)$. Notice that for a constant $c'>0$, we have
\begin{eqnarray*}
&&P(M_n \geq q_\alpha + 4\log p -\log \log p) \\
&&\quad\geq P\left(M_n^o \geq \frac{c'}{3}\sqrt{\frac{M_n^o}{\log p}}+q_\alpha + 4\log p -\log \log p\right) - P\left(M_n^o - M_n \geq \frac{c'}{3}\sqrt{\frac{M_n^o}{\log p}}\right),
\end{eqnarray*}
Consider the event $A = \{M_{n}^o \geq c'+q_{\alpha} + 4\log p - \log \log p\}$, then we have $P(A) \to 1$. Since
\begin{align*}
P\left(M_n^o - M_n \geq \frac{c'}{3}\sqrt{\frac{M_n^o}{\log p}}\right) \to 0,
\end{align*}
and
\begin{eqnarray*}
 &&P\left(M_n^o \geq \frac{c'}{3}\sqrt{\frac{M_n^o}{\log p}}+q_\alpha + 4\log p -\log \log p\right) \\
 &&\quad\geq  P\left(M_n^o \geq \frac{c'}{3}\sqrt{\frac{M_n^o}{\log p}}+q_\alpha + 4\log p -\log \log p|A\right)P(A) \to 1.
\end{eqnarray*}
The results thus follow.~~~~$\blacksquare$

\noindent\textbf{Proof of Theorem \ref{optimal}}

This theorem is essentially proved in \citet{cai2013two}, we skip the proof here.


\noindent\textbf{Proof of Theorem \ref{ther5}}

WLOG, assume $a_{i,i}^{(g)} = 1$ for $g = 1, 2$ and $i = 1, \cdots, p$. Similar to the proof of Theorem \ref{ther1}, define the oracle quantities,
\begin{align*}
	(\hat{a}_{i,j}^{o,(g)}) & = \hat{\A}^{o,(g)}= \frac{1}{nq}\sum\limits_{k=1}^{n} \X_{k}^{(g)} (\textbf{B}^{(g)})^{-1} (\X_{k}^{(g)})^{'},\\
	(\hat{r}_{i,j}^{o,(g)})& = \hat{\R}_{A}^{o,(g)} = \left(\frac{\hat{a}_{i,j}^{o,(g)}}{(\hat{a}_{i,i}^{o,(g)}\hat{a}_{j,j}^{o,(g)})^{1/2}}\right),\\ \hat{\theta}_{i,j}^{o,(g)}&=\frac{1}{n_{g}q}\sum\limits_{k=1}^n\sum\limits_{l=1}^q\Big{[}(\X_k^{(g)}(\B^{(g)})^{-1/2})_{i,l}(\X_k^{(g)}(\B^{(g)})^{-1/2})_{j,l}-\hat{a}_{i,j}^{o,(g)}\Big{]}^2,\\
\hat{\vartheta}_{i,j}^{o,(g)} &= \frac{\hat{\theta}_{i,j}^{o,(g)}}{\hat{a}_{i,i}^{o,(g)}\hat{a}_{j,j}^{o,(g)}},
\end{align*}
for $g = 1, 2$ and define oracle test statistics
\begin{align*}
	\hat{M}_{n}^{*,o} = \max\limits_{1\leq i < j \leq p}\frac{(\hat{r}_{i,j}^{(1)} - \hat{r}_{i,j}^{(2)})^{2}}{\hat{\vartheta}_{i,j}^{o,(1)}/(n_{1}q) + \hat{\vartheta}_{i,j}^{o,(2)}/(n_{2}q)},\\
	M_{n}^{*,o} = \max\limits_{1\leq i < j \leq p}\frac{(\hat{r}^{o,(1)}_{i,j} - \hat{r}^{o,(1)}_{i,j})^{2}}{\hat{\vartheta}_{i,j}^{o,(1)}/(n_{1}q) + \hat{\vartheta}_{i,j}^{o,(2)}/(n_{2}q)}.
\end{align*}
Theorem \ref{ther5} is proved because of the following. Under Conditions (C1$^{*}$)-(C5$^{*}$),
\begin{align}
\label{Stwo1}
&P(M_{n}^{*,o} - 4\log p +\log \log p \leq t) \rightarrow \exp\left(-\frac{1}{\sqrt{8\pi}}\exp(-\frac{t}{2})\right),\\
\label{Stwo2}
&|M_{n}^* - \hat{M}_{n}^{*,o}| = o_{p}(1),\\
\label{Stwo3}
&|M_{n}^{*,o} - \hat{M}_{n}^{*,o}| = o_{p}(1),
\end{align}
where (\ref{Stwo1}) is proved by Lemma \ref{lemma6} and (\ref{Stwo2}) and (\ref{Stwo3}) are proved by Lemma \ref{lemma7} in Section \ref{lemma.sec}.~~~~$\blacksquare$

\noindent\textbf{Proof of Theorem \ref{th4}}

From Theorem \ref{ther1}, we have that, uniformly for $\A \in \mathcal{W}(4)$,
\begin{align*}
	P\left(\max\limits_{(i,j)\notin\Psi} M_{i,j} \geq 4\log p\right)\rightarrow 0.
\end{align*}
By (\ref{22}), Lemma \ref{lemma2} and the inequality
\begin{align*}
\min\limits_{(i,j)\in\Psi}M_{i,j}^o \geq \min\limits_{(i,j)\in\Psi}\frac{1}{2}\cdot\frac{a_{i,j}^2}{\hat{\theta}_{i,j}^o/(nq)} - M_{n}^{o,d},
\end{align*}
we can easily get for any constant $c>0$ and uniformly for $\A \in \mathcal{W}(4)$,
\begin{equation*}\label{24}
P\left(\min\limits_{(i,j)\in\Psi}M_{i,j}^o \geq c+ 4\log p\right)\rightarrow 1,
\end{equation*}
where $M_{i,j}^o = \frac{(\hat{a}_{i,j}^o)^{2}}{\hat{\theta}_{i,j}^o/(nq)}$. Similar to the proof of Theorem \ref{power}, we have
\begin{align*}
P\left(\min\limits_{(i,j)\in\Psi}M_{i,j} \geq 4\log p\right)\rightarrow 1.
\end{align*}
which implies the desired results.~~~~$\blacksquare$

\noindent\textbf{Proofs of Theorems \ref{power*}, \ref{optimal*}, and \ref{th8}}

The proofs of Theorems \ref{power*}, \ref{optimal*}, and \ref{th8} are similar to those of Theorems \ref{power}, \ref{optimal}, and \ref{th4}. We skip the proofs here.

\subsection{Lemmas and Their Proofs}
\label{lemma.sec}

Lemma \ref{lemma2} is the large deviation bound for the oracle estimate of variance $\hat\theta_{i,j}^o$, whose proof is given in \citet{cai2013two}.
\begin{lemma} \label{lemma2}
There exists some constant C $>$ 0 such that
\begin{align*}
&P\left(\max \limits_{i,j}|\hat{\theta}_{i,j}^o - \theta_{i,j}|/(a_{ii}a_{jj}) \geq C\frac{\epsilon_{nq}}{\log p}\right) = O(p^{-1}),
\end{align*}
where $\epsilon_{nq}=\max((\log p)^{1/6}/(nq)^{1/2},(\log p)^{-1}) \rightarrow 0$ as $nq$, $p \rightarrow \infty$.
	
\end{lemma}

Lemma \ref{lemma3} is the large deviation bound for the maximum of the oracle estimate of covariance $\hat{a}_{i,j}^o$ in the one-sample case and $\hat{a}^{o,(g)}_{i,j}$ in the two-sample case, whose proof is given in \citet{cai2013two}.
\begin{lemma} \label{lemma3}
Let $\Lambda$ be any subset of $\{(i, j): 1$ $\leq$ i $\leq$ j $\leq$ p\} and $|\Lambda|$ = Card($\Lambda$). We have for some constant C $>$ 0 that
\begin{align*}
	P\left(\max \limits_{(i,j)\in \Lambda}\frac{(\hat{a}_{i,j}^o - a_{i,j})^{2}}{\theta_{i,j}/(nq)} \geq x^{2}\right) &\leq C|\Lambda|(1-\Phi(x)) + O(p^{-1}),\\
	P\left(\max \limits_{(i,j)\in \Lambda}\frac{(\hat{a}^{o,(1)}_{i,j} - \hat{a}^{o,(2)}_{i,j} -  a^{(1)}_{i,j} + a^{(2)}_{i,j})^{2}}{\theta_{i,j}^{(1)}/(n_{1}q) + \theta_{i,j}^{(2)}/(n_{2}q)} \geq x^{2}\right) &\leq C|\Lambda|(1-\Phi(x)) + O(p^{-1}),
\end{align*}
uniformly for 0 $\leq$ x $\leq$ $(8\log p)^{1/2}$ and $\Lambda \subseteq \{(i, j): 1 \leq i \leq j \leq p\}$.
\end{lemma}

\begin{lemma}\label{Hanson}
Let $X = (X_1, \cdots,X_n)^T$ and $ Y = (Y_1,\cdots,Y_n)^T \in \mathbb{R}^n$ be two random vectors with independent components each and they satisfy $\mathbb{E}X_i = \mathbb{E}Y_i = 0$ and $\|X_i\|_{\psi_2} , \|Y_i\|_{\psi_2} \leq K$. Here $\|\cdot\|_{\psi_2}$ refers to the Orlicz norm  with $\psi_2(x) = e^{x^2}-1$, {where the Orlicz norm of a random variable $X$ is defined as $\|X\|_{\psi} = \inf \{M>0: \mathbb{E}\psi(|X|/M) \leq 1\}$.} Let A be and $n\times n$ matrix. Then, for every $t>0$,
\begin{align*}
P(|X^TAY - \mathbb{E}X^TAY| >t) \leq 2\exp\bigg[-c\min\bigg(\frac{t^2}{K^4\|A\|_{F}^2},\frac{t}{K^2\|A\|_2}\bigg)\bigg],
\end{align*}
{for some constant $c>0$.}
\begin{remark}
	We require $\{X_i, i =1,\cdots,n\}$ are independent with each other and also $\{Y_i, i =1,\cdots,n\}$ are independent with each other but we allow the correlations between $X$ and $Y$. Note that if $Y = X$, this is the famous Hanson-Wright inequality.
\end{remark}
\end{lemma}

Lemma \ref{lemma4} is the oracle version of Theorem \ref{ther1}.
\begin{lemma} \label{lemma4}
Suppose that Conditions (C1) and (C2) hold, then under $H_{0}$, for any $t \in \mathbb{R}$,
\begin{equation*} 
	P(M_{n}^o - 4\log p + \log \log p \leq t) \rightarrow \exp\left(-\frac{1}{\sqrt{8\pi}}\exp(-\frac{t}{2})\right),
\end{equation*}
as $nq$, $p \rightarrow \infty$. Furthermore, under $H_{0}$, the convergence above is uniform for all $\{\X_k,k=1,\ldots,n\}$ satisfying (C1) and (C2).
\end{lemma}

Lemma \ref{lemma5} shows the distance between oracle $M_{n}^o$ and the test statistic $M_n$.
\begin{lemma} \label{lemma5}
Uniformly for all $\{\X_k,k=1,\ldots,n\}$ satisfying Conditions (C1)-(C3), we have that, under $H_0$,
\begin{equation}\label{11}
\centering
|M_{n}^o - \hat{M}_n^{o}| = o_{p}(1).
\end{equation}
\vspace{-0.3in}
\begin{equation}\label{10}
\centering
|M_{n} - \hat{M}_n^{o}| = o_{p}(1),
\end{equation}
\end{lemma}

Lemma \ref{lemma6} is the oracle version of Theorem \ref{ther5}.
\begin{lemma} \label{lemma6}
Suppose that Conditions (C1$^{*}$), (C2$^{*}$), (C4$^{*}$) and (C5$^{*}$) hold, then under $H_{0}^*$, for any $t \in\mathbb{R}$,
\begin{equation} \label{conv.lemma6}
P(M_{n}^{*,o} - 4\log p +\log \log p \leq t) \rightarrow \exp\left(-\frac{1}{\sqrt{8\pi}}\exp(-\frac{t}{2})\right),
\end{equation}
as $nq$, $p \rightarrow \infty$. Furthermore, under $H_{0}^*$, the convergence in (\ref{conv.lemma6}) is uniform for all $\{\X^{(1)}_k,k=1,\ldots,n_1\}$ and $\{\X^{(2)}_k,k=1,\ldots,n_2\}$ satisfying (C1$^{*}$), (C2$^{*}$), (C4$^{*}$) and (C5$^{*}$).
\end{lemma}

Lemma \ref{lemma7} shows the distance between oracle $M_{n}^{*,o}$ and the test statistic $M_n^*$.
\begin{lemma} \label{lemma7}
Uniformly for all $\{\X^{(1)}_k,k=1,\ldots,n_1\}$ and $\{\X^{(2)}_k,k=1,\ldots,n_2\}$ satisfying (C1$^{*}$)-(C5$^{*}$), we have that, under $H_0^*$
\begin{equation}\label{S2}
\centering
|M_{n}^{*,o} - \hat{M}_{n}^{*,o}| = o_{p}(1).
\end{equation}
\vspace{-0.3in}
\begin{equation}\label{S1}
\centering
|M_{n}^* - \hat{M}_{n}^{*,o}| = o_{p}(1),
\end{equation}
\end{lemma}

\noindent\textbf{Proof of Lemma \ref{Hanson}}

WLOG, we assume $K=1$. Following the proofs of Theorem 1.1 in \citet{rudelson2013hanson}, we represent
\begin{align*}
X^TAY - \mathbb{E}X^TAY = \sum_{i}a_{ii}(X_iY_i - \mathbb{E}X_iY_i) + \sum_{i,j:i\neq j}a_{ij}X_iY_j.
\end{align*}
Assume
\begin{align*}
&P(X^TAY - \mathbb{E}X^TAY >t)\\ \leq ~ &P(\sum_{i}a_{ii}(X_iY_i - \mathbb{E}X_iY_i) > t/2) + P(\sum_{i,j:i\neq j}a_{ij}X_iY_j > t/2) =: p_1 + p_2.
\end{align*}
Note that $X_iY_i - \mathbb{E}X_iY_i$ are independent mean-zero subexponential random variables, and
\begin{align*}
\|X_iY_i - \mathbb{E}X_iY_i\|_{\psi_1} \leq 2\|X_iY_i\|_{\psi_1} \leq 2\|X_i\|_{\psi_2}\|Y_i\|_{\psi_2} \leq 2,
\end{align*}
{where $\|\cdot\|_{\psi_1}$ refers to the Orlicz norm  with $\psi_1(x) = e^x - 1$.}
Thus, by using a Bernstein-type inequality we can obtain
\begin{align*}
p_1 \leq \exp\bigg[-c\min\bigg(\frac{t^2}{\|A\|_{F}^2},\frac{t}{\|A\|_2}\bigg)\bigg].
\end{align*}
We use the decoupling method to bound the off-diagonal sum
\begin{align*}
S:= \sum_{i,j:i\neq j}a_{ij}X_iY_j.
\end{align*}
Consider independent Bernoulli random variables $\delta_i \in \{0,1\}$ with $\mathbb{E}\delta_i = 1/2$. We have
\begin{align*}
S = 4\mathbb{E}_{\delta}S_{\delta},
\end{align*}
where
\begin{align*}
S_{\delta} = \sum_{i,j}\delta_i(1-\delta_j)a_{ij}X_iY_j.
\end{align*}
Here $\mathbb{E}_\delta$ denotes the expectation with respect to $\delta = (\delta_1,\cdots,\delta_n).$ For any $\lambda >0$, Jensen’s inequality yields
\yin{\begin{align*}
\mathbb{E}_{X,Y}\exp(\lambda S) \leq \mathbb{E}_{X,Y,\delta}\exp(4\lambda S_\delta),
\end{align*}
where $\mathbb{E}_{X,Y,\delta}$ denotes the expectation with respect to $X,Y$ and $\delta$}. Consider the set of indices $\Lambda_\delta = \{i\in[n]:\delta_i=1\}$, we can express
\begin{align*}
S_\delta = \sum_{i\in\Lambda_\delta, j\in\Lambda_\delta^c}a_{ij}X_iY_j = \sum_{j \in \Lambda_\delta^c}Y_j(\sum_{i\in\Lambda_\delta}a_{ij}X_i).
\end{align*}
Now we condition on $\delta$ and $(X_i)_{i\in \Lambda_\delta}$. We can obtain
\begin{align*}
\mathbb{E}_{(Y_j)_{j \in \Lambda_\delta^c}}\exp(4\lambda S_{\delta}) \leq \exp(C\lambda^2\|S_\delta\|_{\psi_2}^2)\leq \exp(C'\lambda^2\sigma_\delta^2),
\end{align*}
where $\sigma_\delta^2:=\sum_{j\in \Lambda_\delta^c}(\sum_{i\in \Lambda_\delta}a_{ij}X_i)^2$. Taking expectations of both sides again, 
we have
\begin{align*}
\yin{\mathbb{E}_{X,Y}\exp(4\lambda S_\delta) \leq \mathbb{E}_{X,Y}\exp(C'\lambda^2\sigma_\delta^2)=:E_\delta,}
\end{align*}
which holds for any fixed $\delta$. It remains to estimate $E_\delta$. The rest of the proof is the same as proof of Theorem 1.1 in \citet{rudelson2013hanson}. We skip the details here.~~~~$\blacksquare$

\noindent\textbf{Proof of Lemma \ref{lemma4}}

Consider
\begin{equation*}
\widetilde{M_{n}}^{o} = \max\limits_{1\leq i<j \leq p}\frac{(\hat{a}^o_{i,j})^{2}}{\theta_{i,j}/(nq)}.
\end{equation*}
From Lemma \ref{lemma2}, we have $|\hat{\theta}_{i,j}^o/\theta_{i,j} - 1| = o_{p}(\frac{1}{\log p})$, which leads to
\begin{align*}
	|M_{n}^o - \widetilde{M_{n}}^{o}| = o_{p}(\frac{1}{\log p})\widetilde{M_{n}}^{o} = o_{p}(1).
\end{align*}
Thus it suffices to prove that for any $t \in \mathbb{R}$,
\begin{align*}
	P(\widetilde{M_{n}}^{o} - 4\log p + \log \log p \leq t) \rightarrow \exp(-\frac{1}{\sqrt{8\pi}}\exp(-\frac{t}{2}))
\end{align*}
Let $\X_{k}\B^{-1/2} = \Z_{k} =:(Z_{k,i,j})$, where the columns of $\Z_{k}$ are independent. We arrange the indices $\{(i, j): 1 \leq i < j \leq p\}$ in any ordering and set them as $\{(i_{m},j_{m}): m = 1,\cdots,s\}$ with $s = p(p-1)/2$. Let $\theta_{m} = \theta_{i_{m}, j_{m}}$ and define $Y_{k,m,l} = Z_{k,i_m,l}Z_{k,j_{m}, l}$ for $1 \leq k \leq n$ and $1 \leq l \leq q$, $V_{m} = (nq\theta_{m})^{-1/2}\sum_{k=1}^{n}\sum_{l=1}^{q}Y_{k,m,l}$, and $\hat{V}_{m} = (nq\theta_{m})^{-1/2}\sum_{k=1}^{n}\sum_{l=1}^{q}\hat{Y}_{k,m,l}$, where $\hat{Y}_{k,m,l} = Y_{k,m,l}I(|Y_{k,m,l}| \leq \tau_{n}) - \mathbb{E}\{Y_{k,m,l}I(|Y_{k,m,l}| \leq \tau_{n})\}$, and $\tau_{n} = 32\log(p + nq)$. Note that $\widetilde{M_{n}}^{o} = \max_{1 \leq m \leq s}V_{m}^{2}$.

The rest of the proof is completely the same as the proof of Theorem 1 in \citet{xia2017hypothesis}. We skip the details here.~~~~$\blacksquare$

\noindent\textbf{Proof of Lemma \ref{lemma5}}

For a matrix $\A$, denote the element-wise infinity norm as $\|\A\|_\infty = \max_{1\leq i,j \leq p}|a_{i,j}|$ and \yin{denote by $\|\A\|_2$ the spectral norm of $\A$}.

We first show
\begin{align*}
\|\tilde{\B} - \B\|_\infty = O_p[\{\log q /np\}^{1/2}].
\end{align*}
Let $\tilde{\B}=:(\tilde{b}_{i,j})$ and $\B =: (b_{i,j})$ and recall we have assumed $\mathbb{E}\tilde{\B} = \B$. Let $\H_k =: \A^{-1/2}\X_{k}$ and $h_i^{(k)}$ is the i-th column of $\H_k$. It is easy to know the components of $h_i^{(k)}$ are independent. Then, we have
\begin{align*}
\tilde{b}_{i,j} = \frac{1}{np}\sum_{k=1}^{n}(h_i^{(k)})^T\A h_j^{(k)}
\end{align*}
Let $h_i^T:=((h_i^{(1)})^T,(h_i^{(2)})^T,\cdots,(h_i^{(n)})^T)$ and \begin{equation*}
\A^{(n)}:=\left(
\begin{array}{cccc}
\A &  & &\\
 & \A & &\\
 & &\cdots \\
 & & &\A
\end{array}
\right),
\end{equation*}
we obtain
\begin{align*}
\tilde{b}_{i,j} = \frac{1}{np}h_i^T\A^{(n)}h_j.
\end{align*}
By Condition (C2) and Lemma \ref{Hanson}, together with the fact that $\|\A^{(n)}\|_F^2 \leq Cnp$ for some constant $C>0$, we have $\|\tilde{\B} - \B\|_\infty = O_p[\{\log q /np\}^{1/2}]$.
\yin{This yields that $\|\tilde{\B}^{-1}-\B^{-1}\|_2 = \|\B^{-1} (\tilde \B - \B) \tilde\B^{-1}\|_2 = O_{p}[q\{\log q/(np)\}^{1/2}]$, which further implies that
\begin{align}\label{intermatrix}
\|\B^{1/2}(\tilde{\B}^{-1}-\B^{-1})\B^{1/2}\|_{\infty} \leq \|\B^{1/2}(\tilde{\B}^{-1}-\B^{-1})\B^{1/2}\|_2 = O_{p}(q\{\log q/(np)\}^{1/2}).
\end{align}
Notice that
\begin{align*}
||\hat{\A} - \hat{\A}^o||_{\infty} &= ||\frac{1}{nq}\hek\X_{k}\B^{-1/2}\B^{1/2}(\tilde{\B}^{-1}-\B^{-1})\B^{1/2}(\X_{k}\B^{-1/2})^{'}||_{\infty}\\
&\leq \max\limits_{1\leq i,j \leq p} ||\B^{1/2}(\tilde{\B}^{-1}-\B^{-1})\B^{1/2}||_{\infty} \frac{1}{nq}\hek\left(\hel Z_{k,i,l}\right)\left(\hel Z_{k,j,l}\right).
\end{align*}
By \eqref{intermatrix} and independence of $Z_{k,i,l}$, we have
\begin{equation*}
||\hat{\A} - \hat{\A}^o||_{\infty} = O_{p}(q \log\max(p,q,n)\{\log q/(n^{2}p)\}^{1/2}).
\end{equation*}
}
Combined with Condition (C3) we can conclude
\begin{align*}
|\sqrt{M_n^o} - \sqrt{\hat{M}_n^{o}}| &= O_{p}(\sqrt{nq})\max\limits_{1\leq i,j \leq p}|\hat{a}_{i,j} - \hat{a}^o_{i,j}| = o_{p}\{(\log p)^{-1/2}\},
\end{align*}
Notice that when Conditions (C1) and (C2) hold, $M_n^o = O_p(\log p)$. Thus we have
\begin{align*}
|M_n^o - \hat{M}_n^{o}| 
= o_{p}(1),
\end{align*}
then (\ref{11}) is obtained.

To prove \eqref{10}, we first show that
\begin{equation}\label{19}
\max\limits_{1\leq i,j \leq p}|\hat{\theta}_{i,j} - \hat{\theta}_{i,j}^o| = O_{p}(\{q^{3} \log q \log \max(p,q,n)/(np)\}^{1/2}).
\end{equation}
Note that
\begin{align*}
\hat{\theta}_{i,j} - \hat{\theta}_{i,j}^o =\\ \frac{1}{nq}\hek\hel&\left[(\X_{k}\tilde{\B}^{-1/2})_{i,l}(\X_{k}(\tilde{\B}^{-1/2}-\B^{-1/2}))_{j,l}+
(\X_{k}(\tilde{\B}^{-1/2}-\B^{-1/2}))_{i,l}(\X_{k}\B^{-1/2})_{j,l}\right]\\
&\left[(\X_{k}\tilde{\B}^{-1/2})_{i,l}(\X_{k}\tilde{\B}^{-1/2})_{j,l} + (\X_{k} \B^{-1/2})_{i,l}(\X_{k} \B^{-1/2})_{j,l} - (\hat{a}_{i,j} + \hat{a}^o_{i,j})\right],
\end{align*}
it suffices to prove that uniformly for $1\leq i,j \leq p$,
\begin{equation}\label{20}
\begin{split}
\frac{1}{nq}\hek\hel&(\X_{k}\tilde{\B}^{-1/2})_{i,l}
\big(\X_{k}(\tilde{\B}^{-1/2}-\B^{-1/2})\big)_{j,l}
\big((\X_{k} \B^{-1/2})_{i,l}(\X_{k} \B^{-1/2})_{j,l} - \hat{a}^o_{i,j}\big)\\
& = O_{p}(\{q^{3} \log q \log \max(p,q,n)/np\}^{1/2}).
\end{split}
\end{equation}
\yin{Since
\begin{equation}
\label{eq:key-proof}
\|\tilde{\B}^{-1} - \B^{-1}\|_2 = O_p(q\{\log q/(np)\}^{1/2}),
\end{equation}
by Lemma 1 of \cite{chen2021normality}, we have that $\|\tilde{\B}^{-1/2} - \B^{-1/2}\|_2 = O_p(q\{\log q/(np)\}^{1/2})$, then we have that
\begin{align*}
	||\X_{k}(\tilde{\B}^{-1/2}-\B^{-1/2})||_{\infty} = O_{p}(\{q^{3} \log q \log \max(p,q,n)/(np)\}^{1/2}),
\end{align*}
and uniformly in $1\leq i,j \leq p$,
\begin{align*}
	\frac{1}{nq}\hek\hel Z_{k,i,l}^{2}Z_{k,j,l} = O_{p}(1),
\end{align*}
where $Z_{k,i,l}=(\Z_k)_{i,l}$ and $\X_{k}\B^{-1/2} = \Z_{k}$.
Thus it is easy to show that
\begin{align*}
	\frac{1}{nq}\hek\hel Z_{k,i,l}^{2}Z_{k,j,l}(\X_{k}(\tilde{\B}^{-1/2}-\B^{-1/2}))_{j,l} = O_{p}(\{q^{3} \log q \log \max(p,q,n)/np\}^{1/2}),
\end{align*}
and hence
\begin{align*} &\frac{1}{nq}\hek\hel(\X_{k}\tilde{\B}^{-1/2})_{i,l}(\X_{k}(\tilde{\B}^{-1/2}-\B^{-1/2}))_{j,l}(\X_{k} \B^{-1/2})_{i,l}(\X_{k} \B^{-1/2})_{j,l} \\
= &O_{p}(\{q^{3} \log q \log\max(p,q,n)/np\}^{1/2}).
\end{align*}
Similarly,
\begin{align*} \frac{1}{nq}\hek\hel(\X_{k}\tilde{\B}^{-1/2})_{i,l}(\X_{k}(\tilde{\B}^{-1/2}-\B^{-1/2}))_{j,l}\hat{a}^o_{i,j} = O_{p}(\{q^{3} \log q \log\max(p,q,n)/np\}^{1/2}).
\end{align*}
Hence (\ref{20}) is proved and in turn (\ref{19}) is proved.

Under Condition (C3), $\{q^{3} \log q \log\max(p,q,n)/np\}^{1/2} = o({1}/{\log p})$, so  $|\hat{\theta}_{i,j}/\hat{\theta}_{i,j}^o - 1| = o_{p}(\frac{1}{\log p})$. By Lemma \ref{lemma4} and \eqref{11}, we have
$|M_{n} - \hat{M}_n^{o}| = o_{p}(\frac{1}{\log p})\hat{M}_n^{o} = o_{p}(1)$. Hence the proof of (\ref{10}) is completed.~~~~$\blacksquare$
}

\noindent\textbf{Proof of Lemma \ref{lemma6}}

WLOG, assume $a_{i,i}^{(g)} = 1$ for $g = 1, 2$ and $i = 1, \cdots, p$.
Let $n  = \max(n_{1}, n_{2})$ and $A = \{(i, j) : 1 \leq i \leq j \leq p\}$. Define
\begin{align*}
	\widetilde{M_{i,j}}^{*,o} = \frac{(\hat{r}_{i,j}^{o,(1)} - \hat{r}_{i,j}^{o,(2)})^{2}}{\vartheta^{(1)}_{i,j}/(n_{1}q) + \vartheta^{(2)}_{i,j}/(n_{2}q)},
\end{align*}
and
\begin{align*}
	\widetilde{M_{n}}^{*,o} = \max\limits_{1\leq i,j \leq p} \widetilde{M_{i,j}}^{*,o}.
\end{align*}
where $\vartheta_{i,j}^{(g)} = \theta_{i,j}^{(g)}/(a_{i,i}^{(g)}a_{j,j}^{(g)})=\theta_{i,j}^{(g)}$. From Lemma \ref{lemma2}, we have $|\hat{\vartheta}_{i,j}^{o,(g)}/\vartheta_{i,j}^{(g)} - 1| = o_{p}(\frac{1}{\log p})$. Thus it suffices to prove that for any $t \in \mathbb{R}$,
\begin{align*}
	P(\widetilde{M_{n}}^{*,o} - 4\log p + \log \log p \leq t) \rightarrow \exp\left(-\frac{1}{\sqrt{8\pi}}\exp(-\frac{t}{2})\right).
\end{align*}
Let $\X_{k}^{(g)}(\B^{(g)})^{-1/2} = \Z_{k}^{(g)}$, where the columns of $\Z_{k}^{(g)}$ are independent for $g = 1, 2$. Define
\begin{align*}
	V_{i,j} = \frac{U^{(1)}_{i,j} - U^{(2)}_{i,j}}{\left(\vartheta_{i,j}^{(1)}/(n_{1}q)+\vartheta_{i,j}^{(2)}/(n_{2}q)\right)^{1/2}},
\end{align*}
where
\begin{align*}
	U^{(g)}_{i,j} = \frac{1}{n_{g}q}\hekg\hel\left(Z_{k,i,l}^{(g)}Z_{k,j,l}^{(g)} - a_{i,j}^{(g)}\right) = \hat{a}_{i,j}^{o,(g)} - a_{i,j}^{(g)},
\end{align*}
for $g = 1, 2$. Note that $\max_{1 \leq i,j\leq p}|\hat{a}_{i,j}^o - a_{i,j}| = O_p(\sqrt{\frac{\log p}{nq}})$ and $O_p(\sqrt{\frac{\log p}{nq}}) \cdot O_p(\sqrt{\frac{\log p}{nq}}) =  o_{p}\{(nq\log p)^{-1/2}\}$ by the assumption (C1$^{*}$), we can obtain
\begin{align}
	\hat{r}_{i,j}^{o,(g)} - r_{i,j}^{(g)} &= \hat{a}_{i,j}^{o,(g)}(\frac{1-\hat{a}_{i,i}^{o,(g)}\hat{a}_{j,j}^{o,(g)}}{\sqrt{\hat{a}_{i,i}^{o,(g)}\hat{a}_{j,j}^{o,(g)}}(1+\sqrt{\hat{a}_{i,i}^{o,(g)}\hat{a}_{j,j}^{o,(g)}})}) + U_{i,j} \\ \label{diff}
	&= \frac{1}{2}a_{i,j}^{(g)}(1-\hat{a}_{i,i}^{o,(g)} + 1-\hat{a}_{j,j}^{o,(g)})+U_{i,j} + o_{p}\{(nq\log p)^{-1/2}\},
\end{align}
uniformly for  $1 \leq i \leq j \leq p$. Thus we have
\begin{equation} \label{twosam}
	\hat{r}_{i,j}^{o,(g)} - r_{i,j}^{(g)} = U^{(g)}_{i,j} + O_{p}\{(\log p/(nq))^{1/2}\}r_{i,j}^{(g)} + o_{p}\{(nq\log p)^{-1/2}\}.
\end{equation}
Note that $|\hat{\vartheta}_{i,j}^{o,(g)} - \vartheta_{i,j}^{(g)}| = o_{p}(\frac{1}{\log p})$ and that for $(i, j) \in A\setminus A_{\gamma}$, we have $|r_{i,j}^{(g)}| = o\{(\log p)^{-1}\}$. Thus from (\ref{twosam}), it is easy to obtain $\max_{(i,j) \in A \setminus A_{\gamma}}|\sqrt{\widetilde{M_{i,j}}^{*,o}} - V_{i,j}| = o_{p}\{(\log p)^{-1/2}\}$. For $(i, j) \in A_{\gamma}$, by \eqref{diff} we have
\begin{align*}
	\sqrt{\widetilde{M_{i,j}}^{*,o}} = V_{i,j} + t_{i,j}+o_{p}\{(\log p)^{-1/2}\} ,
\end{align*}
where $t_{i,j} = \frac{1}{2}a_{i,j}(\hat{a}_{i,i}^{o,(1)} - \hat{a}_{i,i}^{o,(2)}-(\hat{a}_{j,j}^{o,(1)} - \hat{a}_{j,j}^{o,(2)}))/(\vartheta_{i,j}^{(1)}/(n_{1}q) + \vartheta_{i,j}^{(2)}/(n_{2}q))^{1/2}$. Here $a_{i,j} = a_{i,j}^{(1)} = a_{i,j}^{(2)}$ under the null hypothesis $H_0^*$.
By Condition (C4$^*)$, under the null hypothesis $H_0^*$ we can calculate $\vartheta_{i,j}^{(1)} = (2\kappa_1-1)a_{i,j}^2 + \kappa_1$, $\vartheta_{i,j}^{(2)} = (2\kappa_2-1)a_{i,j}^2 + \kappa_2$ and $\vartheta_{i,i}^{(1)} = \vartheta_{j,j}^{(1)} = 3\kappa_1-1$, $\vartheta_{i,i}^{(2)} = \vartheta_{j,j}^{(2)} = 3\kappa_2-1$. Since
\begin{align*}
\frac{(2\kappa_1-1)a_{i,j}^2 + \kappa_1}{a_{i,j}^2} \geq 3\kappa_1 - 1 ~~~\text{and}~~~ \frac{(2\kappa_2-1)a_{i,j}^2 + \kappa_2}{a_{i,j}^2} \geq 3\kappa_2 - 1,
\end{align*}
we can obtain
\begin{align*}
	|t_{i,j}|\leq \frac{1}{2}\left(\frac{|\hat{a}_{i,i}^{o,(1)} - \hat{a}_{i,i}^{o,(2)}|}{(\vartheta_{i,i}^{(1)}/(n_{1}q) + \vartheta_{i,i}^{(2)}/(n_{2}q))^{1/2}} + \frac{|\hat{a}_{j,j}^{o,(1)} - \hat{a}_{j,j}^{o,(2)}|}{(\vartheta_{j,j}^{(1)}/(n_{1}q) + \vartheta_{j,j}^{(2)}/(n_{2}q))^{1/2}}\right) ,
\end{align*}
and this inequality further implies that
\begin{eqnarray*}
	&&P(\max\limits_{(i,j) \in A_{\gamma}}\widetilde{M_{i,j}}^{*,o} \geq 4\log p - \log \log p + t) \\
	&&\quad\leq \text{Card}(A_{\gamma})\{P(V_{i,j}^{2} \geq (2 - 2\nu)\log p) + P(t_{i,j}^{2} \geq (2 - 2\nu)\log p )\} = o(1),
\end{eqnarray*}
where the last equality is a direct result of Lemma \ref{lemma3} and Condition (C5$^*$). Thus it suffices to prove that for any $t \in \mathbb{R}$,
\begin{align*}
	P(\max\limits_{(i,j) \in A \setminus A_{\gamma}}V_{i,j}^{2} - 4\log p + \log \log p \leq t) \rightarrow \exp(-\frac{1}{\sqrt{8\pi}}\exp(-\frac{t}{2}))
\end{align*}
The rest of the proof is essentially proved in \citet{xia2015testing}. We skip the details here.~~~~$\blacksquare$

\noindent\textbf{Proof of Lemma \ref{lemma7}}

Notice that $\max\limits_{1 \leq i,j \leq p}|\hat{r}_{i,j}^{(g)} - \hat{r}^{o,(g)}_{i,j}| = O_p(\max\limits_{1 \leq i,j \leq p}|\hat{a}_{i,j}^{(g)} - \hat{a}^{o,(g)}_{i,j}|)$ for $g = 1, 2$. Similar to the proof of \eqref{11}, we can get \eqref{S2}. On the other hand, $\max\limits_{1 \leq i,j \leq p}|\hat{\vartheta}_{i,j}^{o,(g)} - \hat{\vartheta}_{i,j}^{(g)}| = O_p(\max\limits_{1 \leq i,j \leq p}|\hat{\theta}_{i,j}^{o,(g)} - \hat{\theta}_{i,j}^{(g)}|)$ for $g=1,2$. Thus similar to the proof of \eqref{10}, we can get \eqref{S1}.~~~~$\blacksquare$

\subsection{Additional Simulations}
\label{add.sim}

\dyP{
In this section, we show some additional simulations. Online Supplement \ref{add.sim.normal} provides additional hypothesis test experiment under normal distribution;  Online Supplement \ref{add.sim.t} lists the simulation results under $t$ distribution; Online Supplement \ref{add.sim.supp} is devoted to the simulation results for support recovery; Pseudo simulation based on real data application is given in Online Supplement \ref{add.sim.pseudo}.
}

\subsubsection{Additional Simulation with Normal Distribution}\label{add.sim.normal}


\dyP{In Section \ref{sim1.sec}, we performed the simulation for one-sample test under normal distribution for $p = \{50, 200\},~ q = \{50, 200\}$, and $n = \{10, 50\}$, whose results is given in Table \ref{Tab02}. Table \ref{Tab02-2} offers additional result on the empirical size when sample size is $n=500$ for our covariance matrix based methods. It can be seen that as sample size gets larger, the proposed tests are no longer undersized.}

\begin{table}[]
	\scriptsize
	\centering
	\begin{tabular}{clcc}
		\toprule
		\multicolumn{2}{c}{}&	
		\multicolumn{2}{c}{n = 500}  \\
		
		\cmidrule(r){3-4}
		
		$p$      &  methods
		
		&  $q=50$   &  $q=200$
		
		  \\
		
		\midrule
		
		&One sample cov: oracle &			6.0(0.8)&		5.5(0.7)	\\
		
		50 &One sample cov: sample-est &		5.1(0.7)&		5.1(0.7)		\\
		
		&One sample cov: banded-est &		5.9(0.7)&		4.9(0.7)		\\

		\midrule

		&One sample cov: oracle&			3.7(0.6)&		4.7(0.7)		\\
		
	200	&One sample cov: sample-est&			3.9(0.6)&		5.0(0.7)		\\
		
		&One sample cov: banded-est&		3.8(0.6)&		4.5(0.7)		\\

		\bottomrule
	\end{tabular}
	\caption{\small The empirical size of the \dyP{testing procedures for the one-sample case under normal distribution based on 1000 replications. The percentages are shown with the standard errors provided in parentheses.} The significant level is $\alpha=5\%$. The number of observations and the dimensions of the matrices vary: $p = \{50, 200\},~ q = \{50, 200\}$, and $n = \{500\}$. }
	
	\label{Tab02-2}
	
\end{table}

\dyP{
In Section \ref{sim2.sec} on two-sample test, we generated the data according to $\X_k^{(g)}\sim \mathcal{MN}_{pq}(\0, \A^{(g)},\B^{(g)} )$, where $\B^{(g)}$ are the autocorrelation matrices of AR(1) process. For the target covariance matrix, under the null, we set $\A^{(1)}=\A^{(2)}=\boldsymbol\Sigma^{(1)}$; under the alterative, we set $(\A^{(1)})^{{-1}} = (\boldsymbol{\Sigma}^{(1)} + \delta\I)/(1 + \delta)$ and $(\A^{(2)})^{{-1}} = (\boldsymbol{\Sigma}^{(1)} + \U + \delta\I)/(1 + \delta)$. In this section, we use $\boldsymbol{\Sigma}^{(2)}$, Model 2 of \citet{cai2013two}, to replace $\boldsymbol{\Sigma}^{(1)}$, Model 1 of \citet{cai2013two}. Here, $\S^{(2)}=\D^{1/2}\S^{*(2)}\D^{1/2}$ and $\S^{*(2)}=(\sigma^{*(2)}_{i,j})$, where $\sigma^{*(2)}_{i,j}=0.5^{|i-j|}$ for $1\le i,j\le p$, and $\D=(d_{i,j})$ is a diagonal matrix with diagonal elements $d_{i,i}=\text{Unif}(0.5,2.5)$ for $i=1,...,p$.

Table \ref{normal-two-test-M2} shows the simulation result for $\S^{(2)}$ corresponding to Table \ref{Tabtwopower} for $\S^{(1)}$. The key message is almost identical to \ref{Tabtwopower}. But sizes in Table \ref{normal-two-test-M2} are generally larger that those in Table \ref{Tabtwopower} and closer to 0.05. This demonstrates that the sizes vary depending on the model setting: $\S^{(1)}$ has exact sparsity while $\S^{(2)}$ has weak sparsity with many nonzero entries but small magnitude.
}

\begin{table}[]
	
	\scriptsize
	
	\centering

	\begin{tabular}{clcccc}
		\toprule
		\multicolumn{2}{c}{}&	
		\multicolumn{2}{c}{n = 10} & \multicolumn{2}{c}{n = 50} \\
		
		\cmidrule(r){3-4} \cmidrule(r){5-6}
		
		$p$      &  methods
		
		&  $q=50$   &  $q=200$
		
		&  $q=50$    &  $q=200$  \\
		
		\midrule
		& &  \multicolumn{4}{c}{Empirical size}\\
		\midrule
		
		&Two sample cov: oracle &		3.4(0.6)&		2.9(0.5)&		3.1(0.6)&		5.4(0.7)\\
		
		&Two sample cov: sample-est &	1.3(0.4)&		0.1(0.1)&		2.6(0.5)&		1.2(0.3)\\
		
		&Two sample cov: banded-est &	1.9(0.4)&		2.7(0.5)&		3.3(0.6)&		3.9(0.6)\\
		
		50&Two sample pre: oracle &		3.2(0.6)&		4.5(0.7)&		4.2(0.6)&		3.8(0.6)\\
		
		&Two sample pre: sample-est&		2.6(0.5)&		0.9(0.3)&		4.9(0.7)&		5.1(0.7)\\
		
		&Two sample pre: banded-est&		3.0(0.5)&		5.6(0.7)&		4.6(0.7)&		3.6(0.6)\\
		
		&Two sample vector &			100(0.0)&		100(0.0)&		1000.0)&		100(0.0)\\
		
		\midrule

		&Two sample cov: oracle&		4.4(0.6)&		5.2(0.7)&		6.1(0.8)&		5.3(0.7)\\
		
		&Two sample cov: sample-est&		3.0(0.5)&		0.7(0.3)&		5.4(0.7)&		3.5(0.6)\\
		
		&Two sample cov: banded-est&	3.2(0.6)&		3.2(0.6)&		5.4(0.7)&		4.9(0.7)\\
		
		200&Two sample pre: oracle&		4.5(0.7)&		5.4(0.7)&		5.1(0.7)&		4.3(0.6)\\
		
		&Two sample pre: sample-est &	4.6(0.7)&		4.3(0.6)&		4.5(0.7)&		4.0(0.6)\\
		
		&Two sample pre: banded-est&		4.5(0.7)&		5.0(0.7)&		4.3(0.6)&		5.1(0.7)\\
		
		&Two sample vector  &			100(0.0)&		100(0.0)&		100(0.0)&		100(0.0)\\

		\midrule
		& &  \multicolumn{4}{c}{Empirical power}\\
		\midrule
		& Two sample cov: oracle  &		77.1(1.3)&		76.0(1.3)&		76.4(1.3)&		80.2(1.3)\\
		
		 &Two sample cov: sample-est&	51.7(1.6)&		1.3(0.4)&		71.6(1.4)&		59.3(1.6)\\
		
		&Two sample cov: banded-est&	56.5(1.6)&		41.9(1.6)&		72.5(1.4)&		75.1(1.4)\\
		
		50 &Two sample pre: oracle   &	  	66.8(1.5)&		48.6(1.6)&		47.5(1.6)&		38.4(1.5)\\
		
		&Two sample pre: sample-est&		45.5(1.6)&		1.6(0.4)&		45.1(1.6)&		27.9(1.4)\\

		&Two sample pre: banded-est&		49.5(1.6)&		26.7(1.4)&		45.8(1.6)&		34.9(1.5)\\
			
		&Two sample vector&			100(0.0)&		100(0.0)&		100(0.0)&		100(0.0)\\
		
		\midrule
		
		&Two sample cov: oracle&		100(0.0)&		98.2(0.4)&		96.8(0.6)&		95.2(0.7)\\
		
		&Two sample cov: sample-est&		100(0.0)&		91.0(0.9)&		96.5(0.6)&		93.6(0.8)\\
		
		&Two sample cov: banded-est&	100(0.0)&		95.7(0.6)&		96.3(0.6)&		94.8(0.7)\\
		
		200&Two sample pre: oracle&		100(0.0)&		91.4(0.9)&		89.0(1.0)&		77.0(1.3)\\
		
		&Two sample pre: sample-est&		100(0.0)&		78.4(1.3)&		89.5(1.0)&		73.4(1.4)\\
		
		&Two sample pre: banded-est&		100(0.0)&		87.1(1.1)&		89.4(1.0)&		75.7(1.4)\\
		
		&Two sample vector&			100(0.0)&		100(0.0)&		100(0.0)&		100(0.0)\\

		\bottomrule
	\end{tabular}
	\caption{\small The empirical size and power of the \dyP{testing procedures for the two-sample case under normal distribution based on 1000 replications. The percentages are shown with the standard errors provided in parentheses. The $\S^{(2)}$ matrix adopts the form of Model 2 in \citet{cai2013two}.} The significant level is $\alpha=5\%$. The number of observations and the dimensions of the matrices vary: $p = \{50, 200\},~ q = \{50, 200\}$, and $\dyP{n_1=n_2=n} = \{10, 50\}$. }
	\label{normal-two-test-M2}
	
\end{table}

\subsubsection{Simulation with $t_3$ Distribution}\label{add.sim.t}


\dyP{We perform simulation study under very heavy tail distribution $t_4$ for both the one-sample and the two-sample cases in this section.

For the one-sample case, we simulate the data according to $\X_k = \A^{1/2}\W_k\B^{1/2}$ for $k=1,\ldots,n$. Here, $\W_k, ~k=1,\ldots,n$, are independent random matrices of dimension $p\times q$, where all of the entries are iid with $t_3$ distribution. This way, the covariance matrix adopts the Kronecker product form $\Cov(\vec(\X_k)) = \B\otimes \A$. $\B$ and $\A$ are designed in the same fashion as in Section \ref{sim1.sec}, except when under the alternative, $\A = (\I + \U + \delta\I)/(1 + \delta)$, where the magnitude of the nonzero entries of $\U$ follow uniform distribution on $[3\{\log p/(nq)\}^{1/2}, 5\{\log p/(nq)\}^{1/2}]$. The remaining settings are identical to Section \ref{sim1.sec}. Table \ref{Tab07} is the counterpart of Table \ref{Tab02} when normal distribution is changed to $t$ distribution. It provides the size of the seven relevant methods and power of our methods with significance level $\alpha = 0.05$. It can be seen that the three methods (iv)-(vi), that are proposed by \cite{xia2017hypothesis}, and the vector-based approach (vii), cannot control the sizes while ours (i)-(iii) can. This magnifies the strong dependence of precision matrix based methods on the Gaussian assumption.
}

\begin{table}[]
	\centering
	\begin{tabular}{clcccc}
		\toprule
		\multicolumn{2}{c}{}&	
		\multicolumn{2}{c}{n = 10} & \multicolumn{2}{c}{n = 50} \\
		
		\cmidrule(r){3-4} \cmidrule(r){5-6}
		
		$p$      &  methods
		
		&  $q=50$   &  $q=200$
		
		&  $q=50$    &  $q=200$  \\
		
		\midrule
		& &  \multicolumn{4}{c}{Empirical size}\\
		\midrule
		& One sample cov: oracle  &	0.5(0.2)&	1.6(0.4)&		1.5(0.4)&	3.7(0.6)
		\\
		
		&One sample cov: sample-est&	0.9(0.3)&	0(0)&	1.8(0.4)&	1.8(0.4) \\

		&One sample cov: banded-est&	1.8(0.4)&	1.4(0.4)&		1.9(0.4)&	3.2(0.6)
		\\
		
		50 &One sample pre: oracle	&	51.6(1.6)&	42.9(1.6)&		41.8(1.6)&	31.6(1.5)
		\\
		
		&One sample pre: sample-est&	11.2(1)&	0(0)&		22.0(1.3)&	6.7(0.8) \\
		
		&One sample pre: banded-est&	20(1.3)&	3.6(0.6)&		23.9(1.3)&	10.8(1.0)
		\\

		&One sample vector&	25.8(1.4)&	31.3(1.5)&		27.9(1.4)&	38.1(1.5)\\
		
		\midrule
		
		&One sample cov: oracle&		0.7(0.3)&	1.2(0.3)&		1.7(0.4)&2.6(0.5)
		\\
		
		&One sample cov: sample-est&	0.7(0.3)&	0.4(0.2)&		1.4(0.4) & 2.0(0.4)
		\\
		
		&One sample cov: banded-est&		1.1(0.3)&	1.2(0.3)&		1.7(0.4) & 2.4(0.5)
		\\
		
		200&One sample pre: oracle&		99.6(0.2)&	98.6(0.4)&		98.8(0.3)& 95.5(0.7)
		\\
		
		&One sample pre: sample-est&		96.7(0.6)&	66.6(1.5)&		96.3(0.6)& 80.7(1.2)
		\\
		
		&One sample pre: banded-est&		98.0(0.4)&	83.5(1.2)&		96.9(0.5)& 82.9(1.2)
		\\
		
		&One sample vector&		32.1(1.5)&	45.6(1.6)&		46.9(1.6)& 56.3(1.6)
		\\
		\midrule
		& &  \multicolumn{4}{c}{Empirical power}\\
		\midrule
		& One sample cov: oracle  &	93.1(0.8)&	90.2(0.9)&		90.6(0.9)&	87.8(1.0)
		\\
		
		50&One sample cov: sample-est&	87.2(1.1)&	8.4(0.9)&	89.1(1.0)&	77.1(1.3) \\

		&One sample cov: banded-est&	95.0(0.7)&	83.5(1.2)&		91.7(0.9)&	87.9(1.0)
		\\

		\midrule
		
		&One sample cov: oracle&		93.3(0.8)&	92.0(0.9)&		91.2(0.9)&89.4(1.0)
		\\
		
		200&One sample cov: sample-est&	91.2(0.9)&	77.8(1.3)&		91.2(0.9) & 86.9(1.1)
		\\
		
		&One sample cov: banded-est&		92.8(0.8)&	91.5(0.9)&		91.4(0.9) & 89.4(1.0)
		\\

		\bottomrule

	\end{tabular}
	\caption{The empirical size and power of the \dyP{testing procedures for the one-sample case under $t_3$ distribution based on 1000 replications. The percentages are shown with the standard errors provided in parentheses.} The significant level is $\alpha=5\%$. The number of observations and the dimensions of the matrices vary: $p = \{50, 200\},~ q = \{50, 200\}$, and $n = \{10, 50\}$. }
	
	\label{Tab07}
	
\end{table}

%
%
%
%
%
%
%
%
%
%
%
%
%
%
%
%
%
%

\dyP{
For the two-sample case, we simulate the data accordingly, $\X_k^{(g)} = (\A^{(g)})^{1/2}\W_k^{(g)}(\B^{(g)})^{1/2}$ for $k = 1, \ldots, n_g, ~ g=1,2$. Again, $\W_k^{(g)}$ for $k=1,\ldots,n_g, ~ g=1,2$ are independent random matrices of size $p\times q$, where all of the entries are iid with $t_3$ distribution. $\B^{(g)}$ is generated in the same way as in Section \ref{sim2.sec}. When $\A^{(g)}$ is generated according to Model 1 in Section \ref{sim2.sec}, Table \ref{t-two-test-M1} presents the results, which is the counterpart of Table \ref{Tabtwopower}; When $\A^{(g)}$ is generated according to Model 2 in Section \ref{add.sim.normal}, Table \ref{t-two-test-M2} presents the results, which is the counterpart of Table \ref{normal-two-test-M2}. Both Tables \ref{t-two-test-M1} and \ref{t-two-test-M2} show again that the precision matrix based methods are not valid with sizes much larger than 0.05. Furthermore, comparing these two tables, it is proved again that the relative sizes of our methods depend on the true model settings. Model 1 tends to lead to more undersized results comparing to Model 2.
}


\begin{table}[]
	
	
	\centering

	\begin{tabular}{clcccc}
		\toprule
		\multicolumn{2}{c}{}&	
		\multicolumn{2}{c}{n = 10} & \multicolumn{2}{c}{n = 50} \\
		
		\cmidrule(r){3-4} \cmidrule(r){5-6}
		
		$p$      &  methods
		
		&  $q=50$   &  $q=200$
		
		&  $q=50$    &  $q=200$  \\
		
		\midrule
		& &  \multicolumn{4}{c}{Empirical size}\\
		\midrule
		
		&Two sample cov: oracle &			1.0(0.3)&		2.5(0.5)&		3.4(0.6)&		2.7(0.5)\\
		
		&Two sample cov: sample-est  &		0.5(0.2)&		0.1(0.1)&		2.4(0.5)&		0.6(0.2)\\
		
		&Two sample cov: banded-est  &		1.3(0.4)&		0.8(0.3)&		2.8(0.5)&		1.7(0.4)\\
		
		50&Two sample pre: oracle  &			39.5(1.5)&		32.6(1.5)&		32.0(1.5)&		30(1.4)\\
		
		&Two sample pre: sample-est &		10.5(1.0)&		1.3(0.4)&		17.2(1.2)&		8.4(0.9)\\
		
		&Two sample pre: banded-est &		14.4(1.1)&		8.8(0.9)&		16.0(1.2)&		8.9(0.9)\\
		
		&Two sample vector  &				100(0.0)&		100(0.0)&		100(0.0)&		100(0.0)\\
		
		\midrule

		&Two sample cov: oracle &			2.2(0.5)&		2.0(0.4)&		1.7(0.4)&		3.3(0.6)\\
		
		&Two sample cov: sample-est &		1.5(0.4)&		0.4(0.2)&		1.1(0.3)&		1.4(0.4)\\
		
		&Two sample cov: banded-est &		2.3(0.5)&		1.5(0.4)&		1.3(0.4)&		3.5(0.6)\\
		
		200&Two sample pre: oracle &			94.0(0.8)&		93.4(0.8)&		92.2(0.8)&		85.0(1.1)\\
		
		&Two sample pre: sample-est  &		76.8(1.3)&		45.8(1.6)&		82.7(1.2)&		58.0(1.6)\\
		
		&Two sample pre: banded-est &		76.6(1.3)&		53.6(1.6)&		83.2(1.2)&		58.9(1.6)\\
		
		&Two sample vector   &				100(0.0)&		100(0.0)&		100(0.0)&		100(0.0)\\

		\midrule
		& &  \multicolumn{4}{c}{Empirical power}\\
		\midrule
		& Two sample cov: oracle  &			52.7(1.6)&		65.4(1.5)&		63.7(1.5)&		68.2(1.5)\\
		
		50 &Two sample cov: sample-est &		36.9(1.5)&		1.3(0.4)&		60.4(1.5)&		52.2(1.6)\\
		
		&Two sample cov: banded-est &		41.2(1.6)&		38.9(1.5)&		60.6(1.5)&		63.2(1.5)\\
		
%
%
%
		
		\midrule
		
		&Two sample cov: oracle &			83.6(1.2)&		83.9(1.2)&		84.2(1.2)&		85.8(1.1)\\
		
		200 &Two sample cov: sample-est &		84.0(1.2)&		75.7(1.4)&		84.9(1.1)&		86.6(1.1)\\
		
		&Two sample cov: banded-est &		84.4(1.1)&		83.6(1.2)&		84.8(1.1)&		87,3(1.1)\\
		
%
%
%

		\bottomrule
	\end{tabular}
	\caption{The empirical size and power of the \dyP{testing procedures for the two-sample case under $t_3$ distribution based on 1000 replications. The percentages are shown with the standard errors provided in parentheses. The $\S^{(1)}$ matrix adopts the form of Model 1 in \citet{cai2013two}.} The significant level is $\alpha=5\%$. The number of observations and the dimensions of the matrices vary: $p = \{50, 200\},~ q = \{50, 200\}$, and $\dyP{n_1=n_2=n} = \{10, 50\}$. }
	\label{t-two-test-M1}
	
\end{table}

\begin{table}[]
	\centering
	\begin{tabular}{clcccc}
		\toprule
		\multicolumn{2}{c}{}&	
		\multicolumn{2}{c}{n = 10} & \multicolumn{2}{c}{n = 50} \\
		
		\cmidrule(r){3-4} \cmidrule(r){5-6}
		
		$p$      &  methods
		
		&  $q=50$   &  $q=200$
		
		&  $q=50$    &  $q=200$  \\
		
		\midrule
		& &  \multicolumn{4}{c}{Empirical size}\\
		\midrule
		& Two sample cov: oracle  &	1.6(0.4)&	2.4(0.5)&		3.8(0.6)&	3.5(0.6)
		\\
		
		&Two sample cov: sample-est&	0.4(0.2)&	0(0)&	2.4(0.5)&	1.6(0.4) \\

		&Two sample cov: banded-est&	1.7(0.4)&	1.8(0.4)&	3.2(0.6)&	3.0(0.5)
		\\
		
		50 &Two sample pre: oracle	&	34.3(1.5)&	32.9(1.5)&		29.4(1.4)&	30.2(1.5)
		\\
		
		&Two sample pre: sample-est&	9.0(0.9)&	1.7(0.4)&	13.3(1.1)&	8.0(0.9) \\
		
		&Two sample pre: banded-est&	11.3(1.0)&	7.0(0.8)&	14.2(1.1)&	8.4(0.9)
		\\

		&Two sample vector&	100(0)&	100(0)&		100(0)&	100(0)\\
		
		\midrule
		
		&Two sample cov: oracle&		2.0(0.4)&	2.5(0.5)&		2.8(0.5)&3.5(0.6)
		\\
		
		&Two sample cov: sample-est&	1.0(0.3)&	0.8(0.3)&	2.2(0.5) & 2.7(0.5)
		\\
		
		&Two sample cov: banded-est&		1.9(0.4)&	1.8(0.4)&	2.4(0.5) & 3.5(0.6)
		\\
		
		200&Two sample pre: oracle&		94.6(0.7)&	92.7(0.8)&		91.1(0.9)& 86.9(1.1)
		\\
		
		&Two sample pre: sample-est&		75.0(1.4)&	46.0(1.6)&	81.1(0.2)& 55.1(1.6)
		\\
		
		&Two sample pre: banded-est&		76.4(1.3)&	53.5(1.6)&	81.5(1.2)& 55.8(1.6)
		\\
		
		&Two sample vector&		100(0)&	100(0)&		100(0)& 100(0)
		\\
		
		\midrule
		& &  \multicolumn{4}{c}{Empirical power}\\
		\midrule
		& Two sample cov: oracle  &	93.9(0.8)&	98.2(0.4)&		98.6(0.4)&	99.3(0.3)
		\\
		
		50&Two sample cov: sample-est&	87.7(1.0)&	11.7(1.0)&	98.5(0.4)&	97.9(0.5) \\

		&Two sample cov: banded-est&	90.5(0.9)&	89.4(1.0)&		98.7(0.4)&	99.3(0.3)
		\\

		\midrule
		
		&Two sample cov: oracle&		99.7(0.2)&	98.4(0.4)&		98.5(0.4)&99.2(0.3)
		\\
		
		200&Two sample cov: sample-est&	99.2(0.3)&	98.2(0.4)&		98.6(0.4) & 99.3(0.3)
		\\
		
		&Two sample cov: banded-est&		99.2(0.3)&	99.0(0.3)&	98.5(0.4) & 99.7(0.2)
		\\

		\bottomrule
	\end{tabular}
	\caption{The empirical size and power of the \dyP{testing procedures for the two-sample case under $t_3$ distribution based on 1000 replications. The percentages are shown with the standard errors provided in parentheses. The $\S^{(2)}$ matrix adopts the form of Model 2 in \citet{cai2013two}.} The significant level is $\alpha=5\%$. The number of observations and the dimensions of the matrices vary: $p = \{50, 200\},~ q = \{50, 200\}$, and $\dyP{n_1=n_2=n} = \{10, 50\}$. }
	\label{t-two-test-M2}
	
\end{table}

\subsubsection{Simulation for Support Recovery}\label{add.sim.supp}

\dyP{In this section, we perform simulation experiments on support recovery for both one-sample and two-sample cases under both normal and $t_3$ distributions. We only demonstrate the performance of our methods (i)-(iii) and the vector-based method (vii) in \citet{cai2013two}. The precision matrix based methods (iv)-(vi) are excluded. This is because the support recovered by them is the locations of the nonzero partial correlations, not the nonzero correlations; hence they are not comparable with ours.

Following \citet{cai2013two}, the accuracy of the support recovery is evaluated by the similarity measure $s(\hat{\Psi},\Psi)$, defined as
\[
s(\hat{\Psi}, \Psi) = \frac{|\hat{\Psi}\cap \Psi|}{\sqrt{|\hat{\Psi}||\Psi|}},
\]
where $\Psi$ is the true support,
$\hat{\Psi}$ is the estimated support, and $|\cdot|$ denotes the cardinality. Note that the similarity measure takes value between zero and one:  $s(\hat{\Psi}, \Psi)=1$ indicates perfect recovery, and $s(\hat{\Psi}, \Psi)=0$ implies complete failure. }

\noindent {\bf One-sample support recovery.}
\dyP{Under normal distribution, we generate the data with the same model as in the one-sample global testing in Section \ref{sim1.sec} except that the covariance matrix $\B$ is the autocorrelation matrix of AR(1) process with coefficient $0.8$ and the symmetric perturbation matrix $\U$ has 50 random nonzero entries of magnitude {$5\{\log p/(nq)\}^{1/2}$}. Under $t_3$ distribution, the same setting of $\A$ and $\B$ is implemented, and the data are produced with $\X_k = \A^{1/2}\W_k\B^{1/2}$ for $k=1,\ldots,n$, where $\W_k$ are iid random matrices with iid entries of $t_3$ distributions.

Tables \ref{Tabonesupp} and \ref{Tabonesupp_tdist} are for the normal and $t_3$ distributions respectively. The means and standard errors of $s(\hat{\Psi}, \Psi)$ for the four methods based on 100 replications are given. Our methods (i)-(iii) have similarity measures that are close to one while the vector-based method (vii) cannot recover the support well. If we further increase the signal strength $5\{\log p/(nq)\}^{1/2}$ to some larger value, our power will be even closer to one.
}

\begin{table}[]
	
	
	\centering

	\begin{tabular}{clcccc}
		\toprule

		\multicolumn{2}{c}{}&	\multicolumn{2}{c}{n = 10} &
		\multicolumn{2}{c}{n = 50}   \\
		
		\cmidrule(r){3-4} \cmidrule(r){5-6}
		
		p      &  methods   &  $q=50$ & $q=200$ &$q=50$ & $q=200$ \\
		
		\midrule

		50&One sample cov: oracle& 99.8(0.1)& 99.5(0.1)& 	99.6(0.1)&	96.6(0.3) \\
		&One sample cov: sample-est& 99.9(0)&70.8(0.8)&	99.5(0.1)&	93.6(0.4)\\
		&One sample cov: banded-est& 99.2(0.1)&98.7(0.2)&98.6(0.1)&	94.7(0.3)\\
		
		\vspace{2ex}
		
		&One sample vector & 51.5(0.3)&45.0(0.3)&45.6(0.3)&	40.0(0.4)\\
		
		200&	One sample cov: oracle&99.9(0)&99.8(0.1)& 98.8(0.1)&	97.7(0.2)\\
		&One sample cov: sample-est&99.9(0) &99.6(0.1)&98.8(0.1)&	97.2(0.2)\\
		&One sample cov: banded-est&99.1(0.1) &98.7(0.1)&97.1(0.2)&	95.2(0.4)\\
		&One sample vector  &21.4(0.1) &18.7(0.1)&17.8(0.1)&	16.0(0.2)\\

		\bottomrule
	\end{tabular}
	\caption{The support recovery performance of our methods (i)-(iii) and the vector-based method (vii) \dyP{for the one-sample case under normal distribution based on 100 replications. The similarity measure is} shown in percentage and the standard errors are provided in parentheses. The number of observations and the dimensions of the matrices vary: $p = \{50, 200\},~ q = \{50, 200\}$, and $n = \{10, 50\}$. }
	\label{Tabonesupp}
	
\end{table}

\begin{table}[]
	
	
	\centering

	\begin{tabular}{clcccc}
		\toprule

		\multicolumn{2}{c}{}&	\multicolumn{2}{c}{n = 10} &
		\multicolumn{2}{c}{n = 50}   \\
		
		\cmidrule(r){3-4} \cmidrule(r){5-6}
		
		p      &  methods   &  $q=50$ & $q=200$ &$q=50$ & $q=200$ \\
		
		\midrule

		50 & One sample cov: oracle & 93.1(0.4) & 95.3(0.4) & 95.6(0.3) & 93.5(0.4) \\
		 & One sample cov: sample-est & 97(0.3) & 67.2(0.8) & 97.1(0.3) & 92(0.4) \\
		 & One sample cov: banded-est & 97.5(0.3) & 98(0.2) & 96.4(0.3) & 93.3(0.3) \\
		 & One sample vector & 51.2(0.3) & 44.4(0.4) & 45.2(0.4) & 38.7(0.5) \\
		
		\\

		200 & One sample cov: oracle & 89.9(0.6) & 93.9(0.4) & 94.6(0.4) & 94.3(0.4) \\
		 & One sample cov: sample-est & 91(0.5) & 94.7(0.3) & 95.1(0.3) & 94.3(0.4) \\
		 & One sample cov: banded-est & 91.3(0.5) & 95.8(0.3) & 94.4(0.4) & 93.9(0.4) \\
		 & One sample vector & 21.1(0.1) & 18.2(0.2) & 18.5(0.2) & 15.7(0.2) \\

		\bottomrule
	\end{tabular}
	\caption{The support recovery performance of our methods (i)-(iii) and the vector-based method (vii) \dyP{for the one-sample case under $t_3$ distribution based on 100 replications. The similarity measure is} shown in percentage and the standard errors are provided in parentheses. The number of observations and the dimensions of the matrices vary: $p = \{50, 200\},~ q = \{50, 200\}$, and $n = \{10, 50\}$. }
	\label{Tabonesupp_tdist}
	
\end{table}

\noindent {\bf Two-sample support recovery.} The data are simulated according to the same $\S^{(1)}$ and $\delta$ as in the two-sample global testing
\dyP{
in Section \ref{sim2.sec}. However, in this case $\A^{(1)} = (\boldsymbol{\Sigma}^{(1)} + \delta\I)/(1 + \delta)$ and $\A^{(2)} = (\boldsymbol{\Sigma}^{(1)} + \U + \delta\I)/(1 + \delta)$, where the perturbation matrix $\U$ has 50 nonzero entries with magnitude {$5\{\log p/(nq)\}^{1/2}$.}
Tables \ref{Tabtwosupp} and \ref{Tabtwosupp_tdist}
provide the means and the standard errors of the same measurement $s(\hat{\Psi}, \Psi)$ over 100 replications for normal and $t_3$ distributions respectively.
Again, they demonstrate the advantage of preserving the matrix structure over the simple vectorization.
}

\begin{table}[]
	
	
	\centering

	\begin{tabular}{clcccc}
		\toprule
		\multicolumn{2}{c}{}&	
		\multicolumn{2}{c}{n = 10} & \multicolumn{2}{c}{n = 50} \\
		
		\cmidrule(r){3-4} \cmidrule(r){5-6}
		
		$p$      &  methods
		
		&  $q=50$   &  $q=200$
		
		&  $q=50$    &  $q=200$  \\
		
		\midrule

		50 & Two sample cov: oracle & 91.0(0.5) & 99.3(0.1) & 98.7(0.2) & 93.2(0.3) \\
		& Two sample cov: sample-est & 90.1(0.4) & 94.6(0.3) & 98.6(0.2) & 91.3(0.4) \\
		& Two sample cov: banded-est & 89.5(0.5) & 98.2(0.2) & 98.6(0.2) & 91.4(0.3) \\
		& Two sample vector & 27.4(0.4) & 30.6(0.4) & 31.1(0.3) & 25.9(0.4) \\
		
		\\
		
		200 & Two sample cov: oracle & 89.8(0.4) & 99.7(0.1) & 99.5(0.1) & 95.8(0.3) \\
		& Two sample cov: sample-est & 89.4(0.4) & 99.9(0.1) & 99.5(0.1) & 95.5(0.3) \\
		& Two sample cov: banded-est & 89.4(0.4) & 99.7(0.1) & 99.4(0.1) & 95.4(0.3) \\
		& Two sample vector & 8.7(0.1) & 11(0.1) & 11.3(0.1) & 8.7(0.1) \\

		\bottomrule
	\end{tabular}
	\caption{The support recovery performance of our methods (i)-(iii) and the vector-based method (vii) \dyP{for the two-sample case under normal distribution based on 100 replications. The $\Sigma^{(1)}$ matrix adopts the form of Model 1 in \citet{cai2013two}. The similarity measure is} shown in percentage and the standard errors are provided in parentheses. The number of observations and the dimensions of the matrices vary: $p = \{50, 200\},~ q = \{50, 200\}$, and $\dyP{n_1=n_2=n} = \{10, 50\}$. }
	\label{Tabtwosupp}
	
\end{table}


\begin{table}[]
	
	
	\centering

	\begin{tabular}{clcccc}
		\toprule
		\multicolumn{2}{c}{}&	
		\multicolumn{2}{c}{n = 10} & \multicolumn{2}{c}{n = 50} \\
		
		\cmidrule(r){3-4} \cmidrule(r){5-6}
		
		$p$      &  methods
		
		&  $q=50$   &  $q=200$
		
		&  $q=50$    &  $q=200$  \\
		
		\midrule

		50 & Two sample cov: oracle & 70.4(0.9) & 90.3(0.5) & 90.1(0.5) & 87.3(0.4) \\
		& Two sample cov: sample-est & 74.3(0.7) & 91(0.4) & 92.6(0.3) & 87.3(0.4) \\
		& Two sample cov: banded-est & 74.1(0.8) & 94.1(0.2) & 92.5(0.3) & 88.5(0.4) \\
		& Two sample vector & 24.4(0.4) & 29.9(0.3) & 29.8(0.3) & 25.3(0.4) \\
		
		\\
		
		200 & Two sample cov: oracle & 62.5(0.7) & 89.4(0.4) & 88.9(0.4) & 88(0.4) \\
		& Two sample cov: sample-est & 63.1(0.7) & 93.1(0.4) & 89.6(0.4) & 88.7(0.4) \\
		& Two sample cov: banded-est & 63.2(0.7) & 92.2(0.4) & 89.6(0.4) & 88.7(0.4) \\
		& Two sample vector & 7.8(0.1) & 10.3(0.1) & 10.2(0.1) & 8.4(0.1) \\

		\bottomrule
	\end{tabular}
	\caption{The support recovery performance of our methods (i)-(iii) and the vector-based method (vii) \dyP{for the two-sample case under $t_3$ distribution based on 100 replications. The $\Sigma^{(1)}$ matrix adopts the form of Model 1 in \citet{cai2013two}. The similarity measure is} shown in percentage and the standard errors are provided in parentheses. The number of observations and the dimensions of the matrices vary: $p = \{50, 200\},~ q = \{50, 200\}$, and $\dyP{n_1=n_2=n} = \{10, 50\}$. }
	\label{Tabtwosupp_tdist}
	
\end{table}

\dyP{
Note that in the experiments above, we set the signal strength, the magnitude of the support, to be $5\{\log p/(nq)\}^{1/2}$, which varies with the sample size $n$ and dimensions $p,~q$. Such a setting was chosen because the theorems in Online Supplement \ref{sec-support} imply the smallest signal over the support should be of rate $\{\log p/(nq)\}^{1/2}$. With this setting, the similarity measures in Tables \ref{Tabonesupp}-\ref{Tabtwosupp_tdist} do not exhibit the features such that they increase with $n$ and $q$ and decrease with $p$.
To make our intuitions right, Table \ref{Tabonesupp_fix} offers the results of support recovery of the one sample case under normal distribution, when the signal strength does not vary with $n,p,q$ and stays fixed at level 0.12. It is clearly seen that when $n$ or $q$ increases, similarity measure increases, and when $p$ increases, similarity measure decreases.
}

\begin{table}[]
	
	
	\centering

	\begin{tabular}{clcccc}
		\toprule

		\multicolumn{2}{c}{}&	\multicolumn{2}{c}{n = 10} &
		\multicolumn{2}{c}{n = 50}   \\
		
		\cmidrule(r){3-4} \cmidrule(r){5-6}
		
		p      &  methods   &  $q=50$ & $q=200$ &$q=50$ & $q=200$ \\
		
		\midrule

		50 & One sample cov: oracle & 11.8(1) & 38.9(0.9) & 51.3(0.9) & 99.6(0.1) \\
		 & One sample cov: sample-est & 7.6(1.2) & 20(0) & 49.3(0.8) & 99.4(0.1) \\
		 & One sample cov: banded-est & 9.8(1) & 39.6(0.9) & 52.9(0.9) & 99(0.1) \\
		 & One sample vector & 6.6(0.4) & 17.9(0.5) & 22.1(0.5) & 46.6(0.3) \\

		200 & One sample cov: oracle & 6(1) & 23.7(0.8) & 29.3(0.9) & 98.6(0.2) \\
		 & One sample cov: sample-est & 8.6(1.1) & 20.3(0.7) & 29.5(0.8) & 98.3(0.2) \\
		 & One sample cov: banded-est & 2.6(0.6) & 21(1.1) & 32.4(1) & 96.7(0.3) \\
		 & One sample vector & 1.4(0.1) & 4.3(0.2) & 5.8(0.2) & 17(0.1) \\

		\bottomrule
	\end{tabular}
	\caption{The support recovery performance of our methods (i)-(iii) and the vector-based method (vii) \dyP{for the one-sample case under normal distribution {\bf with fixed-level signal} based on 100 replications. The similarity measure is} shown in percentage and the standard errors are provided in parentheses. The number of observations and the dimensions of the matrices vary: $p = \{50, 200\},~ q = \{50, 200\}$, and $n = \{10, 50\}$. }
	\label{Tabonesupp_fix}
	
\end{table}

\subsubsection{Pseudo Simulation Based on Real Data Application}\label{add.sim.pseudo}

\dyP{
In this section, {two} pseudo experiments are performed with information from the real data.

Experiment 1: We use the sample sizes and dimensions from the real data to mimic the real world while keeping the model parameters $\A^{(1)},~\A^{(2)}, ~ \B^{(1)},~\B^{(2)}$ the same as in Section \ref{sim2.sec} under $t_3$ distribution.

Table \ref{pseudo} provides the results for Experiment 1. The precision matrix based methods are invalid due to larger empirical sizes than nominal level. For our methods, ``two sample cov: oracle'' and ``two sample: banded-est'' have sizes close to 0.05 under the null and remain powerful under the alternative; ``two sample cov: sample-est'' is severely undersized under the null and not quite powerful under the alternative because with large $q\approx 200$ and small $n\approx 20, p\approx 30$, the sample estimate of $\B\in\R^{200\times 200}$ is not very accurate. Since $\B^{(1)},~\B^{(2)}$ correspond to the covariance matrices of autoregressive time series, they have parsimonious banded structure, which makes banded estimation of $\B$ much more accurate and hence ``two sample: banded-est'' is preferred over ``two sample cov: sample-est'' when $\B$ corresponds to the temporal dimension in practice.
}

\begin{table}[ht]
	\centering
	\begin{tabular}{rlll}
		\hline
		& Method & Empirical size & Empirical power \\
		\hline
		 & Two sample cov: oracle & 2.4(0.5) & 91.3(0.9) \\
		 & Two sample cov: sample-est & 0(0) & 14.5(1.1) \\
		 & Two sample cov: banded-est & 5.4(0.7) & 74.8(1.4) \\
		 & Two sample pre: oracle & 19.9(1.3) & 75.3(1.4) \\
		 & Two sample pre: sample-est & 2.9(0.5) & 10.8(1.0) \\
		 & Two sample pre: banded-est & 6.4(0.8) & 49.5(1.6) \\
		 & Two sample vector & 100(0) & 100(0) \\
		\hline
	\end{tabular}
	\caption{The empirical size and power of the \dyP{testing procedures for the two-sample case under $t_3$ distribution based on 1000 replications. The percentages are shown with the standard errors provided in parentheses. The $\Sigma^{(1)}$ matrix adopts the form of Model 1 in \citet{cai2013two}.} The significant level is $\alpha=5\%$. The number of observations and the dimensions of the matrices are approximate to those of real data, that is $p=30, q=200, \dyP{n_1=n_2=n}=20$. }
	\label{pseudo}
\end{table}


\dyP{
Experiment 2. We estimate $\A^{(1)},~\A^{(2)}, ~ \B^{(1)},~\B^{(2)}$ from the real data, simulate data with these parameters under $t_3$ distribution in the same fashion as in Online Supplement \ref{add.sim.t}, and implement all seven methods for the two-sample hypothesis testing. All seven methods have power one in this case.
}

\subsection{Key References for Table 1}
\label{table.sec}

Table \ref{literature2} provides the key references for Table 1.

\begin{table}
  \begin{center}
\begin{tabular}{|l|l|l|l|}
\hline
          &          & Vector-valued data & Matrix-valued data  \\\hline
Covariance or&One-sample&\cite{chen2010tests}& This article\\
correlation &&\cite{cai2011limiting}&\\
matrix &&\cite{zheng2019test}&\\\cline{2-4}
&Two-sample&\cite{schott2007test}& This article\\
&&\cite{srivastava2010testing}& \\
&&\cite{li2012two}&\\
&&\cite{cai2013two}&\\
&&\cite{cai2013optimal}&\\
&&\cite{cai2016inference}&\\
&&\cite{chang2017comparing}&\\
&&\cite{zheng2019test}&\\
\hline
{Precision } &One-sample&\cite{liu2013gaussian}&\cite{narayan2016mixed}\\
matrix&&\cite{guo2020specification}&\cite{xia2017hypothesis}\\
&&&\cite{chen2019graph}\\\cline{2-4}
          &Two-sample&\cite{xia2015testing}&\cite{xia2018matrix}\\\hline
\end{tabular}
    \caption{Summary of the literature on the hypothesis testing for both vector-valued and matrix-valued data under one-sample and two-sample regimes.}
  \label{literature2}
  \end{center}
\end{table}
\setlength{\bibsep}{0pt}
\bibliographystyle{apalike}
\bibliography{ref}

\end{document}